\documentclass[12pt]{article}
\usepackage{jheppub} 
\pdfoutput=1 
\usepackage{amsfonts,amsmath,amssymb,epsf,physics}
\usepackage{amsmath,braket}
\usepackage{subcaption}
\usepackage{tikz}
\usepackage{circuitikz}
\tikzset{snake it/.style={decorate, decoration=snake}}
\usetikzlibrary{decorations.pathmorphing}
\usepackage{tikz-feynman}

\usepackage{float} 
\tikzset{snake it/.style={decorate, decoration=snake}}

\graphicspath{{figures/} }
\usepackage{graphicx,hyperref,color}
\hypersetup{
    bookmarks=true,         
    unicode=false,          
    pdftoolbar=true,        
    pdfmenubar=true,        
    linktocpage=true,       
    pdffitwindow=false,     
    pdfstartview={FitH},    
    pdfnewwindow=true,      
    colorlinks=true,       
    linkcolor=cyan,          
    citecolor=cyan,        
    filecolor=cyan,      
    urlcolor=cyan           
}
\usepackage{tensor}     
\usepackage{romannum}
\usepackage[dvipsnames]{xcolor}

\numberwithin{equation}{section}

\usepackage{comment}

\newcommand{\be}{\begin{equation}}
	\newcommand{\ba}{\begin{eqnarray}}
		\newcommand{\ea}{\end{eqnarray}}
	\newcommand{\ee}{\end{equation}}

\newcommand{\bea}{\begin{eqnarray}}
	\newcommand{\eea}{\end{eqnarray}}
\newcommand{\bes}{\begin{equation*}}
	\newcommand{\beas}{\begin{eqnarray*}}
		\newcommand{\eeas}{\end{eqnarray*}}
	\newcommand{\bas}{\begin{array*}}
		\newcommand{\eas}{\end{array*}}
	\newcommand{\ees}{\end{equation*}}

\newcommand{\ep}{\epsilon}

      \let\S=\Sigma

\usepackage{amsthm}
\theoremstyle{plain}
\newtheorem{thm}{Theorem}
\newtheorem*{thm*}{Theorem}

\theoremstyle{definition}
\newtheorem{dfn}{Definition}

\theoremstyle{plain}
\theoremstyle{definition}

\theoremstyle{proposition}
\newtheorem{prop}[thm]{Proposition}

\numberwithin{equation}{section}
\theoremstyle{lemma}

\theoremstyle{assumption}
\newtheorem{ass}{Condition}

\theoremstyle{plain}
\newtheorem{conj}[thm]{Conjecture}

\newcommand{\ie}{{\it i.e.,}\ }

\newcommand{\mt}[1]{\textrm{\tiny #1}}

\newcommand{\LAdS}{L_{\mathrm{AdS}}}

\renewcommand{\(}{\left(}

\newcommand{\floor}[1]{\left\lfloor{#1 }\right\rfloor}

\newcommand{\GN}{G_\mt{N}}

\newcommand{\KBb}[2]{{K_{\mathrm{Bb}}\qty(#1,#2)}}

\newcommand{\mA}{\mathcal{A}}
\newcommand{\mB}{\mathcal{B}}

\newcommand{\mH}{\mathcal{H}}

\newcommand{\const}{\text{const.}}

\newcommand{\Comb}[2]{ {{}_{#1}C_{#2}}}

\newcommand{\mO}{\mathcal{O}}
\newcommand{\eneshell}{\mathcal{H}_{E,\Delta E}}
\newcommand{\shelldim}{D_{E,\Delta E}}
\newcommand{\mEeq}{\mathcal{E}_{E,\Delta E}}
\newcommand{\rhomc}{\rho_{\mathrm{mc}}^{(E,\Delta E)}}
\newcommand{\rhocan}{{\rho_{\mathrm{can}}^{(\beta)}}}
\newcommand{\DeltaETH}[2]{\Delta_{\mathrm{diag,ETH}}^{#1}\qty[#2]}
\newcommand{\offdiagset}{{\mathcal{E}'_{E,\Delta E}}}
\newcommand{\offDeltaETH}[2]{\Delta_{\mathrm{off,ETH}}^{#1}\qty[#2]}
\newcommand{\LandauO}[1]{O\qty(#1)}

\newcommand{\eqset}{\mathcal{E}_{\rm{eq}}}

\newcommand{\overlineT}[1]{\overline{#1}^T}
\newcommand{\Prob}[2]{\mathrm{Prob}_{#1}\qty[#2]}
\newcommand{\mcev}[1]{\ev{#1}_{\rm{mc}}^{E,\Delta E}}
\newcommand{\canev}[1]{\ev{#1}_{\rm{can}}^{\beta}}
\newcommand{\Deff}{D_{\rm{eff}}}

\newcommand{\ProbE}[1]{\Prob{\mEeq}{#1}}
\newcommand{\ProboffE}[1]{\Prob{\offdiagset}{#1}}
\newcommand{\EVdiag}[1]{\mathbb{E}_{\mEeq}\qty[{#1}]}
\newcommand{\EVoffdiag}[1]{\mathbb{E}_{\offdiagset}\qty[{#1}]}
\newcommand{\mbO}{\mathbb{O}}
\newcommand{\mbOmc}{\mathbb{O}_{\mathrm{mc}}}
\newcommand{\mbOcan}{\mathbb{O}_{\mathrm{can}}}
\newcommand{\opnormTFD}[1]{\opnorm{#1}^{\mathrm{TFD}}}
\newcommand{\mcWightman}[1]
{G_{{\mathrm{mc},E}}^{(#1)}}
\newcommand{\canWightman}[1]{G_{{\mathrm{can},\beta}}^{(#1)}}
\newcommand{\GE}[2]{G_{\mathrm{E,\beta}}^{(#1)}\qty(#2)}
\newcommand{\tE}{{t_{\mathrm{E}}}}
\newcommand{\xE}{{x_{\mathrm{E}}}}
\newcommand{\XE}{{X_{\mathrm{E}}}}
\newcommand{\Smc}{S_{\mathrm{mc}}}
\newcommand{\Pshelldim}{D_{E,\Delta E}^P}
\newcommand{\Pmcev}[1]{\ev{#1}_{P,\rm{mc}}^{E,\Delta E}}
\newcommand{\DeltaETHP}[2]{\Delta_{P,\mathrm{diag,ETH}}^{#1}\qty[#2]}
\newcommand{\Peqset}{\mathcal{E}^{P}_{E,\Delta E}}

\definecolor{shadecolor}{gray}{0.85}
\newcommand{\grethree}{\mathrm{I}\hspace{-1.2pt}\mathrm{I}
\hspace{-1.2pt}\mathrm{I}}

\newcommand{\opnorm}[1]{\lVert  #1  \rVert_{\rm{op}}}

\newcommand{\bTr}[1]{\mathrm{Tr}\qty[#1]}

\newcommand{\EV}[1]{\mathbb{E}\qty[#1]}

\begin{document}

\subheader{\today}
\title{\boldmath A Strategy for Proving the Strong Eigenstate Thermalization Hypothesis : Chaotic Systems and Holography}

\author[a]{Taishi Kawamoto}

\affiliation[a]{Center for Gravitational Physics and Quantum Information, Yukawa Institute for Theoretical Physics, Kyoto University,\\
	Kitashirakawa Oiwakecho, Sakyo-ku, Kyoto 606-8502, Japan}
\emailAdd{taishi.kawamoto@yukawa.kyoto-u.ac.jp}

\abstract{The strong eigenstate thermalization hypothesis (ETH) provides a sufficient condition for thermalization and equilibration. Although it is expected to hold in a wide class of highly chaotic theories, there are only a few analytic examples demonstrating the strong ETH in special cases, often through methods related to integrability. In this paper, we explore sufficient conditions for the strong ETH that may apply to a broad range of chaotic theories. These conditions are expressed as inequalities involving the long-time averages of real-time thermal correlators. Specifically, as an illustration, we consider simple toy examples which satisfy these conditions under certain mathematical and technical assumptions. This toy models have same properties as holographic theories at least in the perturbation in large $N$. We give a few comments for more realistic holographic models.
}


	\begin{flushright}
		YITP-24-156
		\\
	\end{flushright}
	\maketitle
	\flushbottom
 


\section{Introduction}
A question whether given initial quantum states undergo equilibration and thermalization in closed many-body quantum systems is a deep problem in a wide range of physics, including statistical physics, high energy physics, quantum information, and so on. More concretely, for a given sufficiently high energy shell with energy window $(E,E+\Delta E)$, the general initial state constructed from a superposition of energy eigenstates within the shell reaches a state close to the equilibrium ensemble, like the microcanonical ensemble, through unitary time evolution. It is known that the strong eigenstate thermalization hypothesis (ETH) provides a sufficient condition for this problem \cite{Neumann1929, Srednicki_1999, Deutsch_1991}. See \cite{Gogolin:2015gts, D_Alessio_2016, Mori_2018} for reviews. The statement of the strong ETH is as follows: for any relevant operator $\mO$ in some thermodynamic limit and for any sufficiently high energy eigenstates $\ket{E_n}, \ket{E_m}$ in the energy shell, the matrix elements shows the following suppression for the microcanonical entropy $\Smc(E)$,
\begin{equation}
    \begin{split}
        \text{Diagonal} & \quad \abs{\bra{E_n}\mO\ket{E_n} - \mcev{\mO}} = e^{-\LandauO{\Smc(E)}} \\
        \text{Off-diagonal} & \quad \abs{\bra{E_n}\mO\ket{E_l}} = e^{-\LandauO{\Smc(E)}}.
    \end{split}
\end{equation}
 Here $\mcev{\mO}$ denotes the expectation value of the microcanonical ensemble with energy $E$. There is considerable evidence that the ETH is valid in complex chaotic systems. For example, it is known that a typical Hamiltonian sampled from a Haar random unitary measure shows the above two conditions \cite{Neumann1929, Goldstein:2010ate}. Also, it is known that for typical Hamiltonians with local interactions or some long-range interactions can show strong ETH for the local observables via elaborated random matrix theory \cite{Sugimoto:2020nnw, Sugimoto:2021aao}.
 
 Additionally, there are numerical computations for non-integrable spin chains in the thermodynamic limit where we take the large volume limit \cite{ Kim:2014jfl}. 
 
 Thus, it is natural to expect that sufficiently chaotic quantum systems satisfy the strong ETH for some simple operators in some thermodynamic limit. 

However, it is not an easy task to show the strong ETH for a given single Hamiltonian since doing so requires almost all microstates information. Indeed, we need to compute all of the relevant matrix elements, $\bra{E_n}\mO\ket{E_m}$, which seems hopeless for chaotic systems. Indeed, even though the ETH is expected to be hold in wide class of chaotic theories, there is a only a few concrete illustration for it and such examples are shown because of kinds of integrability. For example, in \cite{Tasaki:2024bvh}, the strong ETH is shown in free fermion chain with specific case of ETH.

As an alternative approach, we may think of the statistical collective behavior of the energy eigenstates in the energy shell. To this end, we introduce the set of energy eigenstates denoted by $\mEeq$ and consider a uniform measure on $\mEeq$. Then, we consider the following probability,
\begin{equation}\label{eq:probE}
\DeltaETH{\ep}{\mO}:= \ProbE{\abs{\bra{E_n}\mO\ket{E_n} - \mcev{\mO}} > \ep},
\end{equation}
for a small positive number $\ep$. This probability evaluates the ratio between the number of energy eigenstates in the shell which violate the diagonal ETH with the order $\ep$ and the whole number of energy eigenstates in the shell. If the probability $\DeltaETH{\ep}{\mO}$ converges to zero in the thermodynamic limit, then we say that the system shows the weak ETH \cite{Biroli_2010erf}. Actually, the weak ETH is too weak as it holds even in integrable systems and does not conclude the thermalization from any initial states \cite{Biroli_2010erf, Mori:2016wet, Iyoda:2017ftm}. However, it provides some information on the thermalization. To see this, we notice that for any quantum many-body systems, any bounded operator $\mO$, and any initial states constructed from the energy shell
\begin{equation}
    \ket{\psi_0} = \sum_{\ket{E_n} \in \eneshell} c_n \ket{E_n},
\end{equation}
and for any positive number $\ep$, it holds that \cite{Mori:2016wet, Mori_2018},
\begin{equation}\label{eq:ratio}
    \begin{split}
        \abs{\lim_{T \to \infty} \frac{1}{T} \int_0^T dt \bra{\psi(t)}\mO\ket{\psi(t)} - \mcev{\mO}} < \opnorm{\mO} \qty(\ep + \sqrt{\frac{\shelldim}{\Deff}\DeltaETH{\ep}{\mO}}).
    \end{split}
\end{equation}
Here we meet only the initial state dependent quantity $\Deff$ called the effective dimension, which counts the effective number of energy eigenstates contributing to the initial state $\ket{\psi_0}$. This inequality states that if 
\begin{equation}
{\frac{\shelldim}{\Deff}\DeltaETH{\ep}{\mO}} \leq e^{-\LandauO{\Smc(E)}}
\end{equation}
holds and if we choose $\ep$ in order of $e^{-\LandauO{\Smc(E)}}$ , we can conclude the thermalization from the given initial state $\ket{\psi_0}$. Thus, for a given quantum many-body system, if we can find ${\frac{\shelldim}{\Deff}\DeltaETH{\ep}{\mO}} \leq e^{-\LandauO{\Smc(E)}}$ for any initial state, then we can conclude thermalization for any initial state.

To confirm this condition, we need to consider the scaling of the involved quantities in the thermodynamic limit. In \cite{PhysRevLett.120.200604}, it is conjectured that for integrable systems, the actual scaling of $\DeltaETH{\ep}{\mO}$ will be 
\begin{equation}
\DeltaETH{\ep}{\mO} \approx e^{-\LandauO{\Smc(E)}},
\end{equation}
but for non-integrable systems, the scaling is conjectured to be
\begin{equation}\label{eq:double_expo0}
\DeltaETH{\ep}{\mO} \approx e^{-\# e^{\LandauO{\Smc(E)}}}, \; \# > 0,
\end{equation}
that is, doubly exponential in the thermodynamic entropy. Note that $\#$ is some constant independent of $\Smc(E)$. If this is true for some non-integrable systems, the condition \eqref{eq:ratio} holds for any initial state, and we can conclude thermalization for any initial state.

In this paper, we try to prove the double exponential scaling \eqref{eq:double_expo0} for some highly chaotic quantum many-body systems. Especially, we mainly consider what are called holographic systems as such chaotic systems. 
Roughly speaking, holographic systems are quantum many-body systems which have higher-dimensional gravity duals via the holographic principle \cite{Susskind:1994vu, tHooft:1993dmi}, such as the AdS/CFT correspondence \cite{Maldacena:1997re, Witten:1998qj, Gubser:1998bc}, where the gravity dual is Einstein gravity in the low energy limit \cite{El-Showk:2011yvt, Heemskerk:2009pn}. 
The AdS/CFT correspondence states that some classes of $d$-dimensional quantum field theories are dual to string theory on $(d+1)$-dimensional anti-de Sitter (AdS) spacetime times some compact manifold. Well-known examples include "four-dimensional $\mathcal{N}=4$ SU$(N)$ super Yang-Mills theory"="Type $\mathrm{IIB}$ superstring theory on AdS$_5 \times \mathbb{S}^5$," "Schwarzian sector of the SYK model"="AdS$_2$ JT gravity" \cite{Kitaev, Maldacena:2016hyu}, and so on. Holographic theories usually have parameters expressing the effective number of degrees of freedom. For example, the central charge $c$ in two-dimensional CFT, the number of gluon indices $\sim N^2$ in SU$(N)$ gauge theories, and the number of Majorana fermions in the SYK models. These parameters are directly related to the thermodynamic entropy. This means that the large degrees of freedom limit, like the large $N$ limit or large central limit, can define the thermodynamic limit for the holographic theories. In this paper, we refer to this thermodynamic limit as the large $N$ limit for all holographic theories. Importantly, the large $N$ limit is the semi-classical limit $G_N \to 0$ of the dual gravity theories, which provides black hole thermodynamics \cite{Bekenstein:1972tm, Bekenstein:1973ur, Gibbons:1976ue, Bardeen:1973gs, Wald_1993}, representing the thermodynamics of the dual holographic theories. Notice that the large $N$ limit defines a different thermodynamic limit from the large volume limit discussed in usual quantum many-body systems like quantum spin chains. The large $N$ limit introduces a different type of thermodynamics and phase transitions, like Hawking/Page transitions dual to confinement/deconfinement phase transitions in large $N$ gauge theories \cite{Hawking:1982dh, Witten:1998zw}. Interestingly, the large $N$ limit induces phase transitions even in $(0+1)$-dimensional quantum mechanics like SYK models \cite{Maldacena:2018lmt}.

Importantly, black holes and holographic theories are considered to be highly chaotic \cite{Hayden:2007cs, Sekino:2008he, Shenker:2013pqa, Maldacena:2015waa, Cotler:2016fpe}. Additionally, from the general relativity insights, massive heavy objects like stars undergo gravitational collapse and form black holes. Since black holes follow thermodynamics and are dual to thermal equilibrium ensembles in the dual holographic theories, these gravitational collapse processes are considered to be equilibration and thermalization in the dual holographic theories \cite{Chesler:2008hg, Bhattacharyya:2009uu, Abajo-Arrastia:2010ajo, Aparicio:2011zy, Balasubramanian:2011ur, Hartman:2013qma, Liu:2013iza, Anous:2016kss}. From these insights, we naively expect that holographic theories show strong ETH or the double exponential scaling \eqref{eq:double_expo0} in some chaotic sector \cite{Fitzpatrick:2015zha}. Related to these insights, there are several investigations of the ETH in large central charge CFTs, for example, \cite{Fitzpatrick:2015dlt, Dymarsky2016SubsystemE, Kraus:2016nwo, Hikida:2018khg, Brehm:2018ipf, Basu:2017kzo, Karlsson:2021duj}, mainly using techniques like the conformal bootstrap and modular invariance specific to conformal field theory. There is also work on SYK models \cite{Sonner:2017hxc}. Note that the problem of the ETH is highly related to the black hole information problems \cite{Hawking:1976bpg, Maldacena:2001kr} in the sense of how single pure states undergo gravitational collapse and mimic thermal equilibrium ensembles.

In this paper, we attempt to prove the double exponential scaling \eqref{eq:double_expo0} from some conditions that we expect to hold in highly chaotic systems. Key mathematical tools are concentration inequalities \cite{Concentration_Inequalities:A_Nonasymptotic_Theory_of_Independence, Vershynin_2018} or the large deviation property of the probability \eqref{eq:probE0} for large microcanonical entropy. 
We find that if the thermal Wightman correlators show the two conditions, quantum mixing conditions and the clustering conditions, then, we can conclude that the  \eqref{eq:probE0} and \eqref{eq:offdiagprob} is sub-Gaussian with an exponentially small parameter for the thermodynamic entropy. Here we refer to the quantum mixing condition as a statement the long time average of the thermal two point functions is exponentially small compared to the entropy. The clustering condition is that the higher order Wightman function can be decomposed to the products of the two point function in the long time average.
As an example of theories showing sub-Gaussianity under some mathematical conjectures, we give large $N$ generalized free fields (GFF) and $1/N$ perturbation for GFF, which are considered a toy model of the AdS/CFT correspondence. With these examples, we give some comments for more realistic holographic systems with wormhole contribution in the gravity path integrals. For examples, it is highly expected that the results in JT gravity \cite{Saad:2019pqd} shows that the matter correlation functions satisfies sufficient conditions for strong ETH. We expect similar mechanisms work for higher dimension holographic theories like some kinds of two dimensional large central charge CFTs or some sector of large $N$ gauge theories. In our discussion, conformal symmetry is not important. Rather it appears as a difficulty for strong ETH since the conformal algebra makes the spectrum highly degenerate and brakes non-resonance conditions. We will give a few comments on this problem. Also we find that the quantum mixing property of the Wightman function appears only when we take the energy window in exponentially small in typical quantum many body systems. This means the Sredniki's ETH ansatz \cite{Srednicki_1999} hold  only in very narrow energy window.\par 
This paper is organized as follows. In section \ref{sec:Review_ETH}, we provide a quick review of the strong and weak eigenstate thermalization hypotheses and the actual scaling of the probability \eqref{eq:probE0}. In section \ref{sec:Double_Expo}, we discuss the sets of sufficient conditions that derive the double exponential scaling. In section \ref{sec:Examples}, we give simple holographic examples that satisfy the conditions. In \ref{sec:Srednicki_ETH} we give a discussion how our discussion for the thermalization and equilibration related to the Srednicki's ETH ansatz. In section \ref{sec:Summary}, we summarize our conclusions and discuss future problems. In Appendix \ref{app:proof}, we provide proofs of the propositions used in the paper. In Appendix \ref{app:equilibartion_inequality_resonance} we discuss some generalization of the theorem in the main part of the paper which is useful for conformal field theories. In Appendix \ref{appendix:Ela^2X^2_bound} and \ref{appendix:1/N_correction_moment}, we give some numerical evidence for the inequality used to show the small deviation from sub-Gaussianity.\par \noindent 
\textbf{[Update in 2026/1/16]} This version of the draft was revised in January 2026 to correct an error in the discussion. We would like to thank Professor Yu Nakayama for bringing this mistake to our attention. The corrections occurs in Section \ref{subsec:Almost Sub-Gaussianity} and \ref{subsubsec:perturbation}.

  \section{Short Review of Strong and Weak ETH}\label{sec:Review_ETH}
Here we give a short review of the ETH and fix the notation in this paper. We consider quantum many-body systems or quantum field theory on $\mathbb{R}\times \mathbb{S}^{d-1}$ and assume that the energy spectrum is discrete in the energy band we focus on. Let us denote the Hamiltonian of this theory as $H=\sum_n E_n \ket{E_n}\bra{E_n}$. Note that since the theory lives on a finite volume spatial manifold, the energy spectrum is discrete. We have a Hilbert space $\mathcal{H}$ on Cauchy slices. We denote the set of all bounded operators on the Hilbert space $\mathcal{H}$ as $\mB(\mH)$. As a first step, let us introduce an energy shell in the total Hilbert space $\mH$,
\begin{equation}
    \eneshell= \bigoplus_{E<E_n<E+\Delta E}\mathbb{C}\ket{E_n}\subset \mathcal{H}
\end{equation}
with dimension $\shelldim=\mathrm{dim}\eneshell$. Also we introduce microcanonical entropy $\Smc(E):=\log\shelldim$. We take the energy width $\Delta E$ to be small compared to the energy $E$ but large enough so that the energy shell contains many energy eigenstates. We define the microcanonical ensemble on this energy shell as
\begin{equation}
    \rhomc = \frac{1}{\shelldim} \Pi_{\eneshell},
\end{equation}
where $\Pi_{\eneshell}$ is a projection onto the energy shell $\eneshell$. We denote the expectation value of some operator $\mO\in\mB(\mH)$ under the microcanonical ensemble as $\mcev{\mO}=\bTr{\mO \rhomc}$. Usually, in some thermodynamic limit, we focus not on all the operators in $\mB(\mH)$ but on some relevant operators that survive in the thermodynamic limit. Let us denote the set of operators we focus on as $\mA\subset \mB(\mH)$. We have a few choices of the thermodynamic limit and this set $\mA$. For example,
\begin{itemize}
    \item large volume limit and $\mA$ is a set of density of macroscopic operators like density of total magnetization \cite{Neumann1929,Goldstein:2010ate,Goldstein2010LongtimeBO,Ogata:2011amo,Tasaki_2016}. This definition leads to the notion called macroscopic thermal equilibrium (short for MATE). Importantly, this definition is also applicable to classical systems.
    \item large volume limit and $\mA$ is a set of any local or few-body operators \cite{Popescu:2005fsm,Popescu2006,Goldstein_2006,Sugita:2006bqs}. This definition leads to the notion called microscopic thermal equilibrium (short for MITE). Also, this definition comes from canonical typicality, that is, the typical states in the energy shell are locally indistinguishable from the microcanonical ensemble. Importantly, the definition is based on quantum entanglement and there is no counterpart for classical systems. Similar notions called sub system ETH are discussed in the context of conformal field theories \cite{Lashkari:2016ethcft}. For the relation to the MATE, see \cite{Goldstein_2017}.
    \item Large $N$ limit and $\mA$ is a set of light single-trace or a few multi-trace operators. This thermodynamic limit is the semi-classical limit of the gravitational theory in the AdS/CFT correspondence and is relevant to black hole thermodynamics.
\end{itemize}
We mainly focus on the large $N$ limit, but our discussion is also useful for other choices.\par
In the standard textbook of statistical physics, the notion of equilibrium states are expressed by the mixed states like microcanonical ensemble or canonical ensemble. However, in the closed systems, we need to define the notion of thermal equilbrium for pure states. Then, for each thermodynamic limit and $\mA$, we define the notion that pure states represent the thermal equilibrium as follows:
\begin{dfn}
We say that a pure state $\ket{\psi}\in \eneshell$ represents equilibrium with respect to a fixed thermodynamic limit and $\mA$ if it holds that for any $\mO \in \mA$
\begin{equation}
\abs{\bra{\psi}\mO\ket{\psi}-\mcev{O}} \leq e^{-\LandauO{\Smc(E)}}
\end{equation}
in the large but finite $\Smc(E)$.
\end{dfn}
There is one caveat about the exponential suppression for our definition of thermal equilibrium. In real quantum many body systems the deviations from the microcanonical ensemble highly depends on the width $\Delta E$ of the energy shells. In this paper, we put $\Delta E =e^{-\LandauO{E}}$. The reason we adapt the exponential suppression is to compare our results to the random matrix theories. Also notably, in the following discussion, we meet a fact that some condition for the thermal correlators can only be consistent only when $\Delta E$ is in this order. We are going to give a more detail discussion in section \ref{sec:Delta_E}. \par 
Actually the thermal equilibrium pure states are typical states in the Hilbert space. Indeed, there is a typicality theorem that states almost all pure states in the energy shell represent equilibrium. Then, it seems natural that given initial atypical non-equilibrium states undergo thermalization and equilibration for chaotic systems that have complex time evolution. However, this is not obvious and it seems that we cannot prove this without another physical condition which depends on the initial states and theory. One of the additional conditions is strong ETH. Before introducing the notion of the ETH, we fix the terminology equilibration and thermalization for a given initial state $\ket{\psi_0}=\sum_{\ket{E_n}\in\eneshell} c_n \ket{E_n}$.
\begin{dfn}[Equilibration]
    For a given initial state $\ket{\psi_0}$, we say the time evolution $\ket{\psi(t)}=e^{-iHt}\ket{\psi_0}$ undergoes equilibration if it holds that for any $\mO\in\mA$
    \begin{equation}\label{eq:equilibration}
\overlineT{\qty(\bra{\psi(t)}\mO\ket{\psi(t)}-\overlineT{(\bra{\psi(t)}\mO\ket{\psi(t)})})^2}\leq e^{-\LandauO{\Smc(E)}}
    \end{equation}
    where we denote the long-time average as 
    \begin{equation}
        \overlineT{A(t)} = \lim_{T\to\infty}\frac{1}{T}\int_{0}^T dt A(t).
    \end{equation}
    This definition of equilibration physically states that the fluctuation of $\bra{\psi(t)}\mO\ket{\psi(t)}$ from the steady long-time average value $\overlineT{\bra{\psi(t)}\mO\ket{\psi(t)}}$ is thermodynamically small.
\end{dfn}
\begin{dfn}[Thermalization]
    For a given initial state $\ket{\psi_0}$, we say the time evolution $\ket{\psi(t)}=e^{-iHt}\ket{\psi_0}$ undergoes thermalization if it holds that for any $\mO\in\mA$,
    \begin{equation}\label{eq:thermalization}
        \abs{\overlineT{\bra{\psi(t)}\mO\ket{\psi(t)}}-\mcev{\mO}} \leq e^{-\LandauO{\Smc(E)}}
    \end{equation}
\end{dfn}
If a given initial state $\ket{\psi_0}$ and theory undergoes equilibration and thermalization, we see that the time-evolved state $\ket{\psi(t)}$ represents equilibrium for most of the time $t$ larger than some relaxation time.
\par
We can think of the sufficient conditions for equilibration and thermalization. The task we should tackle is considering the upper bound for the left-hand side of \eqref{eq:equilibration} and \eqref{eq:thermalization}. Conceptually, there are two ways for the upper bound.
\subsection{Initial State Independent Upper Bound: Strong ETH}
First, let us discuss the upper bound which does not depend on the initial states. We assume the following two conditions for the energy spectrum in the energy shell for simplicity:
\begin{enumerate}
    \item Non-degeneracy
    \begin{equation}\label{eq:non_degeneracy}
        E_n=E_m \; \Rightarrow n=m
    \end{equation}
    \item Non-resonance condition
    \begin{equation}\label{eq:non_resonance}
        E_n-E_m=E_l-E_k \Rightarrow n=l,\; m=k
    \end{equation}
\end{enumerate}
We can loosen these conditions to some extend. 
Then, the left-hand side of \eqref{eq:equilibration} can be written as 
\begin{equation}
    \begin{split}
        \overlineT{\qty(\bra{\psi(t)}\mO\ket{\psi(t)}-\overlineT{(\bra{\psi(t)}\mO\ket{\psi(t)})})^2} &= \sum_{n\neq m} \abs{c_n}^2\abs{c_m}^2 \abs{\bra{E_n}\mO\ket{E_m}}^2\\
        &< \qty(\max_{n\neq m}\abs{\bra{E_n}\mO\ket{E_m}})^2 \sum_{n\neq m}\abs{c_n}^2\abs{c_m}^2\\
        &< \qty(\max_{n\neq m}\abs{\bra{E_n}\mO\ket{E_m}})^2.
    \end{split}
\end{equation}
Thus, if the following inequality called the off-diagonal ETH holds, we see that any initial state undergoes equilibration,
\begin{equation}\label{eq:off_diagonal_ETH}
    \begin{split}
        \text{Off-diagonal ETH:}\quad  &^\forall \mO \in \mA,\; ^\forall \ket{E_n} ,^\forall \ket{E_m}\in \eneshell \; \text{s.t.}\; n\neq l,\\ &\abs{\bra{E_n}\mO\ket{E_m}} \leq e^{-\LandauO{\Smc(E)}}.
    \end{split}
\end{equation}
Next, we can try to consider the upper bound for the left-hand side of \eqref{eq:thermalization}. Let us suppose that the system satisfies the following condition called (diagonal) ETH 
\begin{equation}\label{eq:diagonal_ETH}
    \begin{split}
        \text{Diagonal ETH:}\quad  &^\forall \mO \in \mA,\; ^\forall \ket{E_n}\in \eneshell, \;\\&\abs{\bra{E_n}\mO\ket{E_n}-\mcev{\mO}} \leq e^{-\LandauO{\Smc(E)}}.
    \end{split}
\end{equation}
Then we observe that the left-hand side of \eqref{eq:thermalization} is small for any initial state and we can conclude thermalization from any initial state.\par
We call the two conditions, diagonal and off-diagonal ETH \eqref{eq:diagonal_ETH}, \eqref{eq:off_diagonal_ETH}, the strong ETH, i.e.,
\begin{dfn}[Strong ETH]
    A theory satisfies the strong eigenstate thermalization hypothesis (ETH) on an energy shell $\eneshell$ with respect to a thermodynamic limit and a set of relevant operators $\mA$ if the two conditions, diagonal and off-diagonal ETH \eqref{eq:diagonal_ETH}, \eqref{eq:off_diagonal_ETH}, hold. The strong ETH gives a sufficient condition for thermalization and equilibration from any initial state in the energy shell.
\end{dfn}
There are some numerical verifications of strong ETH \cite{Kim:2014jfl,Beugeling:2014fse}.
However, theoretically, it is a hard task to show strong ETH for a given chaotic Hamiltonian since usually, we do not concretely find the energy eigenstates and not easily compute the amplitude $\bra{E_n}\mO\ket{E_m}$.

 \subsection{Initial State Dependent Upper Bound}
As another upper bound of the left-hand side of \eqref{eq:equilibration} and \eqref{eq:thermalization}, we can consider ones dependent on the initial states, which are useful in our discussion later. To this end, we introduce the set of energy eigenstates denoted by $\mEeq$ and consider a uniform probability measure on $\mEeq$. Then, we consider the following probability,
\begin{equation}\label{eq:probE0}
\DeltaETH{\ep}{\mO}:= \ProbE{\abs{\bra{E_n}\mO\ket{E_n}-\mcev{\mO}}>\ep},
\end{equation}
for a small positive number $\ep=e^{-\LandauO{\Smc(E)}}$ \footnote{To obtain a good upper bound, we need to take care about value of $\ep$. We give an illustration about this point in the later section.}.  This probability evaluates the ratio between the number of energy eigenstates in the shell which violates the diagonal ETH with the order $\ep$ and the whole number of energy eigenstates in the shell. We can also consider a counterpart of $\DeltaETH{\ep}{\mO}$ for the off-diagonal component. For this purpose, let us introduce a set of pairs of energy eigenstates denoted as $\offdiagset$ defined as
\begin{equation}
\offdiagset:= \qty{(\ket{E_n},\ket{E_m}) \in \eneshell\times \eneshell \mid \ket{E_n}\neq \ket{E_m} \:\text{up to un-physical phase shift} }
\end{equation}
Then, we introduce the uniform probability measure on $\offdiagset$ and consider the following probability 
\begin{equation}\label{eq:offdiagprob}
\offDeltaETH{\ep}{\mO}:=\ProboffE{\abs{\bra{E_n}\mO\ket{E_m}}>\ep}.
\end{equation}
We explore the upper bound involving these probabilities. Again, we assume the non-degenerate \eqref{eq:non_degeneracy} and the non-resonance condition \eqref{eq:non_resonance}. Let us start with the one related to thermalization. The following inequality is known \cite{Tasaki_2016}:
\begin{thm}\label{thm:thermalization_inequality}
For any $\ep>0$, for initial state $\ket{\psi_0}=\sum_{\ket{E_n}\in\eneshell} c_n \ket{E_n}$ and bounded operator $\mO \in \mB(\mH)$, it holds that
\begin{equation}\label{eq:thermalization_inequality}
\abs{\overlineT{\bra{\psi(t)}\mO\ket{\psi(t)}}-\mcev{\mO}} <  \opnorm{\mO}\qty(\ep + \sqrt{\frac{2\shelldim}{\Deff}\DeltaETH{\ep}{\mO}}),
\end{equation}
where $\opnorm{\mO}$ denotes the operator norm $\opnorm{\mO} := \sup_{\ket{\psi}\in\mH}\sqrt{\frac{\bra{\psi}\mO^\dag \mO \ket{\psi}}{\braket{\psi|\psi}}}$, $\Deff$ denotes the so-called effective dimension defined in \cite{Tasaki_2016, Mori:2016met} by 
\begin{equation}
\Deff := \qty(\sum_{n}\abs{c_n}^4)^{-1}.
\end{equation}
\end{thm}
The effective dimension for a given initial state expresses the effective number of energy eigenstates in the energy shell contributing to this initial state. For example, a single energy eigenstate has $\Deff=1$. The other example is a maximally coherent initial state with respect to the energy eigenstates basis, $\ket{\psi}=\sum_{\ket{E_n}\in \eneshell}\frac{1}{\sqrt{\shelldim}}\ket{E_n}$ which gives $\Deff= \shelldim$. As an almost trivial proposition, we can prove that the typical initial states have the large effective dimension $\Deff=\LandauO{\shelldim}$ \cite{Linden:2008awz}. Also we note that the operator norm for local operators are generally UV divergent because the two point function of the coincident points are divergent. To avoid this, we consider the operators are always smeared by some function $f$, $\mO[f](t):=\int dx f(x)\mO(x,t), \int dx f(x)=1$ or Euclidean time evolution $\mO(t+ib):=e^{bH}\mO e^{-bH}$. In the later discussion, we omit $f$ or $b$ for notational simplicity. Importantly, we notice that the initial state dependence and the theory dependence of the upper bound are factorized in the inequality \eqref{eq:thermalization_inequality}. The initial state dependence only appears in $\Deff$, and the theory dependence is only included in $\DeltaETH{\ep}{\mO}$. 
From this inequality, we can conclude that for given initial states if for small enough $\ep$ it holds that 
\begin{equation}
\frac{\shelldim}{\Deff}\sup_{\mO\in\mA}\DeltaETH{\ep}{\mO} \leq e^{-\LandauO{\Smc(E)}},
\end{equation}
we can conclude that thermalization holds for the given initial states. Also, since the minimum value of the effective dimension is $\Deff=1$, which is saturated when the initial state is an energy eigenstate, if 
\begin{equation}\label{eq:thermalization_inequality2}
\shelldim\sup_{\mO\in\mA}\DeltaETH{\ep}{\mO} \leq  e^{-\LandauO{\Smc(E)}},
\end{equation}
holds, we can conclude thermalization from any initial states. To prove \eqref{eq:thermalization_inequality2} for a given theory and definition of thermal equilibrium, the actual scaling of $\DeltaETH{\ep}{\mO}$ in the thermodynamic limit is important. For example, it is known that if we take the large volume limit as the thermodynamic limit and $\mA$ is a set of local operators, if the Hamiltonian is translation invariant quantum spin systems on the $d$-dimensional cubic lattice under the periodic boundary conditions and if the scaled cumulant generating function for the relevant operator correlation function is well-defined in the thermodynamic limit, then it holds that there is a positive constant $\gamma(\ep)$ such that
\begin{equation}
\DeltaETH{\ep}{\mO} \leq e^{-\gamma(\ep) \mathrm{Vol}(\mathbb{S}^{d-1})}
\end{equation}
from the large deviation principles \cite{Mori:2016wet}. Also, there are other types of upper bounds for a local operator with big support \cite{Biroli_2010erf,Iyoda:2017ftm}. This inequality holds for a wider class of theories including both integrable and non-integrable or chaotic systems and itself gives less information about thermalization. What we can infer from this inequality is that if the following scaling are true
\begin{equation}\label{eq:largevolumescaling}
\shelldim = e^{s(E)\mathrm{Vol}(\mathbb{S}^{d-1})},\; \Deff = e^{\eta \mathrm{Vol}(\mathbb{S}^{d-1})},
\end{equation}
for a large enough volume and if it holds that $s(E)<\eta+\gamma(\ep)$, then we can prove the thermalization from the initial states whose effective dimension satisfies \eqref{eq:largevolumescaling}.
\par It is actually discussed that the actual scaling of $\DeltaETH{\ep}{\mO}$ is dramatically different between integrable systems and non-integrable systems. In \cite{PhysRevLett.120.200604}, it is conjectured that for integrable systems the actual scaling of $\DeltaETH{\ep}{\mO}$ will be 
\begin{equation}\label{eq:expo_scaling}
\DeltaETH{\ep}{\mO} \approx e^{-\LandauO{\Smc(E)}},
\end{equation}
but for the non-integrable system the scaling is conjectured to be
\begin{equation}\label{eq:double_expo}
\DeltaETH{\ep}{\mO} \approx e^{-\#\;e^{\LandauO{\Smc(E)}}},\; ^\exists \#>0
\end{equation}
that is, doubly exponential for the thermodynamic entropy. This conjecture is numerically verified for quantum spin chains for the XX model as an integrable system and the XXX model with integrability-breaking perturbation as a non-integrable system. Importantly, if this is true for some non-integrable systems, the condition \eqref{eq:thermalization_inequality2} holds, and we can conclude thermalization for any initial states, which is equivalent to the diagonal strong ETH. Note that if there are some ETH violating energy eigenstates like quantum scar states \cite{Bernien:2017ubn,Turner_2018}, the double exponential scaling will be never hold. This is because it should be that $\shelldim \DeltaETH{\ep}{\mO}$ is not double exponentially small in \eqref{eq:thermalization_inequality}.
\par We can do a similar discussion for the off-diagonal elements and equilibration. Indeed, following inequality holds for any quantum many-body systems with non-resonance condition.
\begin{thm}\label{thm:equilibration_inequality}
For any $\ep>0$, any initial state $\ket{\psi_0}$, any bounded operator $\mO\in\mB(\mH)$ and assuming non degeneracy of Hamiltonian and non-resonance condition \eqref{eq:non_resonance}, it holds that 
\begin{equation}\label{eq:off_diagonal_inequality}
\overlineT{\left|\bra{\psi(t)}\mO(t)\ket{\psi(t)} - \overlineT{\bra{\psi(t)}\mO\ket{\psi(t)}}\right|^2} \leq \opnorm{\mO}^2\qty(\ep^2 + \frac{(\shelldim)^{\frac{3}{2}}}{\Deff} \sqrt{\offDeltaETH{\ep}{\mO}}).
\end{equation}
$\offDeltaETH{\ep}{\mO}$ is the probability defined in \eqref{eq:offdiagprob}.
\end{thm}
We provide the proof of this inequality in section \ref{subsec:Proof of Threorem thm:equilibration_inequality}. The large deviation analysis for the off-diagonal elements is discussed in \cite{Arad:2014znf}. 
Similarly to the thermalization case, if an initial states and Hamiltonian satisfy
\begin{align}
    \frac{(\shelldim)^{\frac{3}{2}}}{\Deff} \sqrt{\offDeltaETH{\ep}{\mO}} = e^{-\LandauO{\Smc}}
\end{align}
then we can prove the equilibration from the initial states. Moreover if the hamiltonian satisfies 
\begin{equation}\label{eq:off_diag_double_expo}
\offDeltaETH{\ep}{\mO}\approx e^{-\#' e^{\LandauO{\Smc}}},\; \#'>0
\end{equation}
we can prove the equilibration from any initial states.

\section{Sufficient Condition for the Double Exponential Scaling}\label{sec:Double_Expo}
 Here we describe one way of proving the double exponential scaling \eqref{eq:double_expo}, \eqref{eq:off_diag_double_expo} from some conditions which may hold in highly chaotic systems. A key tool is concentration inequalities \cite{Concentration_Inequalities:A_Nonasymptotic_Theory_of_Independence, Vershynin_2018} and quantum mixing propeties which we explain from now on.
 \subsection{Concentration Inequality and Sub-Gaussian Properties}
Concentration inequalities are probability inequalities of the form
\begin{equation}
    \Prob{}{\abs{X - \EV{X}} > \epsilon} < f(\epsilon),
\end{equation}
where $f(\epsilon)$ is some small function of $\epsilon$. Here, $X$ denotes a random variable, and $\EV{X}$ is the expectation of $X$. For example, the Chebyshev inequality and the Markov inequality are examples of concentration inequalities.

\subsubsection{Sub-Gaussian Random Variables: A Review}
Here, we introduce a popular class of distributions known as sub-Gaussian distributions, which are probability distributions associated with a concentration inequality involving a Gaussian function. More concretely:

\begin{dfn}
A random variable $X$ is $\sigma$-sub-Gaussian if
\begin{equation}\label{eq:def_sub_Gaussian}
    \Prob{}{\abs{X - \EV{X}} > \epsilon} < 2e^{-\frac{\epsilon^2}{\sigma^2}}.
\end{equation}
\end{dfn}

It is known that a wide range of random variables, such as Gaussian, Bernoulli, bounded variables, etc., are sub-Gaussian with a moderate $\sigma$. There are many conditions equivalent to \eqref{eq:def_sub_Gaussian}. Without loss of generality, we can consider a random variable with zero mean, i.e., $\EV{X} = 0$\footnote{This zero-mean condition is for simplicity and can be easily relaxed.}. One equivalent condition is as follows:

\begin{prop}
    Let $X$ be a zero-mean $\sigma$-sub-Gaussian random variable. Then definition \eqref{eq:def_sub_Gaussian} is equivalent to the following conditions:
    \begin{enumerate}
        \item The tails of $X$ satisfy
        \begin{equation}\label{eq:sub_gaussian_1}
            \Prob{}{\abs{X} > t} \leq 2 e^{-\frac{t^2}{K_1^2}} \quad \text{for any } t > 0.
        \end{equation}
        \item The moments of $X$ satisfy
        \begin{equation}\label{eq:sub_gaussian_2}
            \qty(\EV{\abs{X^q}})^{\frac{1}{q}} \leq K_2 \sqrt{q} \quad \text{for any } q \geq 1.
        \end{equation}
        \item The moment generating function of $X^2$ satisfies
        \begin{equation}\label{eq:sub_gaussian_3}
            \EV{\exp{\lambda^2 X^2}} \leq \exp{K_3^2 \lambda^2} \quad \text{for all } \lambda \text{ such that } \abs{\lambda} < \frac{1}{K_3}.
        \end{equation}
        \item The moment generating function of $X^2$ is bounded at some point, namely
        \begin{equation}\label{eq:sub_gaussian_4}
            \EV{\exp{\frac{X^2}{K_4^2}}} \leq 2.
        \end{equation}
        \item For any $\lambda \in \mathbb{R}$, it holds that
        \begin{equation}\label{eq:def_sub_gaussian}
            \EV{e^{\lambda X}} \leq \exp{\frac{\lambda^2 \sigma^2}{2}}.
        \end{equation}
    \end{enumerate}
    Here, $K_1, K_2, \dots, K_4$ are related to $\sigma^2$ up to some numerical constant.
\end{prop}

See \cite{Vershynin_2018} for the proof.

\subsubsection{Quadratic Generalization of sub-Gaussian variables}
Later we will want to allow mild departures from the strictly sub-Gaussian large-deviation profile. A convenient toy model is a bound of the modified Gaussian 
\begin{equation}
    \EV{e^{\lambda X}} \leq \exp{\frac{\lambda^2 \sigma^2}{2}-\sum_{k=2}\frac{t'_k}{4^k}\lambda^{2k}}
\end{equation}
for $\lambda$ in a certain range, or alternatively an explicit upper bound directly on the tail probability:
\begin{equation}\label{eq:quadratic_large_deviation}
\Prob{}{\abs{X}>\epsilon}\leq 2 \exp{-\frac{\epsilon^2}{2\sigma^2}-\sum_{k=1}^M\frac{g_{2k}}{2^k}\ep^{2k}}\, .
\end{equation}
When the coupling $g_{2k}>0$, this quartic in $\epsilon$ correction remains below the purely Gaussian estimate. In particular, $x$ is still sub-Gaussian in the sense that
\begin{equation}
\Prob{}{\abs{X}>\epsilon}\leq 2 \exp{-\frac{\epsilon^2}{2\sigma^2}}\, .
\end{equation}
Moreover, to ensure that the relevant probability integrals are well-defined, we indeed require $g_{2M}>0$, and in what follows we restrict attention to this case. We consider following theorems which we do not have complete proof:
\begin{conj}\label{thm:almost_subgaussian}
    Let $X$ be a random variable with zero mean $\EV{X}=0$. Assume that its moments satisfy the following upper bound for any $1 \leq q \leq L < +\infty$,
    \begin{equation}\label{eq:bound_moment_quadratic}
        \EV{X^{2q}} \leq 2^{q+1} \sigma^{2q} q! \sum_{\{m_k\}: m\leq m_{max}}\frac{\Gamma(q+m)}{\Gamma(q)}\prod_{k=2}^M \frac{(-g_{2k}\sigma^{2k})^{m_k}}{m_k!}
    \end{equation}
    for some finite (sufficiently small) $m_{\max}$ and $g_{2M}>0$. Here we denote $m=\sum_{k=2}^M km_k$.  For any $q \geq L+1$ we only assume the weaker bound
    \begin{equation}
         \EV{X^{2q}} \leq (\sigma')^{2q}.
    \end{equation}
    In addition, suppose that $L$ satisfies $L+1 > \frac{(\sigma')^2}{8\sigma^2}$. Then we can deduce the following bounds for the generating function and the tail probabilities:
    \begin{enumerate}
        \item For some $\abs{\lambda} < \frac{1}{2\sigma}$, we have
        \begin{equation}
        \begin{split}
            \EV{e^{\lambda^2 Z^2}} &\leq 2 e^{4\lambda^2 \sigma^2 - \sum_{k=2}^M 2^kt_k g_{2k}\lambda^{2k}\sigma^{4k}} + F^u(\lambda) \, , \\
            F^u(\lambda) &= \exp(-(L+1)\log\qty(\frac{L+1}{\lambda^2 \sigma'^2 e}))\, .
        \end{split}
        \end{equation}
        \item For any $\lambda > \frac{1}{2\sigma}$, it holds that
        \begin{equation}
            \EV{e^{\lambda X}} \leq 2e^{2\lambda^2 \sigma^2} \qty(2\exp{-\sum_{k=2}^M\frac{t_k}{4^k}g_{2k}\sigma^{2k}} + F^u\qty(\sqrt{\frac{1}{8\sigma^2}})).
        \end{equation}
        \item For $\epsilon > 2\sigma$, we obtain
        \begin{equation}
            \Prob{}{\abs{X} > \epsilon} \leq 2 \exp{-\frac{\epsilon^2}{8\sigma^2}} \qty(4\exp{-\sum_{k=2}^M\frac{t_k}{4^k}g_{2k}\sigma^{2k}} + F^u\qty(\sqrt{\frac{1}{8\sigma^2}})).
        \end{equation}
    \end{enumerate}
\end{conj}
\noindent
\par \noindent \textbf{On the Form of the bound on moment \eqref{eq:bound_moment_quadratic}}\par
Let us briefly explain how the moment inequality \eqref{eq:bound_moment_quadratic} naturally arises. Assume we are given the tail estimate \eqref{eq:quadratic_large_deviation}. Then the $q$-th moment can be controlled as  
\begin{equation}
\begin{split}
        \EV{X^q} &= \int_0^\infty d \epsilon \, \Prob{}{\abs{X}^q > \epsilon}\\
                &= q\int_0^\infty d x \, x^{q-1} \Prob{}{\abs{X} > x}\\
                &= 2q\int_0^\infty d x \, x^{q-1} \Prob{}{\abs{X} > x}\\
                &\leq 2q\int_0^\infty d x \, x^{q-1} \exp\left(-\frac{x^2}{2 \sigma^2} - \sum_{k=2}^M \frac{g_{2k}}{2^k}x^{2k}\right).
\end{split}
\end{equation}
We now view the integral perturbatively in $g$. Since the resulting series is asymptotic and has vanishing radius of convergence, we truncate the expansion at a finite order $m=m_{\max}$. From this perturbative perspective, the upper bound is approximated by
\begin{equation}\label{eq:quadratic_bound_perturbative}
\begin{split}
\EV{X^{2q}} &\leq 2\int_0^\infty d x \, x^{2q-1} e^{-\frac{x^2}{2 \sigma^2}} \sum_{\{m_k\}:m\leq m_{\max}}x^{2m}\prod_{k=2}^M\qty(\frac{1}{m_k!}\qty(-\frac{g_{2k}}{2^k})^{m_k})\\
&=2^{q+1}q!\; \sigma^{2q} \sum_{m\leq m_{\max}}\frac{\Gamma(m+q)}{\Gamma(q)}\prod_{k=2}^M\qty(\frac{(-g_{2k}\sigma^{2k})^{m_k}}{m_k!}).
\end{split}
\end{equation}
Strictly speaking, one should not interchange the order of integration and summation, because the corresponding series is merely asymptotic. Nonetheless, once we truncate at a finite $m_{\max}$, this procedure provides a natural rationale for adopting a moment bound with finite $m_{\max}$. 

In principle, one can bypass the perturbative expansion by evaluating the integral exactly and then comparing the resulting asymptotic series with the exact expression via Borel resummation. We defer such a more refined treatment to future work. Note also that the bounds on the generating function and on tail probabilities can always be relaxed to those of a purely sub-Gaussian variable; however, keeping track of the quartic correction yields sharper estimates. A further remark concerns the modification of the bound for $q<L$: in later applications we will encounter situations where the moment estimate deteriorates for large $q$ when $\sigma$ is small. Past a critical value $q=L$, our control becomes weaker and we retain only a rough estimate with a larger parameter $\sigma'$. Even in such circumstances, provided that $L$ is sufficiently large, we still expect the tail behavior to remain essentially sub-Gaussian. This motivates a careful treatment of finite-$q$ corrections.\par 
We now proceed to the proof of the theorem. Some intermediate steps are not completely rigorous mathematically; however, we support those steps with numerical checks.

\par \noindent \textbf{Upper bound fro $\EV{e^{\lambda^2 X^2}}$}\par 
We first establish an upper bound for $\EV{e^{\lambda^2 X^2}}$ from the assumed moment condition, again adopting a perturbative viewpoint:
\begin{equation}\label{eq:EX^2boundmiddle}
    \begin{split}
        \EV{e^{\lambda^2 X^2}} &= 1+\sum_{q=1}^\infty \frac{\lambda^{2q}}{q!}\EV{X^{2q}}\\
        &\leq 1+\sum_{q=1}^L \lambda^{2q} 2^{q+1} \sigma^{2q} a_q + \sum_{q=L+1}^\infty \frac{(\lambda^2 (\sigma')^2)^q}{q!}\, ,
    \end{split}
\end{equation}
where
\begin{equation}
      a_q =\sum_{m\leq m_{\max}}\frac{\Gamma(m+q)}{\Gamma(q)}\prod_{k=2}^M\qty(\frac{(-g_{2k}\sigma^{2k})^{m_k}}{m_k!}).
\end{equation}
The first two contributions can be bounded by an infinite series:
\begin{equation}
    \begin{split}
        1+\sum_{q=1}^\infty \lambda^{2q} 2^{q+1} \sigma^{2q} a_q &\leq 2 \sum_{q=0}^\infty \lambda^{2q} 2^{q+1} \sigma^{2q} a_q := F_1(\lambda)\, .
    \end{split}
\end{equation}

\par \noindent
\underline{Evaluation of $ F_1(\lambda)$}\par 
We can rewrite $F_1(\lambda)$ as
\begin{equation}
    \begin{split}
        F_1(\lambda)&=  2\sum_{q=0}^{\infty}s^{q}a_q,\;s=2\lambda^2 \;\sigma^2\\
        &=2 \sum_{\{m_k\}:m\leq m_{\max}}\frac{\Gamma(1+m)\;s}{(1-s)^{m+1}} \prod_{k=2}^M\qty(\frac{(-g_{2k}\sigma^{2k})^{m_k}}{m_k!}).
    \end{split}
\end{equation}
If we keep only finitely many terms, then it is legitimate to exchange the order of summation. 

Suppose we can show that, for sufficiently small $m_{\max}$, for $s$ in a small interval $0<s<s_*\sim \frac{1}{2}$, and for small $a_k>0$ and some positive constants $t_k$
\begin{equation}\label{eq:Ela^2X^2_bound}
   \sum_{\{m_k\}:m\leq m_{\max}}\frac{\Gamma(1+m)\;s}{(1-s)^{m+1}} \prod_{k=2}^M\qty(\frac{(-a_k)^{m_k}}{m_k!})\leq \exp{2s-\sum_{k=2}^M t_k a_k s^k}
\end{equation}
where $a_k = g_{2k}\;\sigma^{2k}$.
Then for $\abs{\lambda}<\frac{1}{2\sigma}$, with sufficiently small $m_{\max}$ and $a_k$, we obtain
\begin{equation}\label{eq:EVX^2}
    \EV{e^{\lambda^2 X^2}}\leq 2 \exp{4\lambda^2\sigma^2-\sum_{k=2}^M 2^k t_k g_{2k}\lambda^{2k}\sigma^{4k}}\, .
\end{equation}
 At present we do not have a fully rigorous derivation of \eqref{eq:Ela^2X^2_bound}, but we do have numerical evidence supporting it. Details are deferred to Appendix \ref{appendix:Ela^2X^2_bound}. We choose $t_k=6/2^k$. We see that we have $M\leq m_{\max}\leq 1/a_k$ as usual perturbation theories. Even though the numerical test is difficult for very large $M$ but we expect the inequality is valid for $M=e^{\LandauO{\Smc}}$. This point is highly non-trivial and still open questions.
 
 \par 

\noindent
\underline{Evaluation of the Third Term}\par 
Let us denote the third term in \eqref{eq:EX^2boundmiddle} by $F_2(\lambda)$:
\begin{equation}
     \begin{split}
         F_2(\lambda)&= \sum_{q=L+1}^{\infty} \frac{1}{q!}\qty(\lambda^2(\sigma')^2)^q\\
          &=e^{\lambda^2(\sigma')^2} \frac{\gamma(L+1,\lambda^2(\sigma')^2)}{\Gamma(L+1)}\, ,
     \end{split}
\end{equation}
where $\gamma(a,x)$ is the lower incomplete gamma function,
\begin{equation}
     \gamma(a,x) = \int_0^x d \epsilon\;\epsilon^{a-1} e^{-\epsilon}.
\end{equation}
To obtain a simpler bound, we use the following identity \cite{incomplete_gamma_bound2},
\begin{equation}
    e^x \frac{\gamma(a,x)}{\Gamma(a)}= \frac{x^a}{\Gamma(a)}\sum_{m=0}
^\infty \frac{x^m}{(a)_{m+1}}
\end{equation}
for $a>x$. Together with the estimates
\begin{equation}
    \frac{1}{\Gamma(a)}\leq \qty(\frac{a}{e})^{-(a-1)}
\end{equation}
and 
\begin{equation}
    (a)_{m+1}= a(a+1)\cdots (a+m)\geq a^{m+1},
\end{equation}
we arrive at the simpler bound
\begin{equation}
     e^x \frac{\gamma(a,x)}{\Gamma(a)}= \frac{x^a}{\Gamma(a)}\sum_{m=0}
^\infty \frac{x^m}{(a)_{m+1}}\leq \frac{1}{a-x}e^{a\log{x}-(a-1)\log{\frac{a}{e}}}<e^{-a\log{\frac{a}{xe}}}
\end{equation}
for $x<a$. Consequently, we obtain the bound
\begin{equation}
    F_2(\lambda) \leq  e^{-(L+1)\log{\qty(\frac{L+1}{\lambda^2\sigma'^2 e})}}=:F^{u}(\lambda)
\end{equation}
valid whenever $\lambda^2\sigma'^2< L+1$.

\par
\noindent
\par \noindent \textbf{Inequality for the Generating Function}\par 
Next we consider the generating function $\EV{e^{\lambda X}}$. Assume $\abs{\lambda} > \frac{1}{2\sigma}$. Using the inequality
\begin{equation}
    \lambda x \leq \frac{1}{2}\left(\frac{1}{4\sigma^2}x^2 + 4\sigma^2\lambda^2\right),
\end{equation}
valid for all $\lambda, x \in \mathbb{R}$, we obtain
\begin{equation}\label{eq:ElaX_bound}
     \begin{split}
         \EV{e^{\lambda X}} & \leq e^{2\sigma^2 \lambda^2}\EV{e^{\frac{1}{8\sigma^2} X^2}} \\
         & \leq e^{2\sigma^2 \lambda^2} \left(2\exp{\frac{1}{2} -\sum_{k=2}^M\frac{t_k}{4^k}g_{2k}\sigma^{2k}} + F^u\left(\sqrt{\frac{1}{8\sigma^2}}\right)\right) \\
         & \leq L_1(\lambda) + L_2(\lambda)\, ,
     \end{split}
\end{equation}
where we defined
\begin{equation}
     \begin{split}
         L_1(\lambda) &= 4\exp{(2\sigma^2\lambda^2 +\sum_{k=2}^M\frac{t_k}{4^k}g_{2k}\sigma^{2k})}, \\
         L_2(\lambda) &= \exp{(2\sigma^2\lambda^2)}F^u\left(\sqrt{\frac{1}{8\sigma^2}})\right).
     \end{split}
\end{equation}
In the second line we used \eqref{eq:EVX^2} together with $\lambda^2 \geq \frac{1}{4\sigma^4}$. The first term can also be written as
\begin{equation}
    e^{2\sigma^2 \lambda^2} 2\exp{\frac{1}{2}-\sum_{k=2}^M\frac{t_k}{4^k}g_{2k}\sigma^{2k}} \leq 2\exp{(4\sigma^2\lambda^2 - \sum_{k=2}^M\frac{t_k}{4^k}g_{2k}\sigma^{2k})}.
\end{equation}
On the other hand, the second term becomes
\begin{equation}
 e^{2\lambda^2\sigma^2}F^u\left(\sqrt{\frac{1}{8\sigma^2}}\right) = \exp\qty(2\lambda^2\sigma^2)\exp\qty[-(L+1)\log{\frac{8(L+1)\sigma^2}{e\sigma'^2}}].
\end{equation}

\par
\noindent
\par \noindent \textbf{Inequality for the Probability}\par 
Finally, we derive a tail estimate. By Markov’s inequality, for any $\lambda > 0$ we have
\begin{equation}
     \begin{split}
       \Prob{}{\abs{X} > \epsilon} &= 2\Prob{}{X > \epsilon} = 2\Prob{}{e^{\lambda X} > e^{\lambda \epsilon}} \\
       &\leq 2 e^{-\lambda \epsilon} \EV{e^{\lambda X}}.
     \end{split}
\end{equation}
Using \eqref{eq:ElaX_bound} for $\lambda > \frac{1}{2\sigma}$, we obtain
\begin{equation}
 \begin{split}
     \Prob{}{\abs{X} > \epsilon} &= 2\inf_{\lambda > \frac{1}{2\sigma}} e^{-\epsilon\lambda} \EV{e^{\lambda X}} \\
     &= 2\inf_{\lambda > \frac{1}{2\sigma}} e^{-\epsilon\lambda} (L_1(\lambda) + L_2(\lambda)) \\
     &\leq 2\inf_{\lambda > \frac{1}{2\sigma}} e^{-\epsilon\lambda} L_1(\lambda) + 2\inf_{\lambda > \frac{1}{2\sigma}} e^{-\epsilon\lambda} L_2(\lambda) \\
     &\leq 8 \exp\left(-\sup_{\lambda > \frac{1}{2\sigma}} \left(\epsilon\lambda - 2\lambda^2\sigma^2\right) - \sum_{k=2}^M\frac{t_k}{4^k}g_{2k}\sigma^{2k}\right) \\
     &\quad + 2\exp\left(-\sup_{\lambda > \frac{1}{2\sigma}} \left(\epsilon\lambda - 2\lambda^2\sigma^2\right)\right) F^u\left(\sqrt{\frac{1}{8\sigma^2}}\right).
 \end{split}
\end{equation}
For $\epsilon > 2\sigma$, the supremum is achieved at $\lambda = \frac{\epsilon}{4\sigma^2} > \frac{1}{2\sigma}$, and therefore
\begin{equation}
      \Prob{}{\abs{X} > \epsilon} \leq 2\exp\left(-\frac{\epsilon^2}{8\sigma^2}\right)\left(4\exp\left(-\sum_{k=2}^M\frac{t_k}{4^k}g_{2k}\sigma^{2k}\right) + F^u\left(\sqrt{\frac{1}{8\sigma^2}}\right)\right)
\end{equation}
In later applications, the second term $F^u\left(\sqrt{\frac{1}{8\sigma^2}}\right)$ will be doubly exponentially suppressed in the microcanonical entropy. Hence, for practical purposes, the tail behaves as an almost sub-Gaussian one.

\subsection{Almost Sub-Gaussianity of $\DeltaETH{\ep}{\mO}$ and $\offDeltaETH{\ep}{\mO}$}\label{subsec:Almost Sub-Gaussianity}
One sufficient discussion to prove the double exponential scaling \eqref{eq:double_expo}, \eqref{eq:off_diag_double_expo} may be showing that $\DeltaETH{\ep}{\mO}$ and $\offDeltaETH{\ep}{\mO}$ are sub-Gaussian. Here we discuss some analytic discussion. This is numerically evaluated in the models discussed in \cite{PhysRevLett.120.200604}. To this end, let us write down another expression for $q$-th order moments in terms of the thermal Wightman correlation functions. Notice that we can think of the matrix elements $\bra{E_n}\mO\ket{E_n},\; \bra{E_n}\mO\ket{E_m}$ as random variables of the uniform distribution on $\mEeq, \offdiagset$ respectively \footnote{Please notice that we do not consider any randomness in physics. We just consider single Hamiltonian with fixed energy spectrum and we take fixed operators. The terminology of random variables of matrix elements are just artificial on the uniform distribution introduced by hand.}. 
\par 
 Firstly, let us focus on the computation for the diagonal part $\DeltaETH{\ep}{\mO}$. For the diagonal part, the expectation value of single matrix element, denoted by $\EVdiag{\bra{E_n}\mO\ket{E_n}}$, is obtained just by the one of the microcanonical ensemble,
 \begin{equation}
    \EVdiag{\bra{E_n}\mO\ket{E_n}}= \frac{1}{\shelldim}\sum_{\ket{E_n}\in\eneshell} \bra{E_n}\mO\ket{E_n} = \mcev{\mO}.
 \end{equation}
 To consider the zero mean random variable, we consider the operator which subtracted its one point function,
 \begin{equation}
 \mbOmc:= \mO-\mcev{\mO}
 \end{equation}
 and introduce the microcanonical Wightman function
 \begin{equation}
     \mcWightman{q}(t_1,t_2,\cdots,t_q)= \bTr{(\rhomc)^{\frac{1}{q}}\mbOmc(t_1)(\rhomc)^{\frac{1}{q}}\mbOmc(t_2)\cdots (\rhomc)^{\frac{1}{q}}\mbOmc(t_q)}.
 \end{equation}
 Then similar to the expectation value, the $q$-th moment can be written as 
 \begin{equation}
     \begin{split}
    &\quad \EVdiag{\qty(\bra{E_n}\mbOmc\ket{E_n})^q} \\&= \frac{1}{\shelldim}\sum_{\ket{E_n}\in\eneshell} \bra{E_n}\mbOmc\ket{E_n}^q\\&=
    \frac{1}{\shelldim}\sum_{\ket{E_{n_1}},\cdots\ket{E_{n_q}}\in\eneshell} \bra{E_{n_1}}\mbOmc\ket{E_{n_2}}\bra{E_{n_2}}\mbOmc\ket{E_{n_3}}\cdots \bra{E_{n_q}}\mbOmc\ket{E_{n_1}}\\
    &\times \delta_{n_1 n_2}\cdots \delta_{n_q n_1}
     \end{split}
 \end{equation}
 By assuming the non-degeneracy we have following equality,
 \begin{equation}
     \delta_{nm}= \lim_{T\to \infty}\frac{1}{T}\int_{0}^T dt \;e^{i(E_n-E_m)t}.
 \end{equation}
 By using this and rewriting the phase $i(E_n-E_m)t$ as a real time evolution of the operators, we obtain,
 \begin{equation}
     \begin{split}
    &\quad \EVdiag{\qty(\bra{E_n}\mbOmc\ket{E_n})^q}\\ &= 
    \frac{1}{\shelldim}\sum_{\ket{E_{n_1}},\cdots\ket{E_{n_q}}\in\eneshell} \prod_{i=1}^q \qty(\lim_{T_i\to\infty}\frac{1}{T_i} \int_{0}^{T_i} dt_i)\\
    &\times \bra{E_{n_1}}\mbOmc(t_1)\ket{E_{n_2}}\bra{E_{n_2}}\mbOmc(t_2)\ket{E_{n_3}}\cdots \bra{E_{n_q}}\mbOmc(t_q)\ket{E_{n_1}}\\
    &=\prod_{i=1}^q \qty(\lim_{T_i\to\infty}\frac{1}{T_i} \int_{0}^{T_i} dt_i) \bTr{\qty(\rhomc)^{\frac{1}{q}}\mbOmc(t_1)\cdots \qty(\rhomc)^{\frac{1}{q}}\mbOmc(t_q)}\\
    &=: \overlineT{G^{(q)}_{\mathrm{mc}}(t_1,t_2,\cdots,t_q)}
     \end{split}
 \end{equation}
 where the long time average is taken for any time variables.
 In the end, we find that the $q$-th momentum is obtained as the long time average of the $q$-th connected microcanonical Wightman function $G^{(q)}_{\mathrm{mc}}(t_1,t_2,\cdots,t_{2q})$. Notice that since $\EVdiag{\qty(\bra{E_n}\mbOmc\ket{E_n})^{2q}}$ is clearly positive and thus $\overlineT{G^{(2q)}_{\mathrm{mc}}(t_1,t_2,\cdots,t_{2q})}$ is also positive.
 \par
 For the chaotic system, we expect the following clustering of the $2q$-th moments
 \begin{equation}\label{eq:cluster}
     \overlineT{G^{(2q)}_{\mathrm{mc}}(t_1,t_2,\cdots,t_q)} \leq C\cdot q! \; \qty(a\cdot\overlineT{ G^{(2)}_{\mathrm{mc}}(t_1,t_2)})^p \qty(1+ \LandauO{\qty(\overlineT{\mcWightman{2}})^n}),\;n>0
 \end{equation}
  up to some $q\leq L$, where $C$ and $a$ is $n$ some numerical constant independent of $\Smc(E)$. $L$ is thought to be exponentially large with respect to entropy and the finite $L$ correction only gives negligible corrections to the sub-Gaussian property. This is key condition in our discussion which will be expected to be hold in some simple holographic theories. We will see this point in the next section. Note that this conditions will not be peculiar to the chaotic systems. For example, the free theories satisfy it. The difference between the chaotic and integrable systems are lied in the conditions we discuss in the next subsection. Also we will give some simple examples and explanation about this condition later section. This condition leads us the almost sub Gaussian property of $\DeltaETH{\ep}{\mO}$ almost trivially with $\sigma^2=\overlineT{\mcWightman{2}}$. Moreover, sometimes because of the ensemble equivalence which is going to be discussed in the next sub section, we can replace the condition \eqref{eq:cluster} to the one of the canonical ensemble which is typically easier to prove than the microcanonical one. In the canonical case, we consider following canonical Wightman functions,
  \begin{equation}
      \canWightman{2q}(t_1,t_2,\cdots,t_{2q})=\bTr{\qty(\rhocan)^{\frac{1}{2q}}\mbOcan(t_1)\qty(\rhocan)^{\frac{1}{2q}}\mbOcan(t_2)\cdots \qty(\rhocan)^{\frac{1}{2q}}\mbOcan(t_{2q})}
  \end{equation}
  where $\mbOcan$ is one point function subtracted operator $\mbOcan=\mO-\canev{\mO}$.
 Then, we expect the following clustering property for the long time average of the Wightman functions:
 \begin{ass}(Clustering property for the Wightman functions)\par 
     For the chaotic system and for enough high energy, we expect the following clustering of the microcanonical Wightman functions,
 \begin{equation}\label{eq:cluster}
     \overlineT{G^{(q)}_{\mathrm{mc}}(t_1,t_2,\cdots,t_q)} \leq C\cdot q! \; \qty(a\cdot\overlineT{ G^{(2)}_{\mathrm{mc}}(t_1,t_2)})^q \qty(1+ \LandauO{\qty(\overlineT{\mcWightman{2}})^n}),\;n>0.
 \end{equation}
 Also when we can use the ensemble equivalence, we also expect for the clustering for canonical Wightman function,
 \begin{equation}\label{eq:can_cluster}
     \overlineT{\canWightman{2q}} \leq C'\cdot q! \; \qty(a'\cdot\overlineT{ \canWightman{2}})^q\qty(1+ \LandauO{\qty(\overlineT{\canWightman{2}})^n})
 \end{equation}
up to some $q<L$ where $C'$, $a'$ and  $n>0$ is some numerical constant and $L$ is thought to be be exponentially large with respect to entropy.
 \end{ass}
 
Another property we expect to be hold in chaotic systems is the quantum mixing property for the thermal correlators. We say that the connected canonical Wightman function exhibits quantum mixing if
\begin{equation}
    \canWightman{2}(t_1,t_2) \to 0 \; \text{as} \; \abs{t_1 - t_2} \to \infty
\end{equation}
in the thermodynamic limit. For example, this is expected to be true for light operators beyond the Hawking-Page transition point \cite{Keski-Vakkuri:1998gmz, Maldacena:2001kr, Festuccia:2006sa}. We will see this holographic example later. For large but finite time difference $t_1-t_2$, we expect after some decay time $t_{\mathrm{dec}}$, 
\begin{equation}
    \canWightman{2}(t_1,t_2) \to \frac{1}{\qty(\shelldim)^\gamma} g(t_1,t_2) \; \text{as} \; \abs{t_1 - t_2} \gg t_{\mathrm{dec}},
\end{equation}
with a function $g(t_1, t_2) = \LandauO{1}$ and for some positive constant $\gamma$. Then, we conclude that:

\begin{ass}(Quantum Mixing Condition)\par 
For large but finite $\Smc(E)$, the canonical Wightman function satisfies the following quantum mixing condition:
\begin{equation}\label{eq:quantum_mixing}
    \overlineT{\canWightman{2}(t_1, t_2)} = \LandauO{1}\; \opnorm{\mO}^2\; e^{-\gamma \Smc(E)} = e^{-\LandauO{\Smc(E)}}.
\end{equation}
\end{ass}
This is actually expected to be hold in holographic theories with black hole duals. With these assumptions \eqref{eq:can_cluster}, \eqref{eq:quantum_mixing} and Theorem \ref{thm:almost_subgaussian}, we obtain double-exponential scaling for the diagonal matrix elements, i.e.,
\begin{thm}
By assuming \eqref{eq:can_cluster} and \eqref{eq:quantum_mixing} for the operator $\mO$ and setting $\ep = e^{-\eta \gamma \Smc(E)},\; \eta < \frac{1}{2}$, we find that $\DeltaETH{\ep}{\mO}$ exhibits double-exponential behavior \eqref{eq:double_expo}, that is,
\begin{equation}
    \DeltaETH{\ep}{\mO} \leq 4 e^{-\# \frac{\ep^2}{\overlineT{\canWightman{2}}} + \LandauO{\qty(\overlineT{\canWightman{2}})^n}} = 4 e^{-\# e^{\LandauO{\Smc(E)}}},\; \# > 0.
\end{equation}
\end{thm}

The proof follows by substituting $\sigma^2 = \overlineT{\canWightman{2}}$ and choosing a suitable $g$. The correction term $\qty(\overlineT{\canWightman{2}})^n$ will be discussed in concrete examples in Section \ref{subsubsec:perturbation}.
\subsubsection{Ensemble Equivalence}
Here, we discuss the relevant ensemble equivalence needed above. The microcanonical ensemble correlators are sometimes difficult to analyze, so to study the statistical properties of the systems, we usually use the canonical ensemble. Imprtantly, we can obtain the canonical (real time) Wightman functions from the Euclidean time ordered correlators.  In  holographic theories, the canonical ensemble is easier to handle since its dual is expressed in terms of the AdS black hole with periodicity in the imaginary time direction and we can easily compute the Eucldean time ordered correlation functions via usual dictionary. \par In the large $\shelldim$ limit, we generally expect ensemble equivalence between the microcanonical and canonical ensembles for certain observables \cite{Rulle:1990smr,Lima1972,Brandao:2015eos,TASAKI:2018eoe,Tasaki:2008}. 
 Specifically, we can bound the moment $\EVdiag{\bra{E_n}\mbOmc\ket{E_n}^{2q}}$, or equivalently, the long-time average of the microcanonical ensemble Wightman functions by that of the canonical ensemble under some assumptions. That is, we expect that
\begin{equation}\label{eq:ensemble_equivalance1}
\EVdiag{\bra{E_n}\mbOmc\ket{E_n}^{2q}} = \overlineT{\mcWightman{2q}} \leq \frac{e^{\beta(E+\Delta E)}Z(\beta)}{\shelldim} \overlineT{\canWightman{2q}}
\end{equation}
for some inverse temperature $\beta$.  Moreover, we expect that there exists a positive numerical constant and some temperature $\beta(E)$ such that
\begin{equation}\label{eq:ensemble_equivalance2}
    \frac{e^{\beta(E)(E+\Delta E)}Z(\beta(E))}{\shelldim} \leq \LandauO{\Smc(E)}.
\end{equation}
The $\beta(E)$ is determined by the usual thermodynamic law $\frac{\partial\Smc(E)}{\partial E}=\beta(E)$. This inequality can be roughly understood by the fact that in the thermodynamic limit
\begin{equation}
    Z(\beta) \sim e^{-\beta F(\beta)} \sim e^{-\beta(E-S_{\mathrm{th}}(E))} \sim e^{-\beta E}\shelldim.
\end{equation}
Here, the equivalence $\sim$ means that they are equal up to $\LandauO{\Smc(E)}$ factors. Then, by combining the clustering property \eqref{eq:can_cluster} and quantum mixing property \eqref{eq:quantum_mixing}, we see the scaling of the long time average of the canonical Wightman functions,
\begin{equation}
    \overlineT{\canWightman{2q}} < e^{-\LandauO{\Smc(E)}}.
\end{equation}
for some inverse temperature. Then, from \eqref{eq:ensemble_equivalance1} and \eqref{eq:ensemble_equivalance2}, we can bound the microcanonical one with the same scaling. \par 
Finally, we are going to comment on when we can prove \eqref{eq:ensemble_equivalance1}. For the variance, \ie, $p=1$ case, we can see that \eqref{eq:ensemble_equivalance1} is always true. That is, if the quantum mixing property of the canonical correlation function is true, the microcanonical one also shows mixing,
\begin{equation}
    \overlineT{\mcWightman{2}}<\LandauO{\Smc(E)}\;\overlineT{\canWightman{2}}=e^{-\LandauO{\Smc(E)}}.
\end{equation}
For the higher moment $p>1$, it is not trivial. Here we do not give universal discussion, instead we raise some specific examples which are relevant  for later models. The first example is that when the uniform distribution on $\eqset$ is symmetric distribution around its expectation value $\EV{\bra{E_n}\mO\ket{E_n}}=\mcev{\mO}$. In this case, the odd number of moments vanish and \eqref{eq:ensemble_equivalance1}. The other case is that the longtime average of the microcanonical Wightman correlators are exponentially suppressed. The detail of these discussion about the inequalities \eqref{eq:ensemble_equivalance1} and \eqref{eq:ensemble_equivalance2} is provided in Appendix \ref{subsec:ensemble_equivalnce}.

 
  \subsubsection{Off Diagonal Elements}
   We next move to the discussion of the off-diagonal elements. The expectation value of the matrix elements $\abs{\bra{E_n}\mO\ket{E_m}}^2$ of the uniform measure on $\offdiagset$ is given by 
  \begin{equation}
  \begin{split}
        \EVoffdiag{\abs{\bra{E_n}\mO\ket{E_m}}^2}&= \frac{1}{\shelldim(\shelldim-1)}\sum_{(\ket{E_n},\ket{E_m})\in\offdiagset} \abs{\bra{E_n}\mO\ket{E_m}}^2\\
        &= \frac{1}{\shelldim(\shelldim-1)}\sum_{(\ket{E_n},\ket{E_m})\in\offdiagset} \abs{\bra{E_n}\mbOmc\ket{E_m}}^2\\
         &= \frac{1}{\shelldim(\shelldim-1)}\sum_{\substack{(\ket{E_n},\ket{E_m})\\\in \mEeq\times \mEeq}} \abs{\bra{E_n}\mbOmc\ket{E_m}}^2\\ &-\frac{1}{\shelldim(\shelldim-1)}\sum_{\ket{E_n}\in \mEeq} \abs{\bra{E_n}\mbOmc\ket{E_n}}^2\\
         &= \frac{1}{\shelldim-1}\qty(\bTr{\qty(\rhomc)^{\frac{1}{2}}\mbOmc\qty(\rhomc)^{\frac{1}{2}}\mbOmc}-\overlineT{G_{\rm{mc}}^{(2)}})
  \end{split}
  \end{equation}
  Actually, by using the quantum mixing property the right hand side id bounded by 
  \begin{equation}
      \frac{1}{\shelldim-1}\qty(\bTr{\qty(\rhomc)^{\frac{1}{2}}\mbOmc\qty(\rhomc)^{\frac{1}{2}}\mbOmc}-\overlineT{G_{\rm{mc}}^{(2)}})\leq \frac{1}{\shelldim} \qty(\opnormTFD{\mbOmc})^2
  \end{equation}
  where we define another operator norm $\opnormTFD{\mO}$ on the thermofield double Hilbert space,
  \begin{equation}
      \opnormTFD{\mO}:= \sup_{\ket{\Psi}\in \mathcal{H}\otimes \mathcal{H}}\sqrt{\frac{_{LR}\bra{\Psi}\mO_L\otimes \mO_R\ket{\Psi}_{LR}}{_{LR}\braket{\Psi|\Psi}_{LR}}}.
  \end{equation}
  Note that this is UV finite even we take $\mO$ is genuine local operator on an one point. For light operator the operator norm will be independent of the Hilbert space dimension, we expect,
  \begin{equation}
       \EVoffdiag{\abs{\bra{E_n}\mO\ket{E_m}}^2} = \LandauO{1}e^{-\Smc(E)}
  \end{equation}
  For higher order momenta, we can again rewrite them as the thermal Wightman correlators. 
  \begin{equation}
      \begin{split}
&\quad \EVoffdiag{\abs{\bra{E_n}\mO\ket{E_m}}^{2q}}\\&= \frac{1}{\shelldim(\shelldim-1)}\sum_{\qty(\ket{E_n},\ket{E_m})\in\offdiagset}\abs{\bra{E_n}\mO\ket{E_m}}^{2q}\\&= \frac{1}{\shelldim(\shelldim-1)}\sum_{\qty(\ket{E_n},\ket{E_m})\in\offdiagset}\abs{\bra{E_n}\mbOmc\ket{E_m}}^{2q}\\&=\frac{1}{\shelldim(\shelldim-1)}\sum_{\substack{\qty(\ket{E_n},\ket{E_m})\\\in\mEeq\times \mEeq}}
\abs{\bra{E_n}\mbOmc\ket{E_m}}^{2q}\\
&=A_1-A_2\end{split}
  \end{equation}
  where we define. 
  \begin{equation}
      \begin{split}
          A_1&=\frac{1}{\shelldim(\shelldim-1)}\sum_{\substack{\qty(\ket{E_n},\ket{E_m})\\\in\mEeq\times \mEeq}}
\abs{\bra{E_n}\mbOmc\ket{E_m}}^{2q},\\
A_2&=\frac{1}{\shelldim(\shelldim-1)}\sum_{\ket{E_n}\in\mEeq} \abs{\bra{E_n}\mbOmc\ket{E_n}}^{2q}.
      \end{split}
  \end{equation}
  Since $A_1,A_2$ are positive value, we find
  \begin{equation}
      \EVoffdiag{\abs{\bra{E_n}\mO\ket{E_m}}^{2q}}<A_1+A_2.
  \end{equation}
  The second term $A_2$ is written by the long time average of the $2q$ points microcanonical Wightman function and bounded by above by the inverse powers of the energy shell dimension by using the quantum mixing property \eqref{eq:quantum_mixing} and the clustering property \eqref{eq:can_cluster}. 
  \begin{equation}
      A_2 \leq  \frac{1}{\shelldim-1} \overlineT{G_{(\rm{mc})}^{(2q)}} \leq q!\cdot  \LandauO{1}\;e^{-(q\gamma+1)\Smc(E)}
  \end{equation}
  What is new for the off-diagonal components is the first term. By doing the same trick for the diagonal case, we can rewrite the expression as
  \begin{equation}
      \begin{split}
      & A_1 =\frac{1}{\shelldim(\shelldim-1)} \prod_{i=1}^{q} \lim_{\substack{T_i\to \infty\\S_i\to\infty}} \frac{1}{T_{i}}\frac{1}{S_i}\int_{0}^{T_{i}} dt_{i}\int_{0}^{S_i}ds_i \\ &\times \sum_{\substack{\qty(\ket{E_{n_i}},\ket{E_{m_i}})\\\in(\mEeq\times \mEeq)^q}} \bra{E_{n_i}}\mbOmc\ket{E_{m_i}}\bra{E_{m_i}}\mbOmc\ket{E_{n_i}}e^{i(E_{n_i}-E_{n_{i+1}})t_i}e^{i(E_{m_i}-E_{m_{i+1}})s_i}\\
          &=\frac{1}{\shelldim(\shelldim-1)}\prod_{i=1}^{q}\lim_{\substack{T_i\to\infty\\S_i\to\infty\\\mathfrak{T}_{2q}\to\infty}}
          \frac{1}{T_{i}}\frac{1}{ S_i}\frac{1}{ \mathfrak{T_{2q}}}
          \prod_{k=1}^{2q}\int_{0}^{\mathfrak{T}_k} d\mathfrak{t}_k\; G_{\rm{mc}}^{(2q)}\qty(\mathfrak{t}_1,\mathfrak{t}_2,\cdots,\mathfrak{t}_{2q})\\
          &= \frac{1}{\shelldim(\shelldim-1)}\overlineT{G_{\rm{mc}}^{(2q)}} 
      \end{split}
  \end{equation}
  ,where we determine the time $\mathfrak{t}_k$ by following recursion relation
  \begin{equation}
  \begin{split}
           &-\mathfrak{t}_{2k-1}+\mathfrak{t}_{2k} = s_{k-1}-s_{k}\\
           &-\mathfrak{t}_{2k}+\mathfrak{t}_{2k+1} = -t_{k}+t_{k+1}.
  \end{split}
  \end{equation}
  Also the time $\mathfrak{T}_k$'s are determined by the same manner. Then by using the cluster property \eqref{eq:can_cluster} and the quantum mixing property\eqref{eq:quantum_mixing}, we obtain 
  \begin{equation}
      \begin{split}
    A_1     &<\frac{C}{\shelldim(\shelldim-1)}\;q!\cdot \qty(\overlineT{G_{\rm{can,\beta}}^{(2)}})^q= \LandauO{1}\;q! e^{-(\gamma q+2)\Smc(E)}
      \end{split}
  \end{equation}
  This means that we obtain the double exponential scaling for the off-diagonal elements \eqref{eq:off_diag_double_expo}. 
  \begin{thm}
          By assuming \eqref{eq:can_cluster},\eqref{eq:quantum_mixing} for the operator $\mO$ and set $\ep=e^{-\eta\gamma\Smc(E)},\;\eta<\frac{1}{2}$. Then, we see that $\DeltaETH{\ep}{\mO}$ shows the double exponential behavior \eqref{eq:off_diag_double_expo}, that is,
          \begin{equation}
              \offDeltaETH{\ep}{\mO}\leq 4 e^{-\#' \frac{\ep^2}{\overlineT{\canWightman{2}}}+\LandauO{\qty(\overlineT{\canWightman{2}})^n}} = 4e^{-\#' e^{\LandauO{\Smc(E)}}},\;\#'>0.
          \end{equation}
      \end{thm}
      We can freely include the correction for the sub-Gaussian in concrete examples. 
 
 \section{Examples: Bottom-Up Holographic Theories}\label{sec:Examples}
Here we discuss concrete examples that satisfy the conditions \eqref{eq:can_cluster} and \eqref{eq:quantum_mixing}, which lead to the double exponential scaling \eqref{eq:double_expo}, with some other assumptions that are expected to hold. We consider simple models of holographic theories, which are dual to QFTs on an AdS black hole. Here, we take the thermodynamic limit as the large $N$ limit and take $\mA$ as the set of light primary operators. Also in this section we do not consider the wormhole contribution for the higher point Wightman functions discussed in \cite{Saad:2019pqd}. In this sense, the model we discussed is not the actual holographic theories. However, we proceed our perturbative effective field theory discussion for the simple illustration. We expect our discussion can be used to more realistic quantum gravity. We give a few comments on this point in section \ref{subsec:back_reaction}. We also provide some comments on more general systems. 

\subsection{Holographic Theories and Large $N$ Generalized Free Field Theories}

\subsubsection{Clustering Property}
One simple or trivial example that shows the clustering property of long-time-averaged thermal Wightman correlators \eqref{eq:can_cluster} is the generalized free field theories with exact large $N$ factorization. Before discussing large $N$ factorization, let us rewrite $\canWightman{2q}$ in the usual form of the thermal correlators:
\begin{equation}
   \canWightman{2q}(t_1,t_2,\cdots,t_{2q}) = \Tr{\rhocan \mbOcan(\tau_1)\mbOcan(\tau_2)\cdots \mbOcan(\tau_{2q})},
\end{equation}
where
\begin{equation}
    \tau_k = t_k - i a_k \beta, \quad a_k = \frac{2q-k}{2q}, \quad k = 1, \cdots, 2q.
\end{equation}
From the general discussion of the Wick rotation \cite{Simmons-Duffin:2019TASI}, we can obtain these Wightman correlators from the Euclidean time-ordered correlators,
\begin{equation}
    G_{\mathrm{E},\beta}^{(2q)}(\tE_1,\tE_2,\cdots,\tE_{2q}) = \bTr{\rhocan\mathcal{T}_{\mathrm{E}}\qty(\mbOcan(\tE_1)\mbOcan(\tE_2)\cdots \mbOcan(\tE_{2q}))},
\end{equation}
via $\tE_k = \frac{2q-k}{2q} \beta + it_k$.

Now, we suppose we can use Wick contraction:
\begin{equation}
\begin{split}
    \canWightman{2q}(t_1,t_2,\cdots,t_{2q}) &= \bTr{\rhocan \mbOcan(\tau_1)\mbOcan(\tau_2)\cdots \mbOcan(\tau_{2q})}\\
    &= \sum_{\text{Pairs}} \prod_{(i,j) \in \text{Pairs}} \bTr{\rhocan \mbOcan(\tau_i)\mbOcan(\tau_j)},
\end{split}
\end{equation}
where we denote $\text{Pairs}$ as the set of pairs of indices appearing in the Wick contraction, and we take the sum over all possible pairs. Of course, we assume that $N$ is large but finite, and that the exact large $N$ factorization does not hold, as it does in the strict large $N$ limit. However, for the moment, let us assume that exact factorization works even for large but finite $N$ as a simple toy example. Later, we will include $1/N$ corrections.

With this large $N$ clustering, we can trivially find the clustering of the long-time-averaged thermal correlators:
\begin{equation}
\begin{split}
    \overlineT{\canWightman{2q}(t_1,t_2,\cdots,t_{2q})} &= \sum_{\text{Pairs}} \prod_{(i,j) \in \text{Pairs}} \overlineT{\bTr{\rhocan \mbOcan(\tau_i)\mbOcan(\tau_j)}}\\
    &= (2q-1)!! \times \qty(\overlineT{\canWightman{2}(\tau_i,\tau_j)})^q\\
    &\leq q! \; \qty(2\overlineT{\canWightman{2}})^q.
\end{split}
\end{equation}

This is exactly the form of \eqref{eq:can_cluster}, and thus we find the sub-Gaussian properties with $\sigma^2 = \overlineT{\canWightman{2}}$.
 \subsubsection{Quantum Mixing}
It is known that the black hole horizon dramatically changes the spectral representation of the thermal two-point functions of the dual holographic theories. We denote the thermal two-point Wightman function as
\begin{equation}
    G_{\mathrm{can},\beta}^{(2)}(t):=  \canev{\mbOcan\qty(t+i\frac{\beta}{2})\mbOcan} := \frac{1}{Z(\beta)}\bTr{\qty(\rhocan)^{\frac{1}{2}}\mbOcan(t)\qty(\rhocan)^{\frac{1}{2}}\mbOcan(0)}
\end{equation}
where $Z(\beta)$ is the thermal partition function, $Z(\beta)= \Tr{e^{-\beta H}}$, and $\rhocan$ is the canonical ensemble at the inverse temperature $\beta$, given by $\rhocan= e^{-\beta H}/Z(\beta)$. It is known that beyond the Hawking-Page transition, i.e., when $\beta<\beta_{\mathrm{HP}}$, the gravitational dual of the canonical ensemble is the black hole. We can always make $\beta$ larger than $\beta_{\mathrm{HP}}$ by taking $E$ large enough. The real-time thermal two-point function in this phase is expressed as 
\begin{equation}
\begin{split}
    G_{\mathrm{can},\beta}^{(2)}(t)&=G_{\text{BH}}^{(2)}(t)+\text{(contributions from other saddles)}.
\end{split}
\end{equation}
Here, $G_{\text{BH}}^{(2)}(t)$ denotes the holographic thermal two-point function evaluated from the effective free field theory on the black hole background. It is known that the leading the blackhole two point function shows ring down behavior of the thermal correlator arises from the black hole horizon \cite{Maldacena:2001kr,Barbon:2003aq,Festuccia:2006sa,Barbon:2014rma}. The horizon induces the continuum spectrum of the QFT on the exterior of the black hole. In terms of von Neumann algebra, the operator algebra of the Rindler wedge of the black hole (or Minkowski space) is of type $\grethree_1$, which exhibits the continuum spectrum of the modular Hamiltonian \cite{Leutheusser:2021frk,Furuya:2023fei,Gesteau:2023rrx}. The discussion above directly generalizes to higher-dimensional AdS$_{d+1}$ black holes or SYK models. Also the decaying structure and pole structure of the thermal correlators are verified in the matrix quantum mechanics \cite{Iizuka:2008hg}.  Then, the actual long time average is coming from the other saddle points. It is discussed in \cite{Saad:2019pqd} that the non-decaying behavior of the late time two pint function is coming from the gravitational path integral with baby universe exchanging. For two dimensional JT  gravity case, the order of the late time behavior is expected $\LandauO{e^{-S_0}}$ where $S_0$ is ground state microcanonical entropy. Even though we do not have concrete discussion for the higher dimensional for the actual gravitational late time behavior of the correlation function, we expect the similar story. In there dimensional case, there are non-wormhole gravitational saddle points which are given by $\mathrm{SL}(2,\mathbb{Z})$ of the non-rotating blackhole and orbifolds. Let us consider BTZ black hole, which has the modular parameter $\tau=i\tau_2=\frac{i\beta}{2\pi}$. Then, the other saddle points are given by the $\mathrm{SL}(2,\mathbb{Z})$ transformation, $\tau'=\frac{a\tau+b}{c\tau+d}$. The late-time behavior of two-point functions and the regularized action with modular parameter $\tau'$ are given by \cite{Kleban:2004rx},
\begin{equation}
\begin{split}
    (\canWightman{2})_{\tau'}&\sim \exp{-2\Delta \frac{\abs{a}\tau_2}{\abs{a\tau+b}^2}t},\\
    I(\tau')&=\frac{\pi}{4\GN}\frac{\tau_2}{\abs{a\tau+b}^2}.
\end{split}
\end{equation}
The contribution from $a=0$ corresponds to thermal AdS$_3$, where the two-point function shows periodic oscillation. Indeed, the exact form is given by
\begin{equation}\label{eq:thermal_AdS}
    G_{\text{Thermal AdS}}^{(2)}(t)= \const\times \sum_{n=-\infty}^\infty \qty(\frac{1}{\cos{(t+in\beta)}+\cos{\phi}})^{\Delta}.
\end{equation}
Here we set the AdS radius to one. Except for the thermal AdS contribution, we find decaying behavior. If we can exchange the long-time average and the geometry sum, we find that 
\begin{equation}
    \overlineT{\canWightman{2}} = e^{-I_{\mathrm{BH}}+I_{\mathrm{Thermal AdS}}} \;\overlineT{G^{(2)}_{\mathrm{Thermal AdS}}}=\LandauO{1}\opnorm{\mO}^2\;e^{-\gamma \Smc(E)}.
\end{equation}
In the case of three-dimensional pure Einstein gravity,
\begin{equation}\label{eq:gamma_three_dim}
    \gamma= \frac{1}{2}\qty(1-\qty(\frac{\Smc(E)}{2\pi E})^2).
\end{equation}
However, notice that if we consider the micorcanonical correlation functions, the thermal AdS geometry is not included in the saddle points summation since it is just obtained from the vacuum AdS and have energy below the blackhole threshold. Thus, for the late time correlation function which only including the enough highenergy saddles is basically governed by the wormhole geometry. 
Combining the discussion of the ring down behavour and the late time behavour from the wormhole geometries, we expect the quantum mixing behavior \eqref{eq:quantum_mixing} for generic holographic theories which have the blackhole as a leading saddle. Note that below the Hawking-Page transition, there is no quantum mixing since the correlation function \eqref{eq:thermal_AdS} shows recurrence with a recurrence time $t_{\mathrm{rec}}^{\qty(\mathrm{Thermal AdS})} = \LandauO{1}$. 

With these discussions, we see 
\begin{equation}
\begin{split}
    \overlineT{\canWightman{2}} = \overlineT{G^{(2)}_{\mathrm{BH}}}+G_{\mbO}\;e^{-\gamma \Smc(E)}=e^{-\LandauO{\Smc(E)}}
\end{split}
\end{equation}
where $G_{\mbO}=\LandauO{1}\opnorm{\mO}^2$. \par 

There is a caveat regarding conformal symmetry. Due to conformal symmetry and its representation, the non-degeneracy and non-resonance conditions \eqref{eq:non_resonance} do not hold. For example, there are many pairs of energy eigenstates whose energy difference is one. There are three options to resolve this problem. One is to modify the inequality in Theorem \ref{thm:thermalization_inequality} and Theorem \ref{thm:equilibration_inequality}. We discuss the modified inequality in Appendix \ref{sec:WIthout_non-resonance}. The other option is to deform the CFT slightly by some irrelevant operators, which break conformal symmetry a little. Since irrelevant deformations may not change the bulk interior, especially the structure of the black hole, the structure of the quantum mixing property (and also the clustering property) will be preserved.
The third option is that we are just focusing on the conformal primary operators and primary states.  In conformal field theories, we can discuss ETH just for primary operators \cite{Lashkari:2016ethcft}\footnote{ In the paper, they take the large volume limit as the thermodynamic limit.}. To this end, let us introduce a sets of primary energy eigenstates with certain energy or equivalently conformal dimension $E^P$ such that $E\leq E^P \leq E+\Delta E$ and denotes the sets $\mathcal{E}^{P}_{E,\Delta E}$. We then consider the microcanonical ensemble jut with primary states. The expectation value of such a primary microcanonical ensemble is given by $\Pmcev{\mO}=\frac{1}{\Pshelldim}\sum_{\ket{E_n^P}\in\mathcal{E}^{P}_{E,\Delta E}}\bra{E_n^P}\mO\ket{E_n^P}$.
Then, we consider the following ETH for primary operators
\begin{equation}
    \begin{split}
        \text{Diagonal} & \quad \abs{\bra{E_n^P}\mO^P\ket{E_n^P} - \mcev{\mO}} = e^{-\LandauO{\Smc(E)}} \\
        \text{Off-diagonal} & \quad \abs{\bra{E_n^P}\mO^P\ket{E_l^P}} = e^{-\LandauO{\Smc(E)}}.
    \end{split}
\end{equation}
for any primary states $\ket{E^P_n},\ket{E^P_m}\in \Peqset$ and some light primary operators $\mO^P$. Then, similar to the usual ETH, it holds that 
\begin{equation}
    \begin{split}
       \abs{\bra{E^P_n}\mO\ket{E^P_n}- \Pmcev{\mO}} <\opnorm{\mO}\qty(\ep+\sqrt{\Pshelldim\DeltaETHP{\ep}{\mO}})
    \end{split}
\end{equation}
where $\Pshelldim=\abs{\mathcal{E}^{P}_{E,\Delta E}}$ and $\DeltaETHP{\ep}{\mO}$ is defined as
\begin{equation}
    \DeltaETHP{\ep}{\mO}:= \Prob{\mathcal{E}^{P}_{E,\Delta E}}{\abs{\bra{E_n^P}\mO\ket{E_n^P}-\Pmcev{\mO}}>\ep}.
\end{equation}
We can write similar inequality for the off-diagonal components. We can show that the second order of the moments on the uniform distributions for primary energy eigenstates are bounded above by the one of the distribution on the full energy eigenstates and we can find the quantum mixing for the primary microcanoncal correlators,
\begin{equation}
\begin{split}
     \frac{1}{\Pshelldim}\sum_{\ket{E_n^P}\in\Peqset}\qty(\bra{E_n^P}\qty(\mO^P-\Pmcev{\mO^P})\ket{E_n^P})^2&\leq \EVdiag{\bra{E_n}\mbOmc\ket{E_n}^2}\\&= \overlineT{\mcWightman{2}}=e^{-\LandauO{\Smc(E)}}.
\end{split}
\end{equation}
The proof is almost same as the one of the ensemble equivalence in Appendix \ref{app:proof}. For the higher order moments, if we see the two microcanonical averages of the one point functions are exponentially close, we can use the results from the full energy eigenstates microcanical ensemble. However, we do not have good justification for it and we do not delve in this any more here.
\par 
As a comment for more generic systems with a general thermodynamic limit, in quantum many-body systems with a discrete spectrum, the thermal two-point functions exhibit Poincaré recurrence. Here we consider the following recurrence:
\begin{equation}
    G_{\mathrm{can},\beta}^{(2)}(t+t_{\mathrm{rec}}) =  G_{\mathrm{can},\beta}^{(2)}(t) + e^{-\LandauO{\Smc(E)}}.
\end{equation}
Then the long-time average is 
\begin{equation}
    \overlineT{G_{\mathrm{can},\beta}^{(2)}} = \frac{1}{t_{\mathrm{rec}}}\int_{0}^{t_{\mathrm{rec}}} dt\;  G_{\mathrm{can},\beta}^{(2)}+e^{-\LandauO{\Smc(E)}}.
\end{equation}
Typically, around the decay time $t_{\mathrm{dec}}$, which is the time scale at which the connected thermal correlator becomes thermodynamically small, it is expected that 
\begin{equation}
    G_{\mathrm{can},\beta}^{(2)} = \LandauO{1}\times  e^{-\frac{t}{t_\mathrm{dec}}} + e^{-\LandauO{\Smc(E)}}.
\end{equation}
Thus, we see
\begin{equation}
    \overlineT{G_{\mathrm{can},\beta}^{(2)}} =  \LandauO{\frac{t_{\mathrm{dec}}}{t_{\mathrm{rec}}}}+e^{-\LandauO{\Smc(E)}}.
\end{equation}
Typical chaotic systems show $t_{\mathrm{dec}}=\LandauO{\beta}$ and $t_{\mathrm{rec}}=e^{\LandauO{\Smc}}$. Thus, chaotic systems show quantum mixing. \par 

In summary, we find that if the system has a gravity dual with a semi-classical black hole solution, then the system shows quantum mixing behavior of the thermal two-point function, at least in the canonical ensemble.

 \subsection{Adding $1/N$ Corrections}\label{subsec:1/N_correction}
In the last subsection, we focused only on the generalized free part in the large $N$ factorization. As discussed this is correct when we take strict large $N$ limit. However, in realistic holographic systems, we usually have sub-leading corrections in the large $1/N$ expansion. Since we are interested in the scaling of quantities with large but finite $N$, we seriously take care about these $1/N$ corrections. These corrections arise from the weak interactions of the fields. This means that we have non-trivial connected correlation functions that are not just products of two-point functions. Here, we discuss whether these interactions affect the assumption in \eqref{eq:can_cluster}.

\subsubsection{Corrections to the Two-Point Function}
Now we discuss the $1/N$ corrections to the two-point functions. Suppose we have a $\Phi^4$ interaction in the bulk,
\begin{equation}
    S_{\mathrm{EFT}}= \int d^{d+1}x \sqrt{-G}\qty[\partial_M \Phi \partial^M \Phi+m^2 \Phi^2+\frac{g_4}{4!}\Phi^4]
\end{equation}
where $g_4= 1/\mathrm{Poly}(N)$. We could also consider other polynomial couplings $\Phi^n$ with $n=\LandauO{N^0}$. Additionally, we do not concern ourselves with gravitational corrections here. Instead, we focus purely on the $\Phi^4$ theory in a thermal geometry for illustration. Using standard techniques in QFT, we can compute the two-point function in terms of the unperturbed one and the 1PI diagrams. Even in the three-dimensional case, it seems difficult to compute the quantum corrections in the black hole background exactly. However, we can estimate the late-time behavior of the exact two-point function using perturbation theory. The following corrections arise:
\begin{itemize}
    \item From diagrams like the sunset diagram, we obtain the following expression at each order in perturbation theory \cite{Festuccia:2006sa}:
    \begin{equation}
    \begin{split}
        \canWightman{2}(t) &= \sum_{n=0}^{n_{\max}} \canWightman{2,n}(t) \\
        \canWightman{2,n}(t) &= \sum_{k=0}^n c_{n,k}t^k \canWightman{2,0}(t)
    \end{split}
    \end{equation}
    where $\canWightman{2,0}(t)$ is the two-point function of the unperturbed theory. Thus, the total two-point function exhibits quantum mixing.
    \item Loop diagrams introduce UV divergences, as is typical in quantum field theory. After renormalization, the conformal dimension acquires anomalous corrections \cite{Banados:2022nhj}. This implies that the quasi-normal modes are renormalized or corrected as well.
\end{itemize}

 \subsubsection{Perturbative Computation and Almost Sub-Gaussianity}\label{subsubsec:perturbation}
In this subsection, we discuss the Wick diagram computation for the long-time average of the thermal correlators in perturbation theory for $1/N$ corrections. 

\begin{figure}[h]
    \centering
    \includegraphics[width=0.85\linewidth]{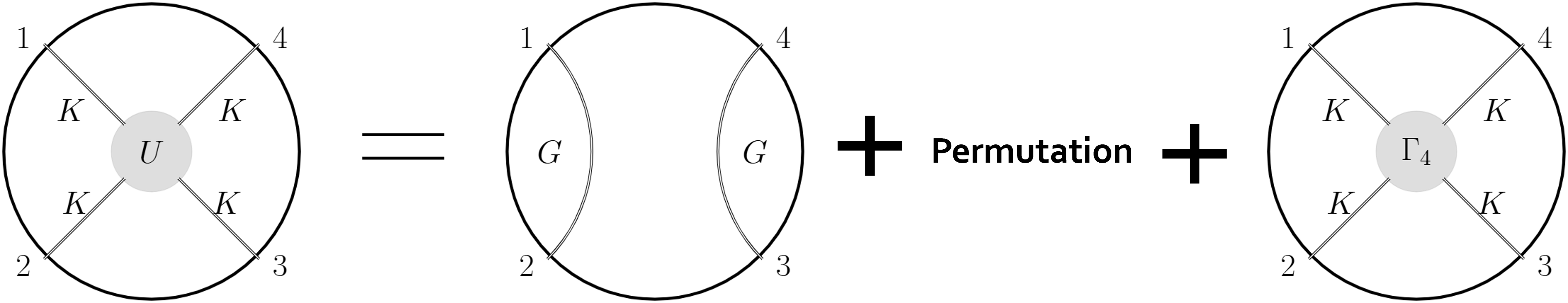}
    \caption{Sketch of the Witten diagram for the four-point function. We denote $K$ as the bulk-to-boundary propagator, and $G$ denotes the boundary correlators. $\Gamma_4$ represents the sum of the 1PI diagrams for the four-point vertex.}
    \label{fig:witten_diagram}
\end{figure}

First, let us consider the case $q=2$, i.e., we consider four-point functions. Suppose we can compute the connected Euclidean thermal correlator using the Witten diagram:
\begin{equation}
    \begin{split}
        \GE{4}{\tE_1,\tE_2,\tE_3,\tE_4} &= \prod_{k=1}^4 \int d^{d+1}\XE_k \KBb{\xE_1}{\XE_1} \KBb{\xE_2}{\XE_2} \\
        &\times \KBb{\xE_3}{\XE_3} \KBb{\xE_4}{\XE_4} U(\XE_1,\XE_2,\XE_3,\XE_4).
    \end{split}
\end{equation}
Here, we denote $\KBb{\xE}{\XE}$ as the bulk-to-boundary propagator. The function $U$ represents all of the amputated Witten diagrams with four external legs. For large $N$, we have a perturbation series:
\begin{equation}
\begin{split}
\GE{4}{\tE_1,\tE_2,\tE_3,\tE_4} &= \canWightman{2}(\tE_1,\tE_2)\canWightman{2}(\tE_3,\tE_4)+\text{(permutation)} \\
&\quad + G_{\beta,\mathrm{conn}}^{(4)}({\tE_1,\tE_2,\tE_3,\tE_4}) \\
G_{\beta,\mathrm{conn}}^{(4)}({\tE_1,\tE_2,\tE_3,\tE_4}) &:= \prod_{k=1}^4 \int d^{d+1}\XE_k \KBb{\xE_1}{\XE_1} \KBb{\xE_2}{\XE_2} \\
       &\quad \times \KBb{\xE_3}{\XE_3} \KBb{\xE_4}{\XE_4} \\
       &\quad \times \Gamma_4(\XE_1,\XE_2,\XE_3,\XE_4).
\end{split}
\end{equation}
Here, $\Gamma_4$ denotes the 1PI vertex. See Figure \ref{fig:witten_diagram} and Figure \ref{fig:1p1} for intuition.

In the semi-classical limit, the bulk-to-boundary propagator is also have similar structure to the boundary two point function:
\begin{equation}
\begin{split}
     \KBb{x}{X}=K_{\mathrm{Bb}}^{\mathrm{BH}}({x},{X})  e^{-I_{\mathrm{BH}}} + e^{-\gamma \Smc(E)} K_{\mbO}(X_{\mathrm{E}}).
\end{split}
\end{equation}
After the Wick rotation, since the boundary points always lie outside the black hole, the first term shows similar decay to the boundary correlators:
\begin{equation}
    \overlineT{ K_{\mathrm{Bb}}^{\mathrm{BH}}({x},{X_{\mathrm{E}}})}=0.
\end{equation}
Then, similar to the boundary correlators, we assume:
\begin{equation}
    \overlineT{\KBb{x}{X_{\mathrm{E}}}} = e^{-\gamma \Smc(E)} K_{\mbO}(X_{\mathrm{E}}),
\end{equation}
with some order $\LandauO{1}$ function $K_{\mbO}(X)$. Using these expressions for the bulk-to-boundary propagators, we find:
\begin{equation}
    \begin{split}
        \overlineT{G_{\beta,\mathrm{conn}}^{(4)}} &= \qty(\prod_{k=1}^4 \int d^{d+1}{X_{\mathrm{E}}}_k \overlineT{K_{\mathrm{Bb}}({x_k},{X_k})})\Gamma_4(X_1,X_2,X_3,X_4) \\
       &= e^{-4 \gamma \Smc(E)}\qty(\prod_{k=1}^4 \int d^{d+1}X_k K_{\mbO}(X_k))\Gamma_4(X_1,X_2,X_3,X_4).
    \end{split}
\end{equation}
Thus, the connected part is of order $\LandauO{\qty(\overlineT{\canWightman{2}})^4}$ and is exponentially small compared to the generalized free part. In total,
\begin{equation}
\begin{split}
     \overlineT{\canWightman{4}} &= 3 \qty(\overlineT{\canWightman{2}})^2 + \Gamma_4 \qty(\overlineT{\canWightman{2}})^4, \\
     \Gamma_4 &= \frac{1}{(G_{\mbO})^4}\qty(\prod_{k=1}^4 \int d^{d+1}X_k K_{\mbO}(X_k))\Gamma_4(X_1,X_2,X_3,X_4)).
\end{split}
\end{equation}
\noindent
\textbf{General $2q$ point function}\par 
We now extend this discussion to the general $2q$-point function $\overlineT{\canWightman{2q}}$ and its decomposition into exact two-point functions and 1PI diagrams \footnote{We note that the discussion in the previous version lacks contribution from many diagrams. We would like to thank Yu Nakayama pointing out this. }. From the Wick contraction analysis, we obtain the following decomposition:

\begin{equation}\label{eq:2q_point}
\begin{split}
     \overlineT{\canWightman{2q}} &= \sum_{\{n_k\}:\sum_{k=1}^q k n_k=q } \qty(\overlineT{\canWightman{2}})^{n_1} \frac{(2q)!}{\prod_{k=1}^q (n_k)! ((2k)!)^{n_k}} \prod_{k=2}^q \qty(\Gamma_{2k}\qty(\overlineT{\canWightman{2}})^{2k})^{n_k}.
\end{split}
\end{equation}
where the formula is derived by the combinations. $\Gamma_{2k}$ are the $2k$-order 1PI diagrams obtained by similar manner to $\Gamma_4$. For example, for $\Phi^4$ theory, the leading diagrams of $\Gamma_6$ is given by loop diagrams with three $\Gamma_4$.

We assume $\Gamma_{2k}$ are small in large $N$ and $\Gamma_{2k}$ are monotonically decrease in $k$. The power of $\overlineT{\canWightman{2}}$ is determined as follows: First we have $n_{1}$ pairs of external lines which are not connected in each other. Next we consider connect parts of left $(2q-2n_1)$ lines in $n_2$ number of 4 vertex diagrams. We continue this procedure up to $2q$ vertex diagrams.
\begin{figure}[h]
    \centering
    \resizebox{0.75\textwidth}{!}{%
\begin{circuitikz}
\tikzstyle{every node}=[font=\LARGE]
\filldraw[lightgray,opacity=0.5]  (3.25,7.5) circle (0.75cm);
\filldraw[black]  (6.5,7.5) circle (0.15cm);
\filldraw[black]  (11,7.5) circle (0.15cm);
\filldraw[black]  (9.5,7.5) circle (0.15cm);
\draw(3.25,7.5)node{$\Gamma_4$};
\draw(4.75,7.5)node{\Huge{$=$}};
\draw(6.5,8)node[above]{{$g_4$}};
\draw(8,7.5)node{{\Huge{$+$}}};
\draw(12.5,7.5)node{{\Huge{$+$}}};
\draw(13.5,7.5)node{{\Huge{$\cdots$}}};
\draw[dashed] (5.5,8.5)--(7.5,6.5);
\draw[dashed] (5.5,6.5)--(7.5,8.5);
\draw[dashed] (8.5,8.5)--(9.5,7.5);
\draw[dashed] (8.5,6.5)--(9.5,7.5);
\draw[black] (9.5,7.5)to [out=60,in=120] (11,7.5);
\draw[black] (9.5,7.5)to [out=-60,in=-120] (11,7.5);
\draw[dashed] (11,7.5)--(12,8.5);
\draw[dashed] (11,7.5)--(12,6.5);
\end{circuitikz}}
    \caption{Sketch of the 1PI diagram $\Gamma_4$ in the perturbation series of $g_4$.}
    \label{fig:1p1}
\end{figure}

Next, we are going to compare \eqref{eq:bound_moment_quadratic} with \eqref{eq:2q_point}. In the comparizon, it is natural to identify 
\begin{equation}
    \sigma^2 = \overlineT{\canWightman{2}},\; \Gamma_{2k}= g_{2k}.
\end{equation}
Under this identification, we discuss following mathemathical propasitions,
\begin{conj}\label{prop:1/N_correction_moment}  Consider the two positive functions for vectors $x=(x_1,x_2,x_3,\cdots,x_{L})$ , 
    \begin{equation}
        \begin{split}
            G^{(2q)}(x)&:= \sum_{\{n_k\}:\sum_{k=1}^qkn_k=q} \frac{(2q)!}{\prod_{k=1}^q ((2k)!)^{n_k}(n_k)!} x_1^{n_1-q} \; \prod_{k=2}^q \qty(x_k)^{n_k} \\
            E^{(2q)}(x)&:= 2^{q+1}q!\; \sum_{\{m_k\}:m\leq m_{\max}} \frac{\Gamma(m+q)}{\Gamma(q)}\prod_{k=2}^{M} \frac{1}{m_k!}(-x_k)^{m_k}.
        \end{split}
    \end{equation}
    We put $x_1 = \overlineT{\canWightman{2}}$ and $x_{k}=\Gamma_{2k}\qty(\overlineT{\canWightman{2}})^{2k}$.
    For sufficiently small $x$ and for some positive integer $m_{\max}$, we have 
    \begin{equation}\label{eq:1/N_correction_moment}
        G^{(2q)}(x) \leq E^{(2q)}(x).
    \end{equation}
\end{conj}
In Appendix \ref{appendix:1/N_correction_moment}, we provide numerical evidence supporting this proposition for the case $\ep_{k}=-1$.
 In our application, we have , which is extremely small and we take $x_k= y^k , k\geq 2$ with $y=\frac{\overlineT{\canWightman{2}}}{S}$. The numerics suggest that the inequality holds for $M=m_{\max}=\floor{\frac{q}{2}}$, and up to $y\sim 1/q^c$ with $c\sim 1$. Since $y=\frac{1}{\mathrm{Poly}(\Smc)}e^{-\LandauO{\Smc}}$, we may thus use the inequality for $q$ up to $q=\LandauO{(\Smc(E))^{\frac{1}{c}}}e^{\frac{2}{c}\gamma\Smc(E)}$. Notice that we do not prove the inequality for really large $q,M$ and $m_{\max}$ and thus still open questions. Possibly we may wonder the correlation functions become large because of many diagram but we expect we can bounded them by taking large enough $m_{\max}$. 
For arbitrarily large $q$, we can still employ a simple, albeit loose, bound:
 \begin{equation}
     \begin{split}
         \EVdiag{\bra{E_n}\mbO\ket{E_n}^{2q}}&=\frac{1}{\shelldim}\sum_{\ket{E_n}\in\eneshell}\bra{E_n}\mbO\ket{E_n}^{2q}\\&\leq (\opnorm{\mbO})^{2q}\\&\leq (2\opnorm{\mO})^{2q}.
     \end{split}
 \end{equation}
 \par 
We now invoke Conjecture \eqref{appendix:Ela^2X^2_bound}. Setting $\sigma^2=\overlineT{\canWightman{2}}, g_{2k}=\Gamma_{2k} $, $M=m_{\max}=\floor{\frac{q}{2}}, \sigma'=2\opnorm{\mO}$ and $L=\LandauO{(\Smc(E))^{\frac{1}{c}}}e^{\frac{2}{c}\gamma\Smc(E)}$, we obtain the scaling
  \begin{equation}
      \begin{split}
          \frac{\sigma'^2}{(L+1)\sigma^2}\sim \frac{1}{\Smc(E)^\frac{1}{c}}e^{-\qty(\frac{2}{c}-1)\Smc(E)}=e^{-\LandauO{\Smc(E)}},
      \end{split}
  \end{equation}
and by choosing $\ep=\sigma^\lambda,\;\lambda<1$ we find
 \begin{equation}
     \DeltaETH{\ep}{\mO}\leq e^{-\#e^{\LandauO{\Smc(E)}}}\qty(e^{-\#'e^{-\LandauO{\Smc(E)}}}+e^{-\#''e^{\LandauO{\Smc(E)}}}),\;\#,\#',\#''>0.
 \end{equation}
 Note that $M$ is originally independent from $q$ and possibly relate to $L$. However, for computational reason we set $M$ depends on $q$. Therefore, even after including perturbative $1/N$ corrections, we still recover the double exponential scaling of $\DeltaETH{\ep}{\mO}$ if the inequality \eqref{eq:1/N_correction_moment} is correct. An entirely analogous analysis applies to the off-diagonal quantity $\offDeltaETH{\ep}{\mO}$. In conclusion, under the assumptions of non-degeneracy and non-resonance (which might be further relaxed) and with assumptions that the conjectured inequalities are valid, this simple holographic toy model, dual to a weakly interacting scalar theory in AdS, continues to exhibit strong ETH.
\subsubsection{Discussion: Including Back Reaction and wormhole}\label{subsec:back_reaction}

 In above section, our discussion does not including gravitational back reaction. However, in realistic AdS/CFT correspondence with large but finite $N$ case, we need to include it. Especially, for large $q=\LandauO{\Smc(E)}$, we need to think that the product of the light operators as the heavy operators. In such cases, it is highly challenging to discuss the correlation functions since we do not have a nice effective field theory picture in the bulk black hole with temperature $\beta$. However, it may be possible to tame this kind of huge multi-point correlation functions by replacing the vacuum of the generalized free fields with another vacuum. Let me clarify this point. Suppose we have weakly coupled QFTs on the AdS black hole with temperature $\beta$. In terms of AdS/CFT language, this corresponds to considering the Fock space constructed by the GNS construction from the canonical ensemble with inverse temperature $\beta$ \cite{Leutheusser:2021frk}. Thus, the effective weakly coupled field theory picture is highly state-dependent. However, we have another point of view for the black hole which can be seen as the excited states from the AdS vacuum. In the AdS vacuum, we have another Fock space and black holes will be obtained from a huge number of insertions of operators to this vacuum.\par
 Here we can do similar things for the higher point function for the finite temperature, $\canWightman{2q}(t_1,t_2,t_3,\cdots,t_{2q})$. Suppose we take $t_1,t_2\gg t_3,\cdots,t_{2q}$ and divide the sets of the operators into the first two and the rest $(2q-2)$. Like the above arguments, we consider the rest $(2q-2)$ as huge excitation for the black hole and recognize the first two operators as a probe field on this excited black hole. In concrete terms, we suppose the following clustering property appears:
 \begin{equation}
     \begin{split}
\canWightman{2q}(t_1,t_2,t_3,\cdots,t_{2q})\approx G^{(2)}_{\mathrm{Excited}}(t_1,t_2)\canWightman{2q-2}(t_3,\cdots,t_{2q-2}),
     \end{split}
 \end{equation}
 where $G^{(2)}_{\mathrm{Excited}}(t_1,t_2)$ is computed from the black hole with excitation. We will perform the concrete computation here from now on. To express the huge excitation, we consider the AdS Vaidya metric with fluid matter whose stress tensor has the order $\LandauO{\frac{1}{\GN}}$. The AdS Vaidya metric is given by:
 \begin{equation}
     ds^2=-\qty(\frac{r^2}{\LAdS^2}-M(v))dv^2+2 dv dr + r^2 d\phi^2.
 \end{equation}
 The profile of $M(v)$ is determined if we assign the concrete form of the bulk fluid stress tensor, but the detail is not important. Since we expect that at late time we put the probe operators, the black hole settles down to another equilibrium and we just use the correlation function of the other equilibrium. Thus, the two-point function $G^{(2)}_{\mathrm{Excited}}(t_1,t_2)$ shows the exponential decay. So,
 \begin{equation}\label{eq:long_time_factorization}
 \begin{split}
      &\lim_{T_2\to \infty}\lim_{T_1\to \infty}\frac{1}{ T_1T_2}\int_{0}^{T_1}\int_{0}^{T_2} dt_1 dt_2 \canWightman{2q}(t_1,t_2,\cdots,t_{2q})\\
      &= \overlineT{G_{\mathrm{can},\beta'}^{(2)}} \canWightman{2q-2}(t_3,\cdots,t_{2q-2})
 \end{split}
 \end{equation}
 where $\beta'$ is the new temperature after settling down. By doing similar things for $\canWightman{2q-2}(t_3,\cdots,t_{2q-2})$, we obtain the clustering property \eqref{eq:can_cluster}. Note that this discussion based on AdS/CFT correspondence is highly heuristic and not proven at all. We will address this point in future work.
\begin{figure}[!htbp]
\centering
    \begin{tikzpicture}
        \draw[black](-2,-2)--(-2,2);
        \draw[snake it](-2,2)--(3,2);
        \draw[black](3,2)--(3,-2);
        \draw[black,dashed](3,2)--(-1,-2);
        \draw[black,dashed](2,2)--(-2,-2);
        \filldraw[yellow](3,-2)--(-1,2)--(0.5,2)--(3,-0.5)--cycle;
        \draw[red](-2,-2)--(0.5,0.5);
        \draw[red](3,2)--(1.75,0.75);
        \draw[red](0.5,0.5)to[out=30,in= -150](1.75,0.75);
    \end{tikzpicture}
    \caption{The picture for the energy injection changes the shape of the black horizon. The yellow region corresponds to the shock wave like fluid by the huge number of the operators. We compute the two point function of the probe operators far above the yellow region.}
        \label{fig:AdSbrane}
\end{figure}
Also we may thin about the similar recursion relations for the thermal correlators from the symmetry perspective. Suppose we are again interested in the long time behavior of $t_1,t_2\gg t_3,\dots,t_{2q}$. In the frequency space after the Fourie transformation, the behavior of the long time limit is governed by the analytic stricture around zero frequency $\omega \sim 0$. Then, the factorization structure like \eqref{eq:long_time_factorization} have similarity to the soft theorem in the quantum field theory typically with massless gauge symmetry. In our case, we may deduce the recursion relations for the thermal correlators from the soft theorem of the symmetry appears in the effective field theory in the long time regime. One possible symmetry will be the symmetry emerges in hydrodynamics and the asymptotic symmetries of the blackhole horizon which recently discussed in \cite{Knysh:2024asf}. This is very interesting direction to be the another future work. \par 
Also we give a few comments about the wormhole contribution which we do not take into account seriously. If we taken into account these wormhole, the late time correlation function is dominated by these wormhole contribution and the long time average of the averaged correlation functions may not take the form of \eqref{eq:2q_point}. This situation can be easily understood from the configuration like the Figure \ref{fig:wormhole}. However, as discussed in \cite{Saad:2019pqd}, the non-decaying corelation functions comes from the worldline which intersects the wormhole cycles and these term have the order of $\LandauO{e^{-3\Smc(E)}}$ and we obtain the \eqref{eq:cluster}. For more late times more wormhole solutions contributes and the actual correaltion functions shows plateau which will be order of $\LandauO{e^{-2\Smc(E)}}$.  which These discussion is naive and we need to do the concrete computation especially for higher dimension. This is also the important future problem.
\begin{figure}[!ht]
\centering
\resizebox{0.5\textwidth}{!}{%
\begin{circuitikz}
\tikzstyle{every node}=[font=\LARGE]
\draw [dashed] (-9.25,15) ellipse (4cm and 2cm);
\draw[black] (-12,15)to [out=80,in=180] (-11,17)to [out=0,in=100](-10,15);
\draw[black] (-11.5,15)to [out=80,in=180] (-11,16)to [out=0,in=100](-10.5,15);
\draw[black] (-8.25,15)to [out=80,in=180] (-7.25,17)to [out=0,in=100](-6.25,15);
\draw[black] (-7.75,15)to [out=80,in=180] (-7.25,16)to [out=0,in=100](-6.75,15);
\draw(-12.5,16.5)node[above]{$\mathbb{O}_1$};
\filldraw[black] (-12.5,16.25) circle (0.15cm);
\draw(-12.5,13.6)node[below]{$\mathbb{O}_2$};
\filldraw[black] (-12.5,13.8) circle (0.15cm);
\draw(-6,13.6)node[below]{$\mathbb{O}_3$};
\filldraw[black] (-6,13.8) circle (0.15cm);
\draw(-6,16.5)node[above]{$\mathbb{O}_4$};
\filldraw[black] (-6,16.25) circle (0.15cm);
\draw[red,dashed] (-12.5,16.25)to [out=0,in=90] (-11,15)to [out=-90,in=0](-12.5,13.8);
\draw[red,dashed] (-6,13.8)to [out=180,in=-90] (-7.25,15)to [out=+90,in=180](-6,16.25);
\end{circuitikz}
}%

\caption{Sketch of the wormhole contribution to the for point function. The red line corresponds to the worldline whose contribution to the correlation functions are not decaying.}
\label{fig:wormhole}
\end{figure}

 \section{Relation to the Standard Form of ETH ansatz}\label{sec:Srednicki_ETH}
 Some reader might wonder how our results related to the usual form of the ETH called, Srednicki's ETH ansatz,
 \begin{equation}\label{eq:Srednicki_ETH}
     \bra{E_n}\mO\ket{E_m} = \mO(E) \delta_{nm}+ e^{-\frac{1}{2}S_{\mathrm{mc}}(E)} f_{\mO}(E,\omega)R_{nm}+\LandauO{e^{-S_{\mathrm{mc}}(E)}}.
 \end{equation}
 Actually, we can derive \eqref{eq:Srednicki_ETH} from our results. Let us start from the diagonal component. Let us put the initial state $\ket{\psi_0}$ in the Theorem \ref{thm:thermalization_inequality} as a single energy eigenstate in the energy shell $\ket{E_n}\in \eneshell$. Then, by definition, its effective dimension is just one, $\Deff=1$. Then, the inequality states that
 \begin{equation}
     \abs{\bra{E_n}\mO\ket{E_n}-\mcev{\mO}} < \opnorm{\mO}\qty(\ep + \sqrt{2 \shelldim \DeltaETH{\ep}{\mO}})
 \end{equation}
 for any positive $\ep>0$. If we take $\ep=\sigma^\eta\sim e^{-\frac{\eta\gamma}{2}\Smc(E)},\eta<1$ and if the theory satisfies the double exponential scaling \eqref{eq:double_expo}, then 
 \begin{equation}
     \abs{\bra{E_n}\mO\ket{E_n}-\mcev{\mO}} = \LandauO{e^{-\frac{\eta\gamma}{2}\Smc(E)}}.
 \end{equation}
 This is the exactly the diagonal part of \eqref{eq:Srednicki_ETH}. 
 For off-diagonal components, we take care of the theorem \ref{thm:equilibration_inequality}. In the theorem, by putting the initial state $\ket{\psi_0}$ as $\ket{\psi_0}=\frac{1}{\sqrt{2}}\qty(\ket{E_n}+\ket{E_m}),\;n\neq m$ and again assuming the non-resonance conditions, we see
 \begin{equation}
 \abs{\bra{E_n}\mO\ket{E_m}}^2 \leq 4 \opnorm{\mO}\qty(\ep^2 + 2\qty(\shelldim)^{\frac{3}{2}}\sqrt{\offDeltaETH{\ep}{\mO}}).
 \end{equation}
 If we take $\ep=\sigma^\eta\sim e^{-\frac{\eta\gamma}{2}\Smc(E)}$ and if the theory satisfies the double exponential scaling \eqref{eq:off_diag_double_expo}, we obtain 
 \begin{equation}
     \abs{\bra{E_n}\mO\ket{E_m}} = \LandauO{e^{-\frac{\eta\gamma}{2}\Smc(E)}}.
 \end{equation}
 This is exactly the off-diagonal part of \eqref{eq:Srednicki_ETH}. In the context of random matrix theory or typical form of ETH, we expect the sub-leading term is order $\LandauO{e^{-\frac{1}{2}\Smc(E)}}$, \cite{Beugeling:2014fse}. In our pure gravity computation like \eqref{eq:gamma_three_dim} shows $\eta\gamma<1$ and there is a deviation from the form of \eqref{eq:Srednicki_ETH}. However, it is enough to conclude the thermalization and equilibration from any initial states in the energy shell. 

 \subsection{Consistent order of  $\Delta E$ with Quantum Mixing}\label{sec:Delta_E}
As mentioned in Section \ref{sec:Review_ETH}, the order of width of energy shell $\Delta E$ is actually important for late time behavior of the correlation functions.  Indeed, as we will see from now on, the order of long time average of the two point microcanonical Wightman function of non-integrable systems is determined by the polynomial of $\Delta E$ \footnote{Thank you for R.Hamazaki gives a very insightful comments and discussion about the width of energy shell in these contexts.}.  In \cite{Srednicki:1995pt,Ikeda_2015}, it is discussed that the energy eigenstate expectation value of a few-body observable $\mO$ in non-integrable systems are known to be behave as
 \begin{equation}
     \bra{E_n}\mO\ket{E_n}= f\qty(\frac{E_n}{c_{\mathrm{eff}}})+\delta\mO_n
 \end{equation}
 where $f(x)$ is a smooth function and $\delta\mO_n$ represents random fluctuation around it. $c_{\mathrm{eff}}$ denotes the parameter we take the thermodynamic limit like system size or $N$ in holographic examples. By using this ansatz, let us evaluate $\EVdiag{\bra{E_n}\mbO_{\mathrm{mc}}\ket{E_n}^2}$. 
 \begin{equation}
     \begin{split}
         \EVdiag{\bra{E_n}\mbO_{\mathrm{mc}}\ket{E_n}^2}&= \frac{1}{\shelldim}\sum_{\ket{E_n}\in\eneshell}\bra{E_n}\mO\ket{E_n}^2-\qty(\frac{1}{\shelldim}\sum_{\ket{E_n}\in\eneshell}\bra{E_n}\mO\ket{E_n})^2\\
         &=(\Delta \mO^2)_{f}+(\Delta \mO^2)_{\mathrm{cross}}+(\Delta \mO^2)_{\mathrm{rand}}
     \end{split}
 \end{equation}
 where
 \begin{equation}
     \begin{split}
         (\Delta \mO^2)_{f}&=  \frac{1}{\shelldim}\sum_{\ket{E_n}\in\eneshell}f\qty(\frac{E_n}{c_{\mathrm{eff}}})^2-\qty(\frac{1}{\shelldim}\sum_{\ket{E_n}\in\eneshell}f\qty(\frac{E_n}{c_{\mathrm{eff}}}))^2\\
          (\Delta \mO^2)_{\mathrm{cross}}&=  \frac{2}{\shelldim}\sum_{\ket{E_n}\in\eneshell}f\qty(\frac{E_n}{c_{\mathrm{eff}}})\delta\mO_n-\frac{2}{(\shelldim)^2}\sum_{\ket{E_n}\in\eneshell}f\qty(\frac{E_n}{c_{\mathrm{eff}}})\sum_{\ket{E_m}\in\eneshell}\delta\mO_m\\
           (\Delta \mO^2)_{\mathrm{rand}}&=  \frac{1}{\shelldim}\sum_{\ket{E_n}\in\eneshell}\delta\mO_n^2-\qty(\frac{1}{\shelldim}\sum_{\ket{E_n}\in\eneshell}\delta\mO_n)^2.
     \end{split}
 \end{equation}
 We can estimate the first two term by doing the Taylor expansion for $f\qty(\frac{E_n}{c_{\mathrm{eff}}})$ around $E$,
 \begin{equation}
     f\qty(\frac{E_n}{c_{\mathrm{eff}}})= f\qty(\frac{E}{c_{\mathrm{eff}}})+\frac{E_n-E}{L}f'\qty(\frac{E}{c_{\mathrm{eff}}})+\frac{1}{2}\qty(\frac{E_n-E}{c_{\mathrm{eff}}})^2f''\qty(\frac{E}{c_{\mathrm{eff}}})+\cdots.
 \end{equation}
 The first term $(\Delta\mO^2)_f$ is then evaluated as 
 \begin{equation}
     \begin{split}
         (\Delta\mO^2)_f&= \frac{1}{2}\frac{\mcev{H^2}-\qty(\mcev{H})^2}{c_{\mathrm{eff}}^2}f''\qty(\frac{E}{c_{\mathrm{eff}}})+\cdots = \LandauO{\qty(\frac{\Delta E}{c_{\mathrm{eff}}})^2}.
     \end{split}
 \end{equation}
 Sometimes or with fine tuned parameters, we can make $\mcev{H^2}-\qty(\mcev{H})^2$ is exactly zero. However, in such a case, the higher order corrections of the Taylor expansions become just a leading term. Since each terms include the expectation values of polynomial of the Hamiltonian, the resulting order of $(\Delta \mO^2)_f$ is just a polynomial of $\Delta E$.
 The third term $(\Delta \mO^2)_{\mathrm{rand}}$ is variance of $\delta\mO_n$ and discussed that it will be in the order of $e^{-\Smc(E)}$ \cite{Beugeling:2014fse,Ikeda_2015}. Similar order estimations for the random fluctuation shows that
 \begin{equation}
     (\Delta\mO^2)_{\mathrm{cross}}\sim \Delta E\opnorm{\mO}\times \frac{1}{\shelldim}\sum_{\ket{E_n}\in\eneshell}\delta\mO_{n} \sim \Delta E \; e^{-\frac{1}{2}\Smc(E)}
 \end{equation}
 and thus
 \begin{equation}
     e^{-\LandauO{L}}=(\Delta\mO^2)_{\mathrm{rand}}\ll (\Delta\mO^2)_{\mathrm{cross}}\ll (\Delta\mO^2)_{f}=\LandauO{\qty(\frac{\Delta E}{c_{\mathrm{eff}}})^2}.
 \end{equation}
 In total, for not too small $\Delta E$,
 \begin{equation}
     \EVdiag{\bra{E_n}\mbO_{\mathrm{mc}}\ket{E_n}^2}\sim (\Delta\mO^2)_{f}=\LandauO{\qty(\frac{\Delta E}{c_{\mathrm{eff}}})^2}.
 \end{equation}
 By combing these the mixing condition \eqref{eq:quantum_mixing} is consistent when 
 \begin{equation}
     \Delta E = e^{-\LandauO{\Smc(E)}}.
 \end{equation}
 The loophole of our discussion is that $f''\qty(\frac{E}{c_{\mathrm{eff}}})$ is exponentially suppressed. Actually this is a case for Haar random unitary discussion for ETH \cite{Neumann1929}. However, even in holographic CFTs, the function form of $f(\frac{E}{c_{\mathrm{eff}}})$ does not take such a form. Indeed, for large $c$ two dimensional CFTs, the form of $f(\frac{E}{c_{\mathrm{eff}}})$ is given by \cite{Kraus:2016nwo},
 \begin{equation}
     f\qty(\frac{E}{c_{\mathrm{eff}}}) \propto \qty(\frac{12E}{c}-1)^{E_{\mO}}\exp{-2\pi E_\chi \sqrt{\frac{12 E}{c}-1}}
 \end{equation}
 whose second derivative is not exponentially suppressed. Our results here is consistent with the discussion about the strong ETH and behavior of OTOC in realistic quantum many body systems with usual width of energy shell \cite{PhysRevLett.120.080603,Huang:2017fng}. These results shows that the form of Srednicki ansatz \cite{Srednicki:1995pt} only holds in very narrow energy window in realistic quantum many body systems where we typically set $\Delta E = E^{\frac{1}{2}}$. 
 \section{Summary and Discussion}\label{sec:Summary}
  Finally, we give a summary of the discussion. In the first half of the paper, we provide the sufficient conditions for the double exponential scaling \eqref{eq:double_expo}. One condition is the clustering condition \eqref{eq:can_cluster} which guarantees the sub-Gaussian property of the probability that the energy eigenstates violate the ETH. The second condition is the quantum mixing property of the thermal two-point function $\canWightman{2}$. We have argued the examples that satisfy these two conditions. One trivial example is the generalized free field theories. Because of the exact large $N$ factorization property, we obtain the clustering property. Also because of the horizon structure which induces the complete spectrum of the Lehmann spectral representation of the thermal correlators, we meet the quantum mixing property. As noted, the GFF trivially satisfies the clustering property. However, notice that the clustering property is just for the time-averaged thermal correlators and not for the correlators themselves. As a slightly non-trivial example, we can add small interactions corresponding to the $1/N$ correction. In this case, we have non-trivial connected correlators but we see the clustering satisfied because of the complete spectrum of the bulk-to-boundary propagators and the connected part of  the Wightman functions are highly suppressed after long time average.  In the discussion of the Witten diagram we do not consider any back reaction and wormhole contributions for the higher order correalation functions. In section \ref{subsec:back_reaction}, we give heuristic discussion for the case, we include huge back reaction. To make some physical view, we divided the higher-point function into the probe part and huge excitation part and replaced it with the probe correlation function in the back-reacted geometry. In these computations, we implicitly ignored the baryon-like configuration of the large $N$ gauge theory. The complete rigorous discussion will be future work.
  \par
  Our results states that it seem to be true that the bottom-up holographic theories satisfy the strong ETH by using the inequality \eqref{eq:thermalization_inequality2} for the light single trace scalar operators.  However, there are some discussions, like in the AdS/CFT correspondence, there are quantum scar states which violate the strong ETH \cite{Dodelson:2022eiz, Milekhin:2023was}. The gravity dual of these scar states is expressed as the star-like solutions. Moreover, top-down holographic theories sometimes show integrability in some sectors which may bring us the quantum scar states. It is highly important to generalize our discussion to the case we have internal symmetry or conformal symmetry. Also, it is interesting to discuss the scaling of $\DeltaETH{\ep}{\mO}$ and $\offDeltaETH{\ep}{\mO}$ for heavy operators.
  \par
  Finally, our results also have some implications for the failure of thermalization in integrable or many-body localized systems. For example, for free theories, we have the clustering property \eqref{eq:can_cluster} but they fail to the quantum mixing \eqref{eq:quantum_mixing}. In the integrable case, the long-time averaged two-point function is roughly given by the inverse of the Poincaré recurrence time which is $\LandauO{1}$ for the entropy. Then, we see that the small $\Deff$ initial states may fail to thermalize. 
  \par
  Also, we discuss the off-diagonal ETH and we found that the same arguments give a double exponential scaling \eqref{eq:off_diag_double_expo} which ensures the equilibration for all initial states.
\acknowledgments
We are grateful to Tadashi Takayanagi, Julian Sonner, Vijay Balasubramanian, Hal Tasaki,  Gabriele Di Ubaldo for useful discussions. Also, we appropriate Ryusuke Hamazaki for very insightful comments and discussion about the width of the energy shell. We also appreciate Yu Nakayama for checking the draft and pointing out mistakes in the previous version.
This work is supported by Grant-in-Aid for JSPS Fellows No. 23KJ1315.
\appendix
\section{Proof of Inequalities}\label{app:proof}
In this section, we give mathematical proof of the theorem we discuss so far. 
 \subsection{Proof of Theorem \ref{thm:thermalization_inequality}}
 Here we give a proof of the inequality \eqref{eq:thermalization_inequality}.
 The proof the proposition is following: Let us denote the sets of the energy eigenstates satisfy $\abs{\bra{{E_n}}\mO\ket{E_n}-\mcev{\mO}}<\epsilon \opnorm{\mO}$ as $\mEeq^{(\ep)}$. Then, by simple algebra, we find
\begin{equation}
    \begin{split}
&\abs{\overlineT{\bra{\psi(t)}\mO\ket{\psi(t)}}-\mcev{\mO}}\\&= \abs{\sum_{\ket{E_n}\in \mEeq} \abs{c_n}^2\qty(\bra{E_n}\mO\ket{E_n}-\mcev{\mO})}\\
     &\leq \sum_{\ket{E_n}\in \mEeq} \abs{c_n}^2 \abs{\bra{E_n}\mO\ket{E_n}-\mcev{\mO}}\\
     &\leq \sum_{\ket{E_n}\in \mEeq-\mEeq^{(\ep)}} \abs{c_n}^2 \abs{\bra{E_n}\mO\ket{E_n}-\mcev{\mO})}\\&+ \sum_{\ket{E_n}\in \mEeq^{(\ep)}} \abs{c_n}^2 \abs{\bra{E_n}\mO\ket{E_n}-\mcev{\mO})}\\
     &\leq \sqrt{\sum_{\ket{E_n}\in \mEeq-\mEeq^{(\ep)}}\abs{c_n}^4 \sum_{\ket{E_n}\in \mEeq-\mEeq^{(\ep)}} \abs{\bra{E_n}\mO\ket{E_n}-\mcev{\mO})}^2} + \epsilon \opnorm{\mO}\\
     &\leq \sqrt{\sum_{\ket{E_n}\in \mEeq}\abs{c_n}^4 \sum_{\ket{E_n}\in \mEeq-\mEeq^{(\ep)}} \abs{\bra{E_n}\mO\ket{E_n}-\mcev{\mO})}^2} + \epsilon \opnorm{\mO}.
    \end{split}
\end{equation}
Recalling $\sum_{\ket{E_n}\in \mEeq}\abs{c_n}^4= \qty(\Deff)^{-1}$ and 
 \begin{align}
     \abs{\bra{E_n}\mO\ket{E_n}-\mcev{\mO})}  &\leq \abs{\bra{E_n}\mO\ket{E_n}} + \abs{\mcev{\mO}} \leq 2 \opnorm{\mO}\\
     \sum_{\ket{E_n}\in \mEeq-\mEeq^{(\ep)}}&= \shelldim \times \DeltaETH{\ep}{\mO},
 \end{align}
 we obtain the inequality.
\subsection{Proof of Theorem \ref{thm:equilibration_inequality}}\label{subsec:Proof of Threorem thm:equilibration_inequality}
 Here we give a proof of \ref{thm:equilibration_inequality} which is the inequality for the off-diagonal component.
\begin{proof}
    By assuming the non-degeneracy condition, we see
    \begin{equation}
        \overlineT{\left|\bra{\psi(t)}\mO\ket{\psi(t)} - \overlineT{\bra{\psi(t)}\mO\ket{\psi(t)}}\right|^2} = \sum_{(\ket{E_n},\ket{E_m})\in \offdiagset} \abs{c_n}^2 \abs{c_m}^2 \abs{\bra{E_n}\mO\ket{E_m}}^2
    \end{equation}
    Next, we define a subset of $\offdiagset$ by 
    \begin{equation}
       \offdiagset^{(\ep)} = \{(\ket{E_n},\ket{E_m})\in \offdiagset; \abs{\bra{E_n}\mO\ket{E_m}}^2 \leq \ep^2 \opnorm{\mO}^2\} \subset \offdiagset
    \end{equation}
    Then, the summation can be decomposed into two parts,
    \begin{equation}
        \begin{split}
            &\sum_{(\ket{E_n},\ket{E_m})\in \offdiagset} \abs{c_n}^2 \abs{c_m}^2 \abs{\bra{E_n}\mO\ket{E_m}}^2 \\
            &= \sum_{\substack{(\ket{E_n},\ket{E_m})\in  \offdiagset^{(\ep)}}}\abs{c_n}^2 \abs{c_m}^2 \abs{\bra{E_n}\mO\ket{E_m}}^2 \\&+ \sum_{\substack{(\ket{E_n},\ket{E_m})\in \offdiagset- \offdiagset^{(\ep)}}} \abs{c_n}^2 \abs{c_m}^2 \abs{\bra{E_n}\mO\ket{E_m}}^2 \\
            &\leq \ep^2 \opnorm{\mO}^2 + \sqrt{\left(\sum_{\substack{(\ket{E_n},\ket{E_m})\\\in \offdiagset- \offdiagset^{(\ep)}}}\abs{c_n}^4\abs{c_m}^4 \right)\left(\sum_{\substack{(\ket{E_n},\ket{E_m})\\\in \offdiagset- \offdiagset^{(\ep)}}}\abs{\bra{E_n}\mO\ket{E_m}}^4\right)}
        \end{split}
    \end{equation}
    In the second line, we use the definition of $\offdiagset^{(\ep)}$ and the Cauchy-Schwarz inequality. Next, we discuss the upper bound of the terms in the square root. The first one is bounded above by the inverse of the effective dimension.
    \begin{equation}
        \sum_{(\ket{E_n},\ket{E_m})\in \offdiagset- \offdiagset^{(\ep)}} \abs{c_n}^4\abs{c_m}^4 \leq \sum_{\ket{E_n},\ket{E_m}\in \mEeq^2} \abs{c_n}^4\abs{c_m}^4 = \frac{1}{(\Deff)^2}
    \end{equation}
    The second one is bounded above by the operator norm.
    \begin{equation}
        \begin{split}
            \abs{\bra{E_n}\mO\ket{E_m}} &= \abs{\Tr_{\eneshell}\qty[{(\ket{E_m}\bra{E_n})}\mO]}\\
            &\leq \sqrt{\Tr_{\eneshell}\qty[{(\ket{E_m}\bra{E_n})(\ket{E_n}\bra{E_m})}]\Tr_{\eneshell}\qty[\mO^\dag \mO]}\\
            &=\sqrt{\Tr_{\eneshell}\qty[\mO^\dag \mO]}\\
            &= \sqrt{\sum_{\ket{E_n}\in \eneshell}\bra{E_n}\mO^\dag \mO\ket{E_n}}\\
            &=\sqrt{\shelldim}\opnorm{\mO}
        \end{split}
    \end{equation}
    In the second line, we use the Cauchy-Schwartz inequality for the Hilbert-Schmidt inner product on the energy shell. Then, we see 
    \begin{equation}
        \begin{split}
            \sum_{\substack{(\ket{E_n},\ket{E_m})\\\in \offdiagset- \offdiagset^{(\ep)}}}\abs{\bra{E_n}\mO\ket{E_m}}^4 &\leq (\shelldim)^2 \opnorm{\mO}^4 \left(\sum_{\substack{(\ket{E_n},\ket{E_m})\\\in \offdiagset- \offdiagset^{(\ep)}}} 1\right) \\&= (\shelldim)^3 \opnorm{\mO}^4 \offDeltaETH{\ep}{\mO}
        \end{split}
    \end{equation}
    By combining these inequalities, we obtain the inequality.
    \end{proof}
\subsection{Discussion about \eqref{eq:ensemble_equivalance1} and Proof of \eqref{eq:ensemble_equivalance2}}\label{subsec:ensemble_equivalnce}
 Here we give a discussion about the inequalities \eqref{eq:ensemble_equivalance1} and \eqref{eq:ensemble_equivalance2}. This part includes the same discussion in \cite{Tasaki:2008,TASAKI:2018eoe} but we put the arguments for the self consistency. Let us start with $p=1$ case where the inequality is always true without any assumption. It is known that for any probability distribution, expectation value and for any $a\in \mathbb{R}$, it holds that 
 \begin{equation}\label{eq:moment_inequality_shift}
     \EV{(X-\EV{X})^2}\leq \EV{(X-\EV{X}+a)^2}.
 \end{equation}
 Then, we by setting $a=\canev{\mO}-\mcev{\mO}$, we obtain,
 \begin{equation}
     \EVdiag{\bra{E_n}\mbO_{\mathrm{mc}}\ket{E_n}^2}\leq \EVdiag{\bra{E_n}\mbO_{\mathrm{can}}\ket{E_n}^2}
 \end{equation}
 where $\mbO_{\mathrm{can}}:=\mO-\canev{\mO}$. By elaborating the right hand side, we see
 \begin{equation}
     \begin{split}
         \EVdiag{\bra{E_n}\mbO_{\mathrm{can}}\ket{E_n}^2}&= \frac{1}{\shelldim}\sum_{\ket{E_n}\in\eneshell}\bra{E_n}\mbO_{\mathrm{can}}\ket{E_n}^2\\
         &\leq \frac{e^{\beta (E+\Delta E)}}{\shelldim}\sum_{\ket{E_n}\in\eneshell}e^{-\beta E_n}\bra{E_n}\mbO_{\mathrm{can}}\ket{E_n}^2\\
         &\leq \frac{Z(\beta)e^{\beta (E+\Delta E)}}{\shelldim}\frac{1}{Z(\beta)}\sum_{\ket{E_n}\in \mH}e^{-\beta E_n}\bra{E_n}\mbO_{\mathrm{can}}\ket{E_n}^2\\
         &= \frac{Z(\beta)e^{\beta (E+\Delta E)}}{\shelldim}\overlineT{\canWightman{2}(t_1,t_2)}.
     \end{split}
 \end{equation}
 This is exactly the \eqref{eq:ensemble_equivalance1} with $p=1$. Next let us consider the higher order moments. In this case, simple generalizations of \eqref{eq:moment_inequality_shift} are not always true. \begin{equation}\label{eq:shift_higher_order}
     \EVdiag{(X-\EVdiag{X})^{2q}}\leq \EVdiag{(X-\EVdiag{X}+a)^{2q}}
 \end{equation}
 for some $a\in\mathbb{R}$. Let us examine whether this inequality is hold or not. 
 \begin{equation}
     \begin{split}
       V^{(2q)} &:=\EVdiag{(X-\EVdiag{X}+a)^{2q}}-\EVdiag{(X-\EVdiag{X})^{2q}}\\ &= \sum_{k=0}^{2q} {}_{2q}C_k \EVdiag{(X-\EVdiag{X})^{k}}a^{2q-k}-\EVdiag{(X-\EVdiag{X})^{2q}}\\
         &=\sum_{k=2}^{2q-1} {}_{2q}C_k \EVdiag{(X-\EVdiag{X})^{k}}a^{2q-k}.
     \end{split}
 \end{equation}
 We may consider some cases such that $V^{(4)}>0$. One simple case is 
 \begin{equation}
     \EVdiag{(X-\EVdiag{X})^{2k-1}}=0.
 \end{equation}
 In this cases,
 \begin{equation}
     V^{(2q)}= \sum_{l=1}^{p-1} \Comb{2q}{2l}\EVdiag{(X-\EVdiag{X})^{2l}}a^{2(p-l)}\geq 0.
 \end{equation}
 This is true for the symmetric distribution. In our $\Phi^4$ theory examples discussed in section \ref{sec:Examples}, this is actually satisfied because of $\mathbb{Z}_2$ symmetry $\Phi\to-\Phi$ where $\Phi$ is dual to $\mbOmc$ or $\mbOcan$. 
 
 The another case is that 
 \begin{equation}\label{eq:non_symmetric}
 \begin{split}
 \EVdiag{(X-\EVdiag{X})^{2l}}&=e^{-2l\gamma' \Smc(E)},\;1\leq l \leq q-1\\, 
   \EVdiag{(X-\EVdiag{X})^{2l+1}}&=\frac{1}{(\Smc(E))^x}e^{-(2l+1)\gamma' \Smc(E)}
 \end{split}
 \end{equation}
 ans $a=1/(\Smc(E))^y$. The reason that we consider the further suppression for odd moments is we expect that the distribution is almost symmetric.  Then let us try to evaluate the very naive lower bound of $V^{(2q)}$:
 \begin{equation}
     \begin{split}
         V^{(2q)} \geq \sum_{l=1}^{q-1}\Comb{2q}{2l}\EVdiag{(X-\EVdiag{X})^{2l}}a^{2q-2l}\qty(1-\frac{2q-2l}{2l+1}\frac{\abs{\EVdiag{(X-\EVdiag{X}})^{2l+1}}}{\EVdiag{(X-\EV{X})^{2l}}}\abs{a}^{-1})
     \end{split}
 \end{equation}
 When $q=\LandauO{1}$, it is clear that the right hand side is positive. When $q$ is very huge, we need to take care about the binomial coefficient. When $l\ll q$, \begin{equation}
    \frac{2q-2l}{2l+1}\frac{\abs{\EVdiag{(X-\EVdiag{X}})^{2l+1}}}{\EVdiag{(X-\EVdiag{X})^{2l}}}\abs{a} \sim q e^{-\gamma'\Smc(E)}(\Smc(E))^{x-y}
 \end{equation}
 The right hand can be smaller than one up to $q\sim \mathrm{Poly}(\Smc(E))e^{\gamma'\Smc(E)} $ when $x>y$. This is the same order of $L$ we meet in section \ref{sec:Examples}. For more large $q\sim l$, the coefficients $\frac{2q-2l}{2l+1}$ is at most order one is absolutely smaller than one. Note that our evaluation is very naive and the situation will be better since some odd moments can and $a$ will be positive. Also notice that conditions \eqref{eq:non_symmetric} gives the quantum mixing and clustering property of the microcanonical Wightman functions. Thus, to be honest, we do not need ensemble equivalence in this case. In this case, we are firstly do the Witten diagram computation with fixed energy blackhole back ground  \cite{Kraus:2016nwo} whose computation gives the microcanonical correlation functions. \par 
 Suppose \eqref{eq:shift_higher_order} is correct in some reasons. Let us write down the definition of the momenta again,
 \begin{equation}
     \begin{split}
         \EVdiag{\bra{E_n}\mbOmc\ket{E_n}^{2q}} &= \frac{1}{\shelldim}\sum_{\ket{E_n}\in\eneshell}\bra{E_n}\mbOmc\ket{E_n}^{2q}\\
         &\leq  \frac{e^{\beta(E+\Delta E)}}{\shelldim}\sum_{\ket{E_n}\in\eneshell}e^{-\beta E_n}\bra{E_n}\mbOmc\ket{E_n}^{2q}\\
          &\leq  \frac{e^{\beta(E+\Delta E)}}{\shelldim}\sum_{\ket{E_n}\in\mH}e^{-\beta E_n}\bra{E_n}\mbOmc\ket{E_n}^{2q}\\
         &= \frac{e^{\beta(E+\Delta E)}Z(\beta)}{\shelldim}\overlineT{\canWightman{2q}}.
     \end{split}
 \end{equation}
 This is the first inequality \eqref{eq:ensemble_equivalance1}. Next we move to the second inequality. Let us discuss on the upper bound on the thermal partition function. Suppose we divide the energy spectrum with the width of $\Delta E$ and label each energy window by index $I\in \mathbb{N}$.
 \begin{equation}
 \begin{split}
     Z(\beta)&= \sum_{E_n} e^{-\beta E_n}\\
     &=\sum_{I=1}^{\infty} \sum_{ 
     \substack{E_n\\ E_0+I\Delta E \leq E_n \leq E_0+(I+1)\Delta E}}e^{-\beta E_n}\\
     &\leq \sum_{I=1}^{\infty} e^{-\beta(E_0+I\Delta E)} D_{E_0+I\Delta E,\Delta E} 
 \end{split}
 \end{equation}
 where we denote the $E_0$ as a ground state energy. Let us introduce one special index $I_*$ which we determine later. Then the right hand side of the final line is decomposed into two parts:
 \begin{equation}
 \begin{split}
          Y_{\mathrm{main}}&:= \sum_{I\leq I_*} e^{-\beta(E_0+I\Delta E)} D_{E_0+I\Delta E,\Delta E}\\
           Y_{\mathrm{rem.}}&:= \sum_{I\geq I_*} e^{-\beta(E_0+I\Delta E)} D_{E_0+I\Delta E,\Delta E}.
 \end{split}
 \end{equation}
 We can easily find the upper bound for the first part $Y_{\mathrm{main}}$. Indeed,
 \begin{equation}
     \begin{split}
         Y_{\mathrm{main}}&\leq I_* \cdot \max_{E}\qty(e^{-\beta E} D_{E,\Delta E}).
     \end{split}
 \end{equation}
 For the upper bound of the second part, we use the following assumption or lemma. 
\begin{ass}
     For any $\Tilde{\beta}$ and $E$, there exists $\Tilde{S}(\beta)=\LandauO{\log{\shelldim}}$, we have the upper bound 
    \begin{equation}\label{eq:shelldim_upper_bound}
        \shelldim \leq e^{\Tilde{S}(\Tilde{\beta})+\Tilde{\beta}E}.
    \end{equation}
\end{ass}
Naively, $\Tilde{S}$ is a 
\begin{equation}
    \Tilde{S}(\Tilde{\beta})= S_{\mathrm{th}}(E_*(\Tilde{\beta}))-\Tilde{\beta}E_*(\Tilde{\beta})
\end{equation}
where $S_{\mathrm{th}}(E)$ is the tehrmodynamic entropy and $E_*(\beta)$ is determined from the thermodynamic relation $\frac{\partial S_{\mathrm{th}}}{\partial E}= \beta$. \eqref{eq:shelldim_upper_bound} is proven for example for any interacting particle systems \cite{Tasaki:2008} and is not the assumption. By using the assumption, we obtain the upper bound by taking $\Tilde{\beta}=\frac{1}{2}\beta$,
\begin{equation}
\begin{split}
    Y_{\mathrm{rem.}} &=\sum_{I\geq I_*} e^{-\beta(E_0+I\Delta E)} D_{E_0+I\Delta E,\Delta E}\\
    &\leq e^{\Tilde{S}-\frac{1}{2}\beta E_0}\sum_{I\geq I_*} e^{-\beta I\Delta E} < 2\; e^{-\beta I_* \Delta E + \Tilde{S}-\frac{1}{2}\beta E_0}
\end{split}
\end{equation}
By taking the $I_*$ large enough, we can always make $Y_{\mathrm{rem.}}<Y_{\mathrm{main}}$. Thus, we obtain,
\begin{equation}
    Z(\beta)< 2 I_*\cdot \max_{E}\qty(e^{-\beta E} \shelldim).
\end{equation}
One simple choice of $I_*$ is such that 
\begin{equation}
    -\beta I_* \Delta E + \Tilde{S}-\frac{1}{2}\beta E_0<0
\end{equation}
,which means at most $I_*= \LandauO{\log{\shelldim}}$. Also, by choosing the $E_{\mathrm{max}}(\beta)$ as the energy which makes maximum of $\max_{E}\qty(e^{-\beta E}\shelldim)=e^{\max_E\qty({-\beta E+\Smc(E)})}$, that is, usual thermodynamic law. If we equate this energy as energy $E$ which is the energy of the energy shell we consider, we can determine $\beta$ as $\beta=\beta(E)$ and obtain
\begin{equation}
     \frac{Z(\beta(E))e^{\beta(E)\qty(E+\Delta E)}}{\shelldim\; }< \LandauO{\log{\shelldim}}.
\end{equation}
\section{Case without non-Resonant Condition}\label{sec:WIthout_non-resonance}
   So far we discuss the non-resonance condition \eqref{eq:non_resonance}. Especially, any conformal field theory brakes the non-resonance condition since the algebra of conformal symmetry creates the harmonic level spacing. For example, in two dimension conformal field theories, for each primary states $\ket{h_p,\Bar{h}_p}$ we have a sets of descendant states with energy $h_p+\Bar{h}_p+N+\Bar{N},\;N,\Bar{N}\in\mathbb{N}\cup \{0\}$ and there are many pairs of states with energy difference $E_n-E_m=1$. \par
   Actually, we can generalize the theorem  \eqref{thm:equilibration_inequality} for the case without the non-resonance conditions.
   \begin{thm}\label{thm:equilibartion_inequality_resonance}
   For any $\ep>0$, any initial state $\ket{\psi_0}\in\eneshell$, any bounded operator $\mO\in\mB(\mH)$ and assuming no-degeneracy of energy spectrum \footnote{Instead we consider operator lies in some super-selection sector of some degenerate Hamiltonian.}, it holds that 
\begin{equation}
\overlineT{\left|\bra{\psi(t)}\mO(t)\ket{\psi(t)} - \overlineT{\bra{\psi(t)}\mO\ket{\psi(t)}}\right|^2} \leq d_G\opnorm{\mO}^2\qty(\ep^2 + \frac{(\shelldim)^{\frac{3}{2}}}{\Deff} \sqrt{\offDeltaETH{\ep}{\mO}})
\end{equation}
where $d_G$ express the maximum number of the degeneracy of energy difference,
\begin{equation}
    d_G:=\max_{(E_n,E_m)\in \mEeq'}\qty{(E_k,E_l)\in \mEeq'|E_n-E_m=E_k-E_l}.
\end{equation}
Also we have another bound, 
\begin{equation}\label{eq:inequlaity_resonance2}
\overlineT{\left|\bra{\psi(t)}\mO(t)\ket{\psi(t)} - \overlineT{\bra{\psi(t)}\mO\ket{\psi(t)}}\right|^2} \leq \inf_{\beta>0}\qty(\mathcal{S}(\beta))\opnorm{\mO}^2\qty(\ep^2 + \frac{(\shelldim)^{\frac{3}{2}}}{\Deff} \sqrt{\offDeltaETH{\ep}{\mO}})
\end{equation}
where
\begin{equation}
   \mathcal{S}(\beta)=e^{2\beta(E+\Delta E)} \sqrt{\overlineT{\qty(\abs{Z(\beta+it)}^2)^2}}.
\end{equation}
where $Z(\beta+it)$ is obtained from analytic continuation of the thermal partition function.
\end{thm}
   The proof of this theorem is very similar to the proof of the another inequality discussed in \cite{Short:2011pvc}. We put the proof of the theorem in Appendix \ref{app:equilibartion_inequality_resonance}. From this result, we can read off that if there are big degeneracy of energy difference, it prevents the equilibration at least by using thin inequality. \par
    As a example, let us evaluate $d_G$ in CFTs. In CFT cases, unless there is a magical mechanism, $d_G$ will be  given by the number of the pairs of energy eigenstates with $E_n-E_m=1$. For each primaries with weight conformal dimension $\Delta_p$, we have conformal families, and each conformal families have the number of degeneracy such that $\floor{E+\Delta E-\Delta_p}$. Then, $d_G$ is evaluated as
    \begin{equation}
        d_G= \sum_{E<\Delta_p<E+\Delta E}d(\Delta_p)\floor{E+\Delta E-\Delta_p}
    \end{equation}
   where $d(\Delta_p)$ is a number of the primary states with weight $\Delta_p$. Let us evaluate the upper bound of $d_G$,
   \begin{equation}
    \begin{split}
        d_G\leq \sum_{E<\Delta_p<E+\Delta E}d(\Delta_p)(E+\Delta E-\Delta_p)< \Delta E\sum_{E<\Delta_p<E+\Delta E}d(\Delta_p).
    \end{split}
   \end{equation}
  By using the known upper bound \cite{Mukhametzhanov:2019pzy},
  \begin{equation}
      d_G < \Delta E \;e^{\Smc(E)+\LandauO{\log{\Smc(E)}}}
  \end{equation}
  for large $E$ and large $c$. In the holographic computation, we can choose $\ep^2= \sigma^{2\eta}=\LandauO{e^{-\eta\gamma\Smc(E)}}$ with $\eta\gamma<\frac{1}{2}$. Thus, if we choose $\Delta E < \LandauO{e^{-\gamma'\Smc(E)}}$ with $\gamma'>\frac{1}{2}$, we can show the equilibration from any initial states in the energy shell. Recall that  condition of $\Delta E$ is such that the energy shell $\eneshell$ includes so many states. This condition is satisfies when $\gamma'<1$ and we take suitable value $\frac{1}{2}<\gamma'<1$. Also we can discuss the second inequality \eqref{eq:inequlaity_resonance2} in a very naive manner. For general chaotic theory, $\abs{Z(\beta+it)}^2$ decays to $Z(2\beta)$ and $\mathcal{S}(\beta)\sim e^{2\beta(E-F)}\sim e^{2\beta \Smc(E)}$ where $F$ denotes the free energy. Inthis case, we need to take $\Delta E$ in more small. \par In both cases, we do not take care about the energy degeneracy. If we consider the operator $\mO$ has zero charge for any symmetry including quantum KdV charge, then our discussion will be fine. However, in real CFTs they have the degeneracy and we need to discuss more general operators. Our discussion above should be generalized for such a case.
   \subsection{Proof of Theorem \ref{thm:equilibartion_inequality_resonance}}\label{app:equilibartion_inequality_resonance}
Here we give a proof of theorem \ref{thm:equilibartion_inequality_resonance}. The proof is almost same as \cite{Short:2011pvc}. Here we do not assume non-resonance condition but assume the non-degeneracy. We evaluate the variance of the long time average
\begin{equation}
\begin{split}
     \overlineT{\abs{\bra{\psi(t)}\mO\ket{\psi(t)}-\overlineT{\bra{\psi(t)}\mO\ket{\psi(t)}}}^2}&=\sum_{\substack{(\ket{E_n},\ket{E_m})\in \mEeq'\\(\ket{E_k},\ket{E_l})\in \mEeq'}} \overlineT{e^{-i((E_n-E_m)-(E_
{k}-E_{l})t}}\\\times &c_{n}c_{m}^* c_{k}c_{l}^* \bra{E_{m}}\mO\ket{E_{n}}\bra{E_{k}}\mO\ket{E_
l}
\end{split}
\end{equation}
Next, we define a set of indices $\alpha= (n,m),n\neq m$ and a matrix and a vector,
\begin{equation}
    \begin{split}
        M_{\alpha\beta} &:= \overlineT{e^{-i((E_n-E_m)-(E_
{k}-E_{l}))t}}\\
v_{\alpha} &=c_{n}c_{m}^* \bra{E_{n}}\mO\ket{E_{m}}.
    \end{split}
\end{equation}
Then, the variance is simple expressd by
\begin{equation}
\begin{split}
     \overlineT{\abs{\bra{\psi(t)}\mO\ket{\psi(t)}-\overlineT{\bra{\psi(t)}\mO\ket{\psi(t)}}}^2}&=\sum_{\alpha,\beta}v_{\alpha}^* M_{\alpha\beta}v_{\beta}\leq \norm{M}\sum_{\alpha}v_{\alpha}^* v_{\alpha}.
\end{split}
\end{equation}
Here $\norm{M}$ denotes the matrix norm of $M_{\alpha\beta}$ and bounded by above
\begin{equation}
    \norm{M}\leq \max_{\beta} \sum_{\alpha} \abs{M_{\alpha\beta}}\leq d_G
\end{equation}
from definition of $d_G$. The sum $\sum_{\alpha}v_\alpha^*v_{\alpha}$ can be bounded above in the same way as Appendix \ref{subsec:Proof of Threorem thm:equilibration_inequality}. Also there is another bound, 
\begin{equation}
    \sum_{\alpha,\beta}v_{\alpha}^* M_{\alpha\beta}v_{\beta}\leq \sqrt{\sum_{\alpha,\beta}\abs{M_{\alpha\beta}}^2}\sum_\alpha v_\alpha v_\alpha^*.
\end{equation}
The norm for $M$ is called the Frobenius norm, 
\begin{equation}
    \begin{split}
\norm{M}_F^2&=\sum_{\alpha,\beta}\abs{M_{\alpha\beta}}^2 \\&\leq \sum_{(\ket{E_n},\ket{E_m},\ket{E_l},\ket{E_k})\in \mEeq^4} \overlineT{e^{-i((E_n-E_m)-(E_
{k}-E_{l}))t}}\overlineT{e^{i((E_n-E_m)-(E_
{k}-E_{l}))t}}\\
&=\lim_{T,T'\to\infty} \frac{1}{TT'}\int_{0}^T\int_{0}^{T'}dtdt' \sum_{(\ket{E_n},\ket{E_m},\ket{E_l},\ket{E_k})\in \mEeq^4} e^{-i((E_n-E_m)-(E_
{k}-E_{l}))(t-t')}\\
&\leq \lim_{T\to\infty} \frac{1}{T}\int_{0}^T dt \;e^{4\beta(E+\Delta E)}\sum_{(\ket{E_n},\ket{E_m},\ket{E_l},\ket{E_k})\in \mH} e^{-\beta(E_n+E_m+E_k+E_l)}e^{-i((E_n-E_m)-(E_
{k}-E_{l}))t}\\
&\leq e^{4\beta(E+\Delta E)} \overlineT{\qty(\abs{Z(\beta+it)}^2)^2}
    \end{split}
\end{equation}
\section{Numerical evidence for the upper bound \eqref{eq:Ela^2X^2_bound}}\label{appendix:Ela^2X^2_bound}
\begin{figure}
    \centering
    \includegraphics[width=0.5\linewidth]{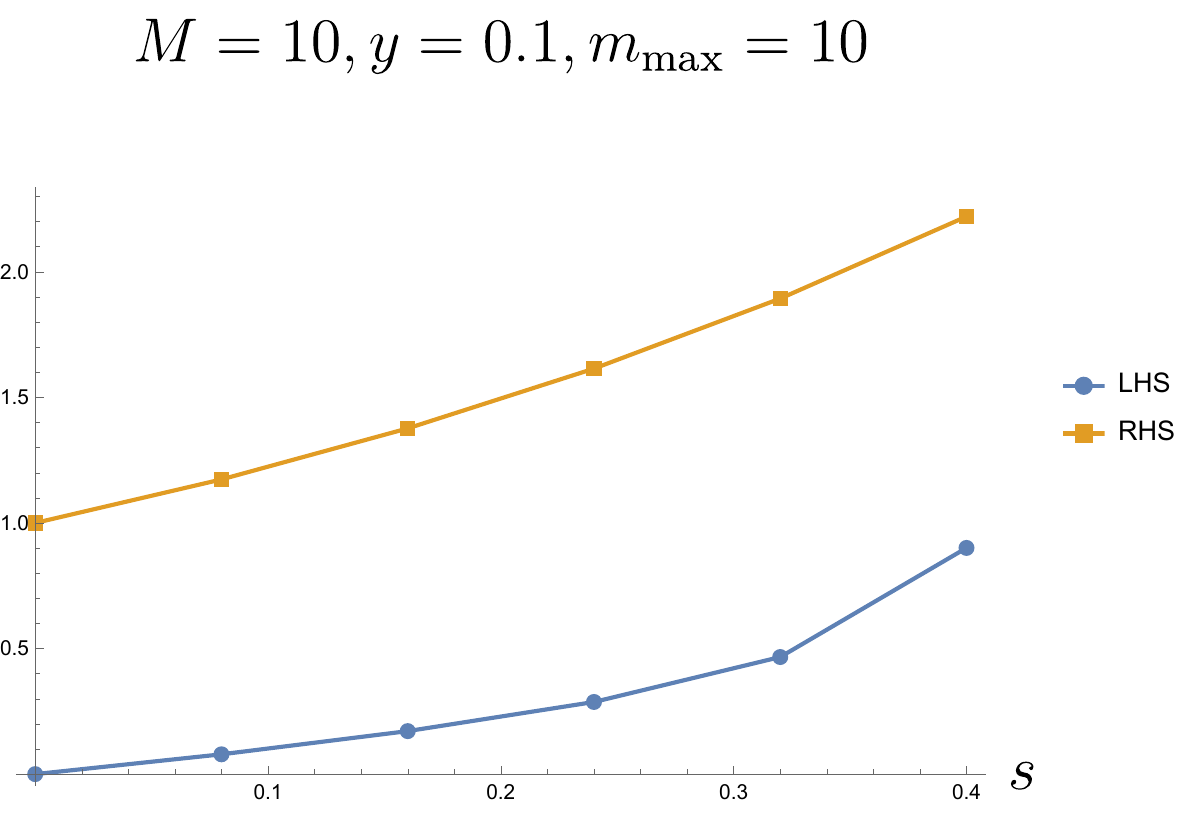}
    \caption{The plots of LHS and RHS with $M=10,y=0.1,m_{\max}=10$. }
    \label{fig:M=10_y=0.1_mmax=10}
\end{figure}

\begin{figure}
    \centering
    \includegraphics[width=0.5\linewidth]{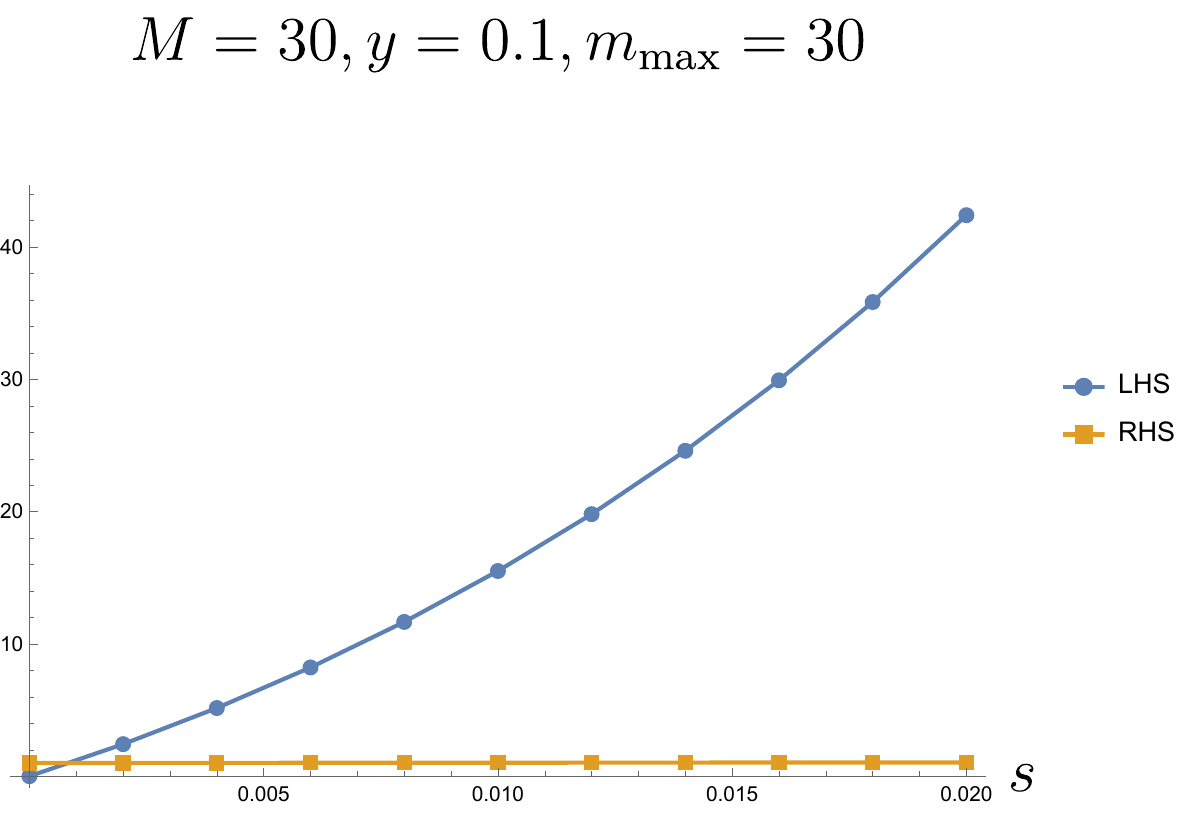}
    \caption{The plots of LHS and RHS with $M=30,y=0.1,m_{\max}=30$.}
    \label{fig:M=10_y=0.1_mmax=30}
\end{figure}

\begin{figure}
    \centering
    \includegraphics[width=0.5\linewidth]{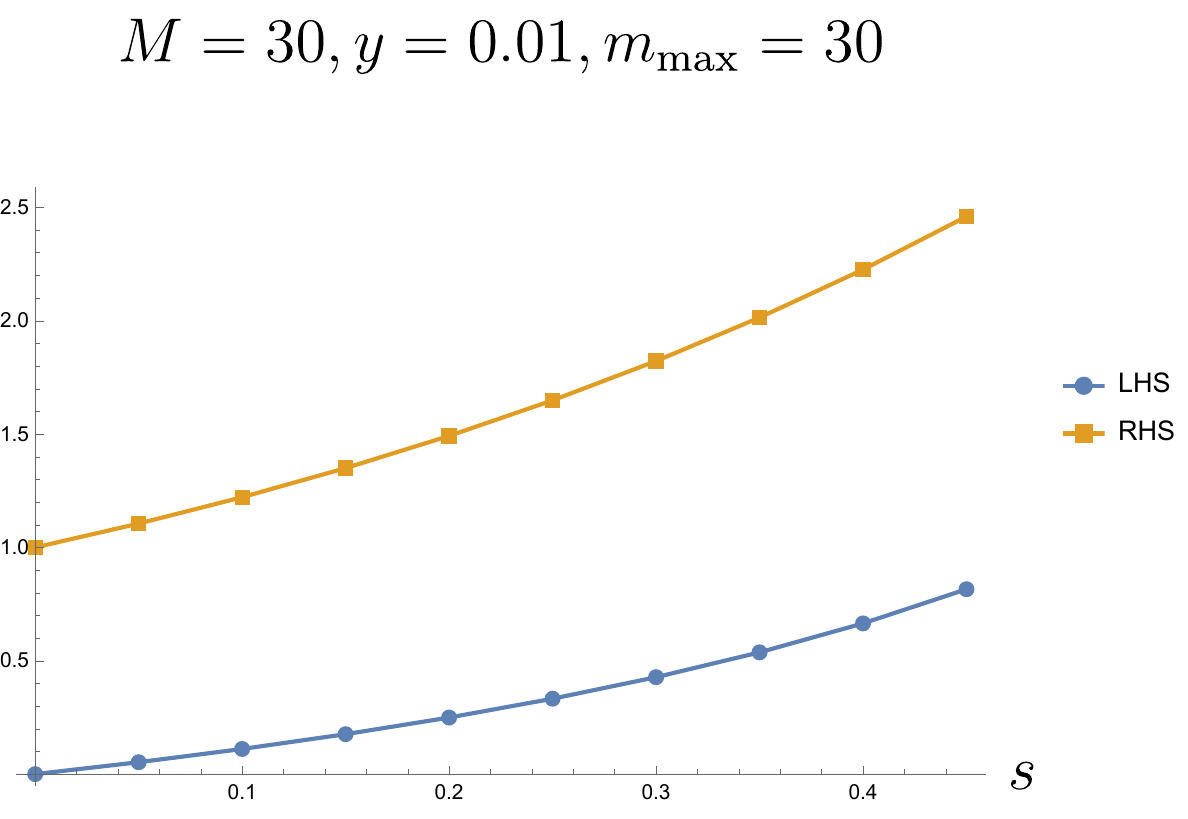}
    \caption{The plots of LHS and RHS with $M=30,y=0.01,m_{\max}=30$.}
    \label{fig:M=30_y=0.01_mmax=30}
\end{figure}

We now comment on the bound \eqref{eq:Ela^2X^2_bound}
\begin{equation}
   \sum_{\{m_k\}:m\leq m_{\max}}\frac{\Gamma(1+m)\;s}{(1-s)^{m+1}} \prod_{k=2}^M\qty(\frac{(-a_k)^{m_k}}{m_k!})\leq \exp{2s-\sum_{k=2}^M t_k a_k s^k}
\end{equation}
Since at $s=0$, the left hand side is zero and right hand side is one we obtain the inequality. The problem is when the inequality break down since the left hand side is asymptotic series. 
We can numerically confirm the inequalities for small $M$ with moderate $a_k,m_{\max}$ and $t_k$.
For simplification we set $a_k= g_{2k}\sigma^{2k}= y^k$ with $y=\frac{\overlineT{G}}{S}$ where $S=\mathrm{Poly}(\Smc)$. Then the left hand side is simplified 
\begin{equation}
\begin{split}
    \text{(LHS)} =\sum_{m=0}^{m_{\max}}\frac{\Gamma(1+m)s}{(1-s)^{m+1}} y^m \qty(\sum_{\{m_k\}:\sum_{k=2}^M km_k=m}\prod_{k=2}^M \qty(\frac{(-1)^{m_k}}{m_k!})).
\end{split}
\end{equation}
With a choice $t_k= 6/2^k$ the right hand side is 
\begin{eqnarray}
       \text{(RHS)} =\exp{2s-6\qty(\frac{ys}{2})^2 \frac{1-\qty(\frac{ys}{2})^{M-1}}{1-\frac{ ys }{2}}}.
\end{eqnarray}
We show some numerical tests in Figure \ref{fig:M=10_y=0.1_mmax=10}, Figure \ref{fig:M=10_y=0.1_mmax=30} and Figure \ref{fig:M=30_y=0.01_mmax=30}. In the first plot, we see the RHS exceeds LHS for $s<1/2$ with $y=0.1$. When we increase $M$ and $m_{\max}$, we observe the region where the inequality is valid is geeing narrow as in Figure \ref{fig:M=10_y=0.1_mmax=30}. Once we make $y$ smaller then the inequality recovered. From the usual discussion of the asymotic series, we expect the inequality is valid for $M<m_{\max}=\LandauO{1/y}$. In the physics discussion, we set $y=e^{-\LandauO{\Smc}}$, and we expect we can set $M$ and $m_{\max}$ to be $e^{\LandauO{\Smc}}$. 

\section{Numerical evidence of the  inequality \eqref{eq:1/N_correction_moment}}\label{appendix:1/N_correction_moment}
\begin{figure}[H]
        \centering
        \includegraphics[width=0.5\linewidth]{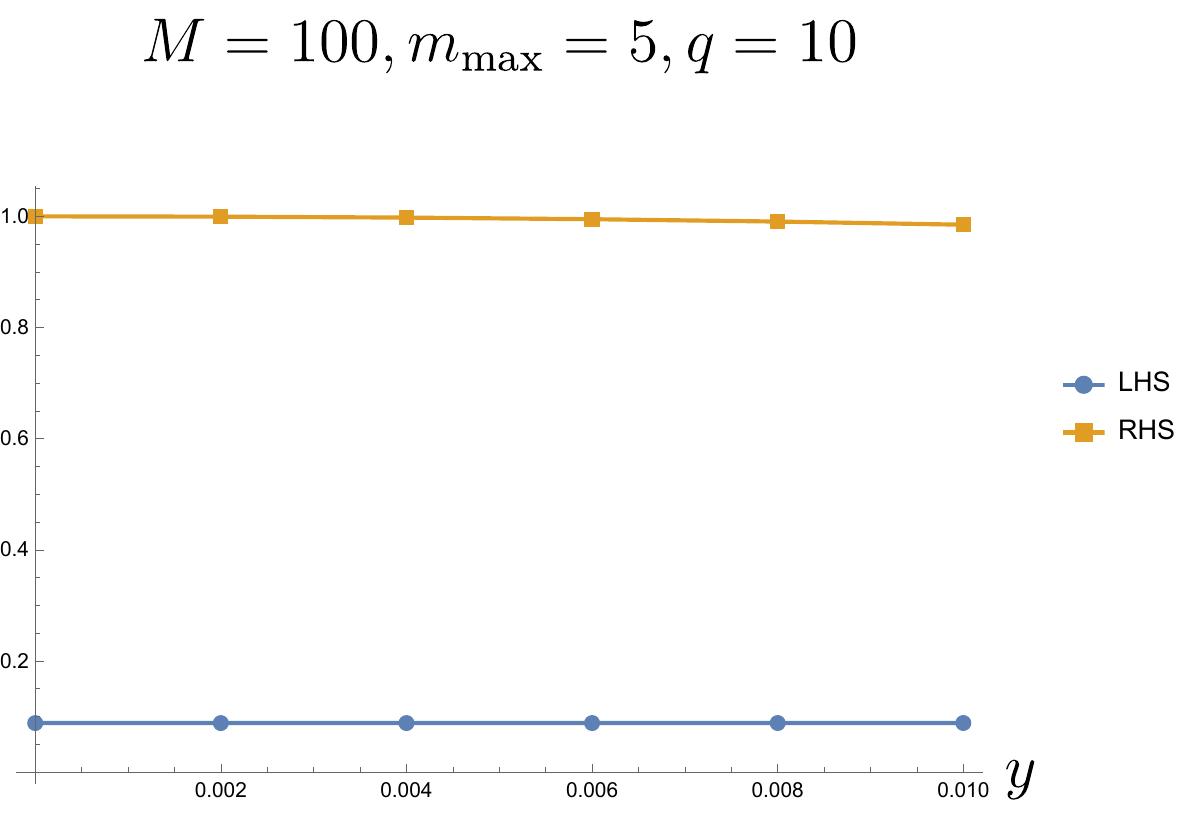}
        \caption{Plots of $G_n^{(2q)}$ and $E^{(2q)}_n$ with $M=100, m_{\max}=5, q=10$.}
        \label{fig:M=100_mmax=5_q=10}
    \end{figure}

\begin{figure}[H]
        \centering
        \includegraphics[width=0.5\linewidth]{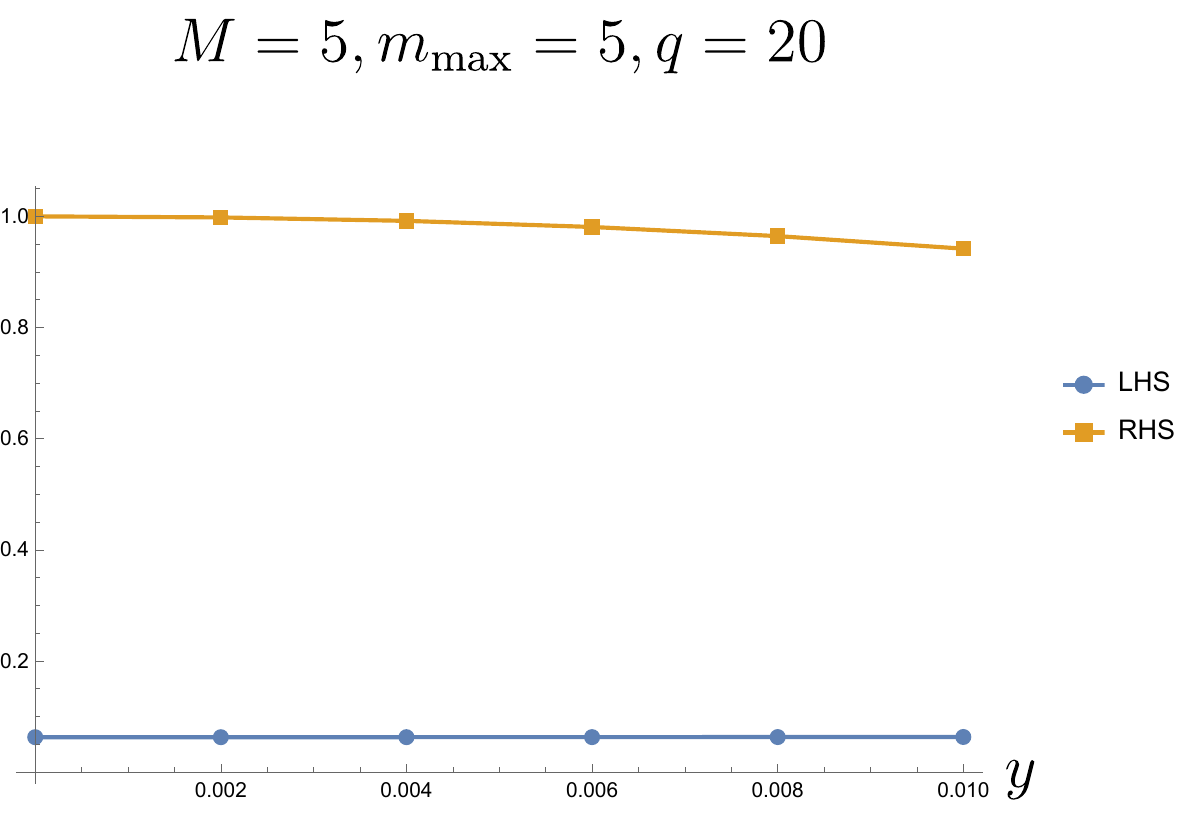}
        \caption{Plots of $G_n^{(2q)}$ and $E^{(2q)}_n$ with $M=2, m_{\max}=2, q=10$.}
        \label{fig:M=10_mmax=5_q=10}
    \end{figure}

\begin{figure}[H]
        \centering
        \includegraphics[width=0.5\linewidth]{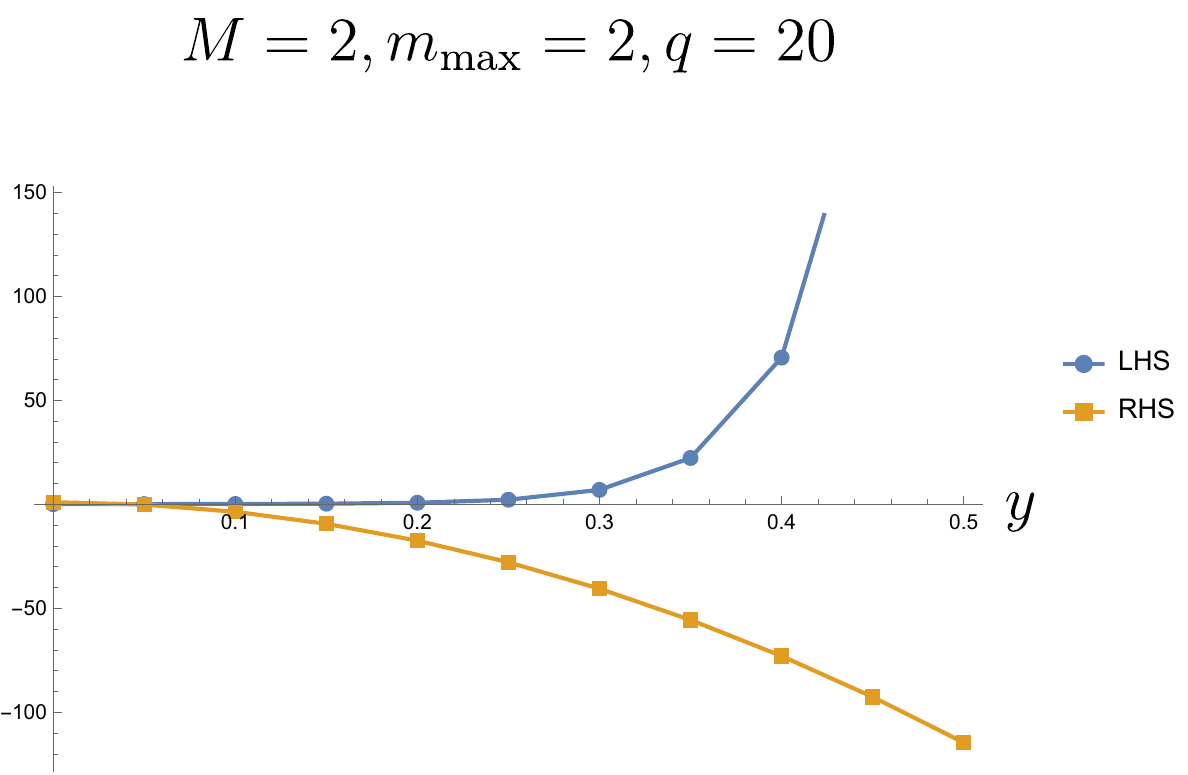}
        \caption{Plots of $G_n^{(2q)}$ and $E^{(2q)}_n$ with $M=2, m_{\max}=2, q=20$.}
        \label{fig:M=10_mmax=5_q=20}
    \end{figure}

\begin{figure}[H]
        \centering
        \includegraphics[width=0.5\linewidth]{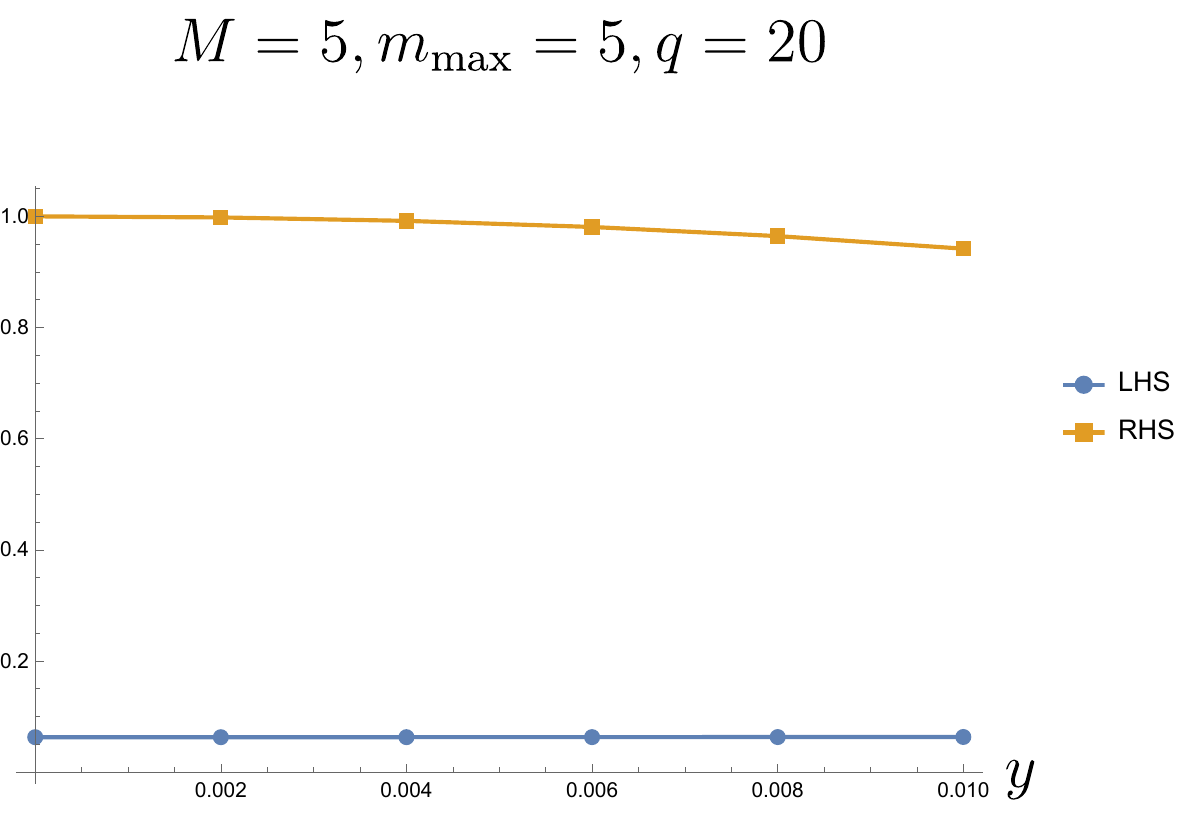}
        \caption{Plots of $G_n^{(2q)}$ and $E^{(2q)}_n$ with $M=5, m_{\max}=5, q=20$.}
        \label{fig:M=5_mmax=5_q=20}
    \end{figure}

We now address the conjectured inequality \eqref{eq:1/N_correction_moment}. The statement is following: 
consider the two positive functions for vectors $x=(x_2,x_3,\cdots,x_{L})$ 
    \begin{equation}
        \begin{split}
            G_n^{(2q)}(x)&:= \frac{1}{2^{q+1}q!}\sum_{\{n_k\}:\sum_{k}kn_k=q} \frac{(2q)!}{\prod_{k=1}^q ((2m)!)^{n_k}(n_k)!} \; \prod_{k=2}^q \qty(x_k)^{n_k} \\
            E_n^{(2q)}(x)&:= \; \sum_{\{m_k\}:m\leq m_{\max}} \frac{\Gamma(m+q)}{\Gamma(q)}\prod_{k=2}^{M} \frac{1}{m_k!}(-x_k)^{m_k}
        \end{split}
    \end{equation}
    where indices $n$ mean the function is normalized by $1/{2^{q+1}q!}$.
    For sufficiently small $x$ and for some positive integer $m_{\max}$, we have 
    \begin{equation}
        G_n^{(2q)}(x) \leq E_n^{(2q)}(x).
    \end{equation}
    For numerical calculation it is better to rewrite the summation to resolve the constraint. For function $G_{n}^{(2q)}(x)$, by using the relation $n_1 = q-\sum_{k=2}^q kn_k$ and for specific case $\Gamma_{2k}=\frac{1}{S^k}$, we have
    \begin{equation}
    \begin{split}
        G_{n}^{(2q)}(y) &= \frac{(2q)!}{2^{q+1}q!} \sum_{n_1'=0}^q \frac{\mathrm{Coeff_G}(n_1')}{((2!)^{q-n_1'} (q-n_1')!} y^{n_1'}\\ 
        \mathrm{Coeff_G}(n_1') &= \sum_{n_2,\cdots,n_q: \sum_{k=2}^q kn_k = n_1'=q-n_1}\frac{1}{\prod_{k=2}^q (\qty(2n_k)!)^{n_k}(n_k)!}
    \end{split}
    \end{equation}
    where $y=\frac{\overlineT{\canWightman{2}}}{S}$ which is very small. Also for $E_n^{(2q)}$, with identification $g_{2k}=-\ep_{k}\Gamma_{2k}=-\frac{\ep_{k}}{S^k}=\frac{1}{S^k}$ and $\sigma^2=\overlineT{\canWightman{2}}$, we have
    \begin{equation}
    \begin{split}
        E_{n}^{(2q)}(y) &=\sum_{m=0}^{m_{\max}} \frac{\Gamma(q+m)}{\Gamma(q)} \mathrm{Coeff_E}(m)y^{m}\\ 
        \mathrm{Coeff_E}(m) &= \sum_{m_2,\cdots,m_M: \sum_{k=2}^M km_k = m}\frac{1}{\prod_{k=2}^q (m_k)!}.
    \end{split}
    \end{equation}
    We see $y=0$ 
    \begin{eqnarray}
        G_{n}^{(2q)}(y=0) = \frac{(2q-1)!!}{2^{q+1}q!}\leq 1 =E_{n}^{(2q)}(y=0).
    \end{eqnarray}
    We set plots in Figure \ref{fig:M=100_mmax=5_q=10}, \ref{fig:M=10_mmax=5_q=10},  \ref{fig:M=10_mmax=5_q=20} and \ref{fig:M=5_mmax=5_q=20}. We see the inequality is valid  for large  enough $m_{\max}$, $M$ and small enough $y$. For fixed $q$, when we small $m_{\max}$ and $M$, we see the earlier violation of inequalities. By making $m_{\max}$ and $M$, we recover the inequalies. We put the plot of crossing points of two function in Figure \ref{fig:ysol_q=40}. We find that the crossing points are given by $y_{\mathrm{sol}}(q)\sim 1/q$. Note that we do not show the inequality for large $q=e^{\LandauO{\Smc}}$ with physical value of $\Smc$ and this is important open question.
  
    \begin{figure}[H]
        \centering
        \includegraphics[width=0.5\linewidth]{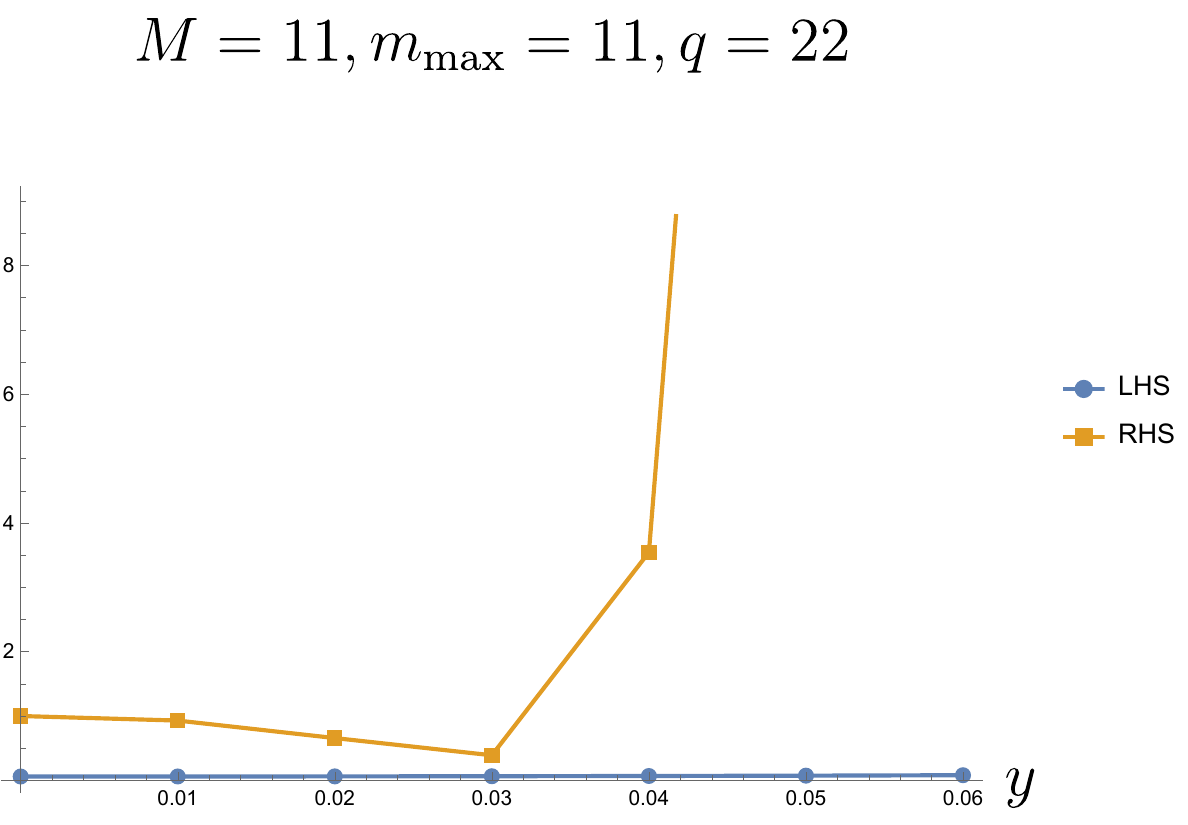}
        \caption{Plots of $G_n^{(2q)}$ and $E^{(2q)}_n$ with $M=11, m_{\max}=11, q=22$.}
        \label{fig:M=11_mmax_11_q=22}
    \end{figure}
    \begin{figure}[H]
        \centering
        \includegraphics[width=0.5\linewidth]{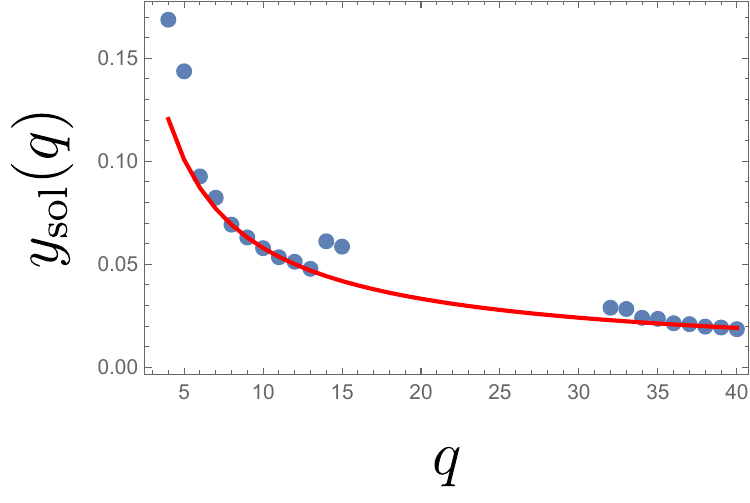}
        \caption{Plots of crossing points of $G_n^{(2q)}$ and $E^{(2q)}_n$ up to $q=40$. We set $m_{\max}=M=\floor{q/2}$.
        The red curve denotes $y_{\mathrm{sol}}(10)\frac{10}{q}$.
        In some $q$ we cannot find solutions. For example there is no solution at $q=22$. See Figure \ref{fig:M=11_mmax_11_q=22}. }
        \label{fig:ysol_q=40}
    \end{figure}
\pagebreak
\bibliographystyle{JHEP}
\bibliography{LDP}

@article{Gogolin:2015gts,
    author = "Gogolin, Christian and Eisert, Jens",
    title = "{Equilibration, thermalisation, and the emergence of statistical mechanics in closed quantum systems}",
    eprint = "1503.07538",
    archivePrefix = "arXiv",
    primaryClass = "quant-ph",
    doi = "10.1088/0034-4885/79/5/056001",
    journal = "Rept. Prog. Phys.",
    volume = "79",
    number = "5",
    pages = "056001",
    year = "2016"
}

@misc{Ogata:2011amo,
      title={Approximating macroscopic observables in quantum spin systems with commuting matrices}, 
      author={Yoshiko Ogata},
      year={2011},
      eprint={1111.5933},
      archivePrefix={arXiv},
      primaryClass={math.OA},
      url={https://arxiv.org/abs/1111.5933}, 
}

@article{Tasaki_2016,
	doi = {10.1007/s10955-016-1511-2},
  
	url = {https://doi.org/10.1007%2Fs10955-016-1511-2},
  
	year = 2016,
	month = {apr},
  
	publisher = {Springer Science and Business Media {LLC}
},
  
	volume = {163},
  
	number = {5},
  
	pages = {937--997},
  
	author = {Hal Tasaki},
  
	title = {Typicality of Thermal Equilibrium and Thermalization in Isolated Macroscopic Quantum Systems},
  
	journal = {Journal of Statistical Physics}
}

@article{Mori:2016met,
  title = {Macrostate equivalence of two general ensembles and specific relative entropies},
  author = {Mori, Takashi},
  journal = {Phys. Rev. E},
  volume = {94},
  issue = {2},
  pages = {020101},
  numpages = {5},
  year = {2016},
  month = {Aug},
  publisher = {American Physical Society},
  doi = {10.1103/PhysRevE.94.020101},
  url = {https://link.aps.org/doi/10.1103/PhysRevE.94.020101}
}

@article{Deutsch_1991,
  title = {Quantum statistical mechanics in a closed system},
  author = {Deutsch, J. M.},
  journal = {Phys. Rev. A},
  volume = {43},
  issue = {4},
  pages = {2046--2049},
  numpages = {0},
  year = {1991},
  month = {Feb},
  publisher = {American Physical Society},
  doi = {10.1103/PhysRevA.43.2046},
  url = {https://link.aps.org/doi/10.1103/PhysRevA.43.2046}
}

@article{Goldstein2010LongtimeBO,
  title={Long-time behavior of macroscopic quantum systems},
  author={Sheldon Goldstein and Joel L Lebowitz and Roderich Tumulka and Nino Zangh{\'i}},
  journal={The European Physical Journal H},
  year={2010},
  volume={35},
  pages={173-200}
}

@misc{Mori:2016wet,
  doi = {10.48550/ARXIV.1609.09776},
  
  url = {https://arxiv.org/abs/1609.09776},
  
  author = {Mori, Takashi},
  
  title = {Weak eigenstate thermalization with large deviation bound},
  
  publisher = {arXiv},
  
  year = {2016},
  
  copyright = {arXiv.org perpetual, non-exclusive license}
}

@article{Arad:2014znf,
    author = "Arad, Itai and Kuwahara, Tomotaka and Landau, Zeph",
    title = "{Connecting global and local energy distributions in quantum spin models on a lattice}",
    eprint = "1406.3898",
    archivePrefix = "arXiv",
    primaryClass = "quant-ph",
    doi = "10.1088/1742-5468/2016/03/033301",
    journal = "J. Stat. Mech.",
    volume = "1603",
    number = "3",
    pages = "033301",
    year = "2016"
}

@article{Popescu:2005fsm,
  doi = {10.48550/ARXIV.QUANT-PH/0511225},
  
  url = {https://arxiv.org/abs/quant-ph/0511225},
  
  author = {Popescu, Sandu and Short, Anthony J. and Winter, Andreas},
  
  title = {The foundations of statistical mechanics from entanglement: Individual states vs. averages},
  
  publisher = {arXiv},
  
  year = {2005},
  
  copyright = {Assumed arXiv.org perpetual, non-exclusive license to distribute this article for submissions made before January 2004}
}

@Article{Popescu2006,
author={Popescu, Sandu
and Short, Anthony J.
and Winter, Andreas},
title={Entanglement and the foundations of statistical mechanics},
journal={Nature Physics},
year={2006},
month={Nov},
day={01},
volume={2},
number={11},
pages={754-758},
issn={1745-2481},
doi={10.1038/nphys444},
url={https://doi.org/10.1038/nphys444}
}

@article{Srednicki_1999,
	doi = {10.1088/0305-4470/32/7/007},
  
	url = {https://doi.org/10.1088%2F0305-4470%2F32%2F7%2F007},
  
	year = 1999,
	month = {jan},
  
	publisher = {{IOP} Publishing},
  
	volume = {32},
  
	number = {7},
  
	pages = {1163--1175},
  
	author = {Mark Srednicki},
  
	title = {The approach to thermal equilibrium in quantized chaotic systems},
  
	journal = {Journal of Physics A: Mathematical and General}
}

@article{Mori_2018,
	doi = {10.1088/1361-6455/aabcdf},
  
	url = {https://doi.org/10.1088%2F1361-6455%2Faabcdf},
  
	year = 2018,
	month = {may},
  
	publisher = {{IOP} Publishing},
  
	volume = {51},
  
	number = {11},
  
	pages = {112001},
  
	author = {Takashi Mori and Tatsuhiko N Ikeda and Eriko Kaminishi and Masahito Ueda},
  
	title = {Thermalization and prethermalization in isolated quantum systems: a theoretical overview},
  
	journal = {Journal of Physics B: Atomic, Molecular and Optical Physics}
}

@article{Goldstein:2010ate,
  title = {Approach to thermal equilibrium of macroscopic quantum systems},
  author = {Goldstein, Sheldon and Lebowitz, Joel L. and Mastrodonato, Christian and Tumulka, Roderich and Zanghi, Nino},
  journal = {Phys. Rev. E},
  volume = {81},
  issue = {1},
  pages = {011109},
  numpages = {9},
  year = {2010},
  month = {Jan},
  publisher = {American Physical Society},
  doi = {10.1103/PhysRevE.81.011109},
  url = {https://link.aps.org/doi/10.1103/PhysRevE.81.011109}
}

@article{Beugeling:2014fse,
  title = {Finite-size scaling of eigenstate thermalization},
  author = {Beugeling, W. and Moessner, R. and Haque, Masudul},
  journal = {Phys. Rev. E},
  volume = {89},
  issue = {4},
  pages = {042112},
  numpages = {9},
  year = {2014},
  month = {Apr},
  publisher = {American Physical Society},
  doi = {10.1103/PhysRevE.89.042112},
  url = {https://link.aps.org/doi/10.1103/PhysRevE.89.042112}
}

@article{Short:2011pvc,
    author = "Short, Anthony J. and Farrelly, Terence C.",
    title = "{Quantum equilibration in finite time}",
    eprint = "1110.5759",
    archivePrefix = "arXiv",
    primaryClass = "quant-ph",
    doi = "10.1088/1367-2630/14/1/013063",
    journal = "New J. Phys.",
    volume = "14",
    number = "1",
    pages = "013063",
    year = "2012"
}

@article{Biroli_2010erf,
   title={Effect of Rare Fluctuations on the Thermalization of Isolated Quantum Systems},
   volume={105},
   ISSN={1079-7114},
   url={http://dx.doi.org/10.1103/PhysRevLett.105.250401},
   DOI={10.1103/physrevlett.105.250401},
   number={25},
   journal={Physical Review Letters},
   publisher={American Physical Society (APS)},
   author={Biroli, Giulio and Kollath, Corinna and Läuchli, Andreas M.},
   year={2010},
   month=dec }

@article{Iyoda:2017ftm,
  title = {Fluctuation Theorem for Many-Body Pure Quantum States},
  author = {Iyoda, Eiki and Kaneko, Kazuya and Sagawa, Takahiro},
  journal = {Phys. Rev. Lett.},
  volume = {119},
  issue = {10},
  pages = {100601},
  numpages = {6},
  year = {2017},
  month = {Sep},
  publisher = {American Physical Society},
  doi = {10.1103/PhysRevLett.119.100601},
  url = {https://link.aps.org/doi/10.1103/PhysRevLett.119.100601}
}

@article{Kim:2014jfl,
    author = "Kim, Hyungwon and Ikeda, Tatsuhiko N. and Huse, David A.",
    title = "{Testing whether all eigenstates obey the eigenstate thermalization hypothesis}",
    doi = "10.1103/PhysRevE.90.052105",
    journal = "Phys. Rev. E",
    volume = "90",
    number = "5",
    pages = "052105",
    year = "2014"
}

@article{Fitzpatrick:2015zha,
    author = "Fitzpatrick, A. Liam and Kaplan, Jared and Walters, Matthew T.",
    title = "{Virasoro Conformal Blocks and Thermality from Classical Background Fields}",
    eprint = "1501.05315",
    archivePrefix = "arXiv",
    primaryClass = "hep-th",
    doi = "10.1007/JHEP11(2015)200",
    journal = "JHEP",
    volume = "11",
    pages = "200",
    year = "2015"
}

@article{Lashkari:2016ethcft,
  doi = {10.48550/ARXIV.1610.00302},
  
  url = {https://arxiv.org/abs/1610.00302},
  
  author = {Lashkari, Nima and Dymarsky, Anatoly and Liu, Hong},
  
  title = {Eigenstate Thermalization Hypothesis in Conformal Field Theory},
  
  publisher = {arXiv},
  
  year = {2016},
  
  copyright = {arXiv.org perpetual, non-exclusive license}
}

@article{Dymarsky2016SubsystemE,
  title={Subsystem ETH},
  author={Anatoly Dymarsky and Nima Lashkari and Hong Liu},
  journal={Physical Review E},
  year={2016}
}

@article{Fitzpatrick:2015dlt,
    author = "Fitzpatrick, A. Liam and Kaplan, Jared",
    title = "{Conformal Blocks Beyond the Semi-Classical Limit}",
    eprint = "1512.03052",
    archivePrefix = "arXiv",
    primaryClass = "hep-th",
    doi = "10.1007/JHEP05(2016)075",
    journal = "JHEP",
    volume = "05",
    pages = "075",
    year = "2016"
}

@article{Hikida:2018khg,
    author = "Hikida, Yasuaki and Kusuki, Yuya and Takayanagi, Tadashi",
    title = "{Eigenstate thermalization hypothesis and modular invariance of two-dimensional conformal field theories}",
    eprint = "1804.09658",
    archivePrefix = "arXiv",
    primaryClass = "hep-th",
    reportNumber = "YITP-18-37, IPMU18-0072",
    doi = "10.1103/PhysRevD.98.026003",
    journal = "Phys. Rev. D",
    volume = "98",
    number = "2",
    pages = "026003",
    year = "2018"
}

@article{Brehm:2018ipf,
    author = "Brehm, Enrico M. and Das, Diptarka and Datta, Shouvik",
    title = "{Probing thermality beyond the diagonal}",
    eprint = "1804.07924",
    archivePrefix = "arXiv",
    primaryClass = "hep-th",
    doi = "10.1103/PhysRevD.98.126015",
    journal = "Phys. Rev. D",
    volume = "98",
    number = "12",
    pages = "126015",
    year = "2018"
}

@article{Basu:2017kzo,
    author = "Basu, Pallab and Das, Diptarka and Datta, Shouvik and Pal, Sridip",
    title = "{Thermality of eigenstates in conformal field theories}",
    eprint = "1705.03001",
    archivePrefix = "arXiv",
    primaryClass = "hep-th",
    doi = "10.1103/PhysRevE.96.022149",
    journal = "Phys. Rev. E",
    volume = "96",
    number = "2",
    pages = "022149",
    year = "2017"
}

@article{Karlsson:2021duj,
    author = "Karlsson, Robin and Parnachev, Andrei and Tadi\'c, Petar",
    title = "{Thermalization in large-N CFTs}",
    eprint = "2102.04953",
    archivePrefix = "arXiv",
    primaryClass = "hep-th",
    doi = "10.1007/JHEP09(2021)205",
    journal = "JHEP",
    volume = "09",
    pages = "205",
    year = "2021"
}

@article{Mukhametzhanov:2019pzy,
    author = "Mukhametzhanov, Baur and Zhiboedov, Alexander",
    title = "{Modular invariance, tauberian theorems and microcanonical entropy}",
    eprint = "1904.06359",
    archivePrefix = "arXiv",
    primaryClass = "hep-th",
    reportNumber = "CERN-TH-2019-043",
    doi = "10.1007/JHEP10(2019)261",
    journal = "JHEP",
    volume = "10",
    pages = "261",
    year = "2019"
}

@article{Tasaki:2024bvh,
    author = "Tasaki, Hal",
    title = "{Macroscopic Irreversibility in Quantum Systems: ETH and Equilibration in a Free Fermion Chain}",
    eprint = "2401.15263",
    archivePrefix = "arXiv",
    primaryClass = "cond-mat.stat-mech",
    month = "1",
    year = "2024"
}

@article{Linden:2008awz,
    author = "Linden, Noah and Popescu, Sandu and Short, Anthony J. and Winter, Andreas",
    title = "{Quantum mechanical evolution towards thermal equilibrium}",
    eprint = "0812.2385",
    archivePrefix = "arXiv",
    primaryClass = "quant-ph",
    doi = "10.1103/PhysRevE.79.061103",
    journal = "Phys. Rev. E",
    volume = "79",
    number = "6",
    pages = "061103",
    year = "2009"
}

@article{Sonner:2017hxc,
    author = "Sonner, Julian and Vielma, Manuel",
    title = "{Eigenstate thermalization in the Sachdev-Ye-Kitaev model}",
    eprint = "1707.08013",
    archivePrefix = "arXiv",
    primaryClass = "hep-th",
    doi = "10.1007/JHEP11(2017)149",
    journal = "JHEP",
    volume = "11",
    pages = "149",
    year = "2017"
}

@article{Milekhin:2023was,
    author = "Milekhin, Alexey and Sukhov, Nikolay",
    title = "{All holographic systems have scar states}",
    eprint = "2307.11348",
    archivePrefix = "arXiv",
    primaryClass = "hep-th",
    month = "7",
    year = "2023"
}

@article{Dodelson:2022eiz,
    author = "Dodelson, Matthew and Zhiboedov, Alexander",
    title = "{Gravitational orbits, double-twist mirage, and many-body scars}",
    eprint = "2204.09749",
    archivePrefix = "arXiv",
    primaryClass = "hep-th",
    reportNumber = "CERN-TH-2022-065",
    doi = "10.1007/JHEP12(2022)163",
    journal = "JHEP",
    volume = "12",
    pages = "163",
    year = "2022"
}

@article{Hawking:1976bpg,
  title = {Breakdown of predictability in gravitational collapse},
  author = {Hawking, S. W.},
  journal = {Phys. Rev. D},
  volume = {14},
  issue = {10},
  pages = {2460--2473},
  numpages = {0},
  year = {1976},
  month = {Nov},
  publisher = {American Physical Society},
  doi = {10.1103/PhysRevD.14.2460},
  url = {https://link.aps.org/doi/10.1103/PhysRevD.14.2460}
}

@article{Saad:2019pqd,
    author = "Saad, Phil",
    title = "{Late Time Correlation Functions, Baby Universes, and ETH in JT Gravity}",
    eprint = "1910.10311",
    archivePrefix = "arXiv",
    primaryClass = "hep-th",
    month = "10",
    year = "2019"
}

@BOOK{Rulle:1990smr,
       author = {{Ruelle}, David},
        title = "{Statistical Mechanics: Rigorous Results}",
         year = 1999,
          doi = {10.1142/4090},
       adsurl = {https://ui.adsabs.harvard.edu/abs/1999smrr.book.....R}
}

@Article{Lima1972,
author={Lima, R.},
title={Equivalence of ensembles in quantum lattice systems: States},
journal={Communications in Mathematical Physics},
year={1972},
month={Sep},
day={01},
volume={24},
number={3},
pages={180-192},
issn={1432-0916},
doi={10.1007/BF01877711},
url={https://doi.org/10.1007/BF01877711}
}

@ARTICLE{TASAKI:2018eoe,
       author = {{Tasaki}, Hal},
        title = "{On the Local Equivalence Between the Canonical and the Microcanonical Ensembles for Quantum Spin Systems}",
      journal = {Journal of Statistical Physics},
         year = 2018,
        month = aug,
       volume = {172},
       number = {4},
        pages = {905-926},
          doi = {10.1007/s10955-018-2077-y},
archivePrefix = {arXiv},
       eprint = {1609.06983},
 primaryClass = {cond-mat.stat-mech},
       adsurl = {https://ui.adsabs.harvard.edu/abs/2018JSP...172..905T}
}

@ARTICLE{Brandao:2015eos,
       author = {{Brandao}, Fernando G.~S.~L. and {Cramer}, Marcus},
        title = "{Equivalence of Statistical Mechanical Ensembles for Non-Critical Quantum Systems}",
      journal = {arXiv e-prints},
         year = 2015,
        month = feb,
          eid = {arXiv:1502.03263},
        pages = {arXiv:1502.03263},
          doi = {10.48550/arXiv.1502.03263},
archivePrefix = {arXiv},
       eprint = {1502.03263},
 primaryClass = {quant-ph},
       adsurl = {https://ui.adsabs.harvard.edu/abs/2015arXiv150203263B}
}

@ARTICLE{Tasaki:2008,
       author = {{Tasaki}, Hal},
        title = "{Mechanics, Statistical, (in Japanese)}",
      journal = {Baifukan},
         year = 2008
}

@article{D_Alessio_2016,
	doi = {10.1080/00018732.2016.1198134},
  
	url = {https://doi.org/10.1080%2F00018732.2016.1198134},
  
	year = 2016,
	month = {may},
  
	publisher = {Informa {UK} Limited},
  
	volume = {65},
  
	number = {3},
  
	pages = {239--362},
  
	author = {Luca D{\textquotesingle}Alessio and Yariv Kafri and Anatoli Polkovnikov and Marcos Rigol},
  
	title = {From quantum chaos and eigenstate thermalization to statistical mechanics and thermodynamics},
  
	journal = {Advances in Physics}
}

@article{Wald_1993,
	doi = {10.1103/physrevd.48.r3427},
  
	url = {https://doi.org/10.1103%2Fphysrevd.48.r3427},
  
	year = 1993,
	month = {oct},
  
	publisher = {American Physical Society ({APS})},
  
	volume = {48},
  
	number = {8},
  
	pages = {R3427--R3431},
  
	author = {Robert M. Wald},
  
	title = {Black hole entropy is the Noether charge},
  
	journal = {Physical Review D}
}

@Article{Neumann1929,
author={Neumann, J. v.},
title={Beweis des Ergodensatzes und desH-Theorems in der neuen Mechanik},
journal={Zeitschrift f{\"u}r Physik},
year={1929},
month={Jan},
day={01},
volume={57},
number={1},
pages={30-70},
issn={0044-3328},
doi={10.1007/BF01339852},
url={https://doi.org/10.1007/BF01339852}
}

@article{Goldstein_2017,
	doi = {10.1002/andp.201600301},
  
	url = {https://doi.org/10.1002%2Fandp.201600301},
  
	year = 2017,
	month = {feb},
  
	publisher = {Wiley},
  
	volume = {529},
  
	number = {7},
  
	pages = {1600301},
  
	author = {Sheldon Goldstein and David A. Huse and Joel L. Lebowitz and Roderich Tumulka},
  
	title = {Macroscopic and microscopic thermal equilibrium},
  
	journal = {Annalen der Physik}
}

@misc{Sugita:2006bqs,
      title={On the Basis of Quantum Statistical Mechanics}, 
      author={Ayumu Sugita},
      year={2006},
      eprint={cond-mat/0602625},
      archivePrefix={arXiv},
      primaryClass={cond-mat.stat-mech},
      url={https://arxiv.org/abs/cond-mat/0602625}, 
}

@article{Goldstein_2006,
	doi = {10.1103/physrevlett.96.050403},
  
	url = {https://doi.org/10.1103%2Fphysrevlett.96.050403},
  
	year = 2006,
	month = {feb},
  
	publisher = {American Physical Society ({APS})},
  
	volume = {96},
  
	number = {5},
  
	author = {Sheldon Goldstein and Joel L. Lebowitz and Roderich Tumulka and Nino Zangh{\`{\i}
}},
  
	title = {Canonical Typicality},
  
	journal = {Physical Review Letters}
}

@article{Bernien:2017ubn,
    author = "Bernien, Hannes and others",
    title = "{Probing many-body dynamics on a 51-atom quantum simulator}",
    eprint = "1707.04344",
    archivePrefix = "arXiv",
    primaryClass = "quant-ph",
    doi = "10.1038/nature24622",
    journal = "Nature",
    volume = "551",
    pages = "579--584",
    year = "2017"
}

@article{Turner_2018,
   title={Quantum scarred eigenstates in a Rydberg atom chain: Entanglement, breakdown of thermalization, and stability to perturbations},
   volume={98},
   ISSN={2469-9969},
   url={http://dx.doi.org/10.1103/PhysRevB.98.155134},
   DOI={10.1103/physrevb.98.155134},
   number={15},
   journal={Physical Review B},
   publisher={American Physical Society (APS)},
   author={Turner, C. J. and Michailidis, A. A. and Abanin, D. A. and Serbyn, M. and Papić, Z.},
   year={2018},
   month=oct }

@article{Knysh:2024asf,
    author = "Knysh, Maria and Liu, Hong and Pinzani-Fokeeva, Natalia",
    title = "{New horizon symmetries, hydrodynamics, and quantum chaos}",
    eprint = "2405.17559",
    archivePrefix = "arXiv",
    primaryClass = "hep-th",
    doi = "10.1007/JHEP09(2024)162",
    journal = "JHEP",
    volume = "09",
    pages = "162",
    year = "2024"
}

@article{Ikeda_2015,
   title={How accurately can the microcanonical ensemble describe small isolated quantum systems?},
   volume={92},
   ISSN={1550-2376},
   url={http://dx.doi.org/10.1103/PhysRevE.92.020102},
   DOI={10.1103/physreve.92.020102},
   number={2},
   journal={Physical Review E},
   publisher={American Physical Society (APS)},
   author={Ikeda, Tatsuhiko N. and Ueda, Masahito},
   year={2015},
   month=aug }

@article{Srednicki:1995pt,
    author = "Srednicki, Mark",
    title = "{Thermal fluctuations in quantized chaotic systems}",
    eprint = "chao-dyn/9511001",
    archivePrefix = "arXiv",
    reportNumber = "UCSB-TH-95-34",
    doi = "10.1088/0305-4470/29/4/003",
    journal = "J. Phys. A",
    volume = "29",
    pages = "L75--L79",
    year = "1996"
}

@article{Sugimoto:2020nnw,
    author = "Sugimoto, Shoki and Hamazaki, Ryusuke and Ueda, Masahito",
    title = "{Test of the Eigenstate Thermalization Hypothesis Based on Local Random Matrix Theory}",
    eprint = "2005.06379",
    archivePrefix = "arXiv",
    primaryClass = "cond-mat.stat-mech",
    reportNumber = "RIKEN-iTHEMS-Report-20",
    doi = "10.1103/PhysRevLett.126.120602",
    journal = "Phys. Rev. Lett.",
    volume = "126",
    number = "12",
    pages = "120602",
    year = "2021"
}

@article{Sugimoto:2021aao,
    author = "Sugimoto, Shoki and Hamazaki, Ryusuke and Ueda, Masahito",
    title = "{Eigenstate Thermalization in Long-Range Interacting Systems}",
    eprint = "2111.12484",
    archivePrefix = "arXiv",
    primaryClass = "cond-mat.stat-mech",
    doi = "10.1103/PhysRevLett.129.030602",
    journal = "Phys. Rev. Lett.",
    volume = "129",
    number = "3",
    pages = "030602",
    year = "2022"
}

@article{PhysRevLett.120.080603,
  title = {Atypicality of Most Few-Body Observables},
  author = {Hamazaki, Ryusuke and Ueda, Masahito},
  journal = {Phys. Rev. Lett.},
  volume = {120},
  issue = {8},
  pages = {080603},
  numpages = {6},
  year = {2018},
  month = {Feb},
  publisher = {American Physical Society},
  doi = {10.1103/PhysRevLett.120.080603},
  url = {https://link.aps.org/doi/10.1103/PhysRevLett.120.080603}
}

@article{Huang:2017fng,
    author = "Huang, Yichen and Brand\~ao, Fernando G. S. L. and Zhang, Yong-Liang",
    title = "{Finite-size scaling of out-of-time-ordered correlators at late times}",
    eprint = "1705.07597",
    archivePrefix = "arXiv",
    primaryClass = "quant-ph",
    doi = "10.1103/PhysRevLett.123.010601",
    journal = "Phys. Rev. Lett.",
    volume = "123",
    number = "1",
    pages = "010601",
    year = "2019"
}

@article{Susskind:1994vu,
    author = "Susskind, Leonard",
    title = "{The World as a hologram}",
    eprint = "hep-th/9409089",
    archivePrefix = "arXiv",
    reportNumber = "SU-ITP-94-33",
    doi = "10.1063/1.531249",
    journal = "J. Math. Phys.",
    volume = "36",
    pages = "6377--6396",
    year = "1995"
}

@article{tHooft:1993dmi,
    author = "'t Hooft, Gerard",
    title = "{Dimensional reduction in quantum gravity}",
    eprint = "gr-qc/9310026",
    archivePrefix = "arXiv",
    reportNumber = "THU-93-26",
    journal = "Conf. Proc. C",
    volume = "930308",
    pages = "284--296",
    year = "1993"
}

@article{Maldacena:1997re,
    author = "Maldacena, Juan Martin",
    title = "{The Large N limit of superconformal field theories and supergravity}",
    eprint = "hep-th/9711200",
    archivePrefix = "arXiv",
    reportNumber = "HUTP-97-A097, HUTP-98-A097",
    doi = "10.4310/ATMP.1998.v2.n2.a1",
    journal = "Adv. Theor. Math. Phys.",
    volume = "2",
    pages = "231--252",
    year = "1998"
}

@article{Witten:1998qj,
    author = "Witten, Edward",
    title = "{Anti-de Sitter space and holography}",
    eprint = "hep-th/9802150",
    archivePrefix = "arXiv",
    reportNumber = "IASSNS-HEP-98-15",
    doi = "10.4310/ATMP.1998.v2.n2.a2",
    journal = "Adv. Theor. Math. Phys.",
    volume = "2",
    pages = "253--291",
    year = "1998"
}

@article{Gubser:1998bc,
    author = "Gubser, S. S. and Klebanov, Igor R. and Polyakov, Alexander M.",
    title = "{Gauge theory correlators from noncritical string theory}",
    eprint = "hep-th/9802109",
    archivePrefix = "arXiv",
    reportNumber = "PUPT-1767",
    doi = "10.1016/S0370-2693(98)00377-3",
    journal = "Phys. Lett. B",
    volume = "428",
    pages = "105--114",
    year = "1998"
}

@article{El-Showk:2011yvt,
    author = "El-Showk, Sheer and Papadodimas, Kyriakos",
    title = "{Emergent Spacetime and Holographic CFTs}",
    eprint = "1101.4163",
    archivePrefix = "arXiv",
    primaryClass = "hep-th",
    doi = "10.1007/JHEP10(2012)106",
    journal = "JHEP",
    volume = "10",
    pages = "106",
    year = "2012"
}

@article{Heemskerk:2009pn,
    author = "Heemskerk, Idse and Penedones, Joao and Polchinski, Joseph and Sully, James",
    title = "{Holography from Conformal Field Theory}",
    eprint = "0907.0151",
    archivePrefix = "arXiv",
    primaryClass = "hep-th",
    reportNumber = "NSF-KITP-09-110",
    doi = "10.1088/1126-6708/2009/10/079",
    journal = "JHEP",
    volume = "10",
    pages = "079",
    year = "2009"
}

@article{Keski-Vakkuri:1998gmz,
    author = "Keski-Vakkuri, Esko",
    title = "{Bulk and boundary dynamics in BTZ black holes}",
    eprint = "hep-th/9808037",
    archivePrefix = "arXiv",
    reportNumber = "CALT-68-2191",
    doi = "10.1103/PhysRevD.59.104001",
    journal = "Phys. Rev. D",
    volume = "59",
    pages = "104001",
    year = "1999"
}

@article{Maldacena:2001kr,
    author = "Maldacena, Juan Martin",
    title = "{Eternal black holes in anti-de Sitter}",
    eprint = "hep-th/0106112",
    archivePrefix = "arXiv",
    reportNumber = "NSF-ITP-01-59",
    doi = "10.1088/1126-6708/2003/04/021",
    journal = "JHEP",
    volume = "04",
    pages = "021",
    year = "2003"
}

@article{Festuccia:2006sa,
    author = "Festuccia, Guido and Liu, Hong",
    title = "{The Arrow of time, black holes, and quantum mixing of large N Yang-Mills theories}",
    eprint = "hep-th/0611098",
    archivePrefix = "arXiv",
    reportNumber = "MIT-CTP-3783",
    doi = "10.1088/1126-6708/2007/12/027",
    journal = "JHEP",
    volume = "12",
    pages = "027",
    year = "2007"
}

@article{Banados:2022nhj,
    author = "Ba\~nados, M\'aximo and Bianchi, Ernesto and Mu\~noz, Iv\'an and Skenderis, Kostas",
    title = "{Bulk renormalization and the AdS/CFT correspondence}",
    eprint = "2208.11539",
    archivePrefix = "arXiv",
    primaryClass = "hep-th",
    doi = "10.1103/PhysRevD.107.L021901",
    journal = "Phys. Rev. D",
    volume = "107",
    number = "2",
    pages = "L021901",
    year = "2023"
}

@misc{Kitaev,
        author  = {Kitaev, Alexei},
        title   = {A simple model of quantum holography},
        howpublished = "\url{https://online.kitp.ucsb.edu/online/entangled15/kitaev/}",
        year    = {2015}
}

@article{Maldacena:2016hyu,
    author = "Maldacena, Juan and Stanford, Douglas",
    title = "{Remarks on the Sachdev-Ye-Kitaev model}",
    eprint = "1604.07818",
    archivePrefix = "arXiv",
    primaryClass = "hep-th",
    doi = "10.1103/PhysRevD.94.106002",
    journal = "Phys. Rev. D",
    volume = "94",
    number = "10",
    pages = "106002",
    year = "2016"
}

@article{Maldacena:2018lmt,
    author = "Maldacena, Juan and Qi, Xiao-Liang",
    title = "{Eternal traversable wormhole}",
    eprint = "1804.00491",
    archivePrefix = "arXiv",
    primaryClass = "hep-th",
    month = "4",
    year = "2018"
}

@article{Furuya:2023fei,
    author = "Furuya, Keiichiro and Lashkari, Nima and Moosa, Mudassir and Ouseph, Shoy",
    title = "{Information loss, mixing and emergent type III$_{1}$ factors}",
    eprint = "2305.16028",
    archivePrefix = "arXiv",
    primaryClass = "hep-th",
    doi = "10.1007/JHEP08(2023)111",
    journal = "JHEP",
    volume = "08",
    pages = "111",
    year = "2023"
}

@article{Leutheusser:2021frk,
    author = "Leutheusser, Samuel Aaron Wehlau",
    title = "{Emergent Times in Holographic Duality}",
    eprint = "2112.12156",
    archivePrefix = "arXiv",
    primaryClass = "hep-th",
    reportNumber = "MIT-CTP/5382",
    doi = "10.1103/PhysRevD.108.086020",
    journal = "Phys. Rev. D",
    volume = "108",
    number = "8",
    pages = "086020",
    year = "2023"
}

@article{Gesteau:2023rrx,
    author = "Gesteau, Elliott",
    title = "{Emergent spacetime and the ergodic hierarchy}",
    eprint = "2310.13733",
    archivePrefix = "arXiv",
    primaryClass = "hep-th",
    month = "10",
    year = "2023"
}

@article{Bekenstein:1972tm,
    author = "Bekenstein, J. D.",
    title = "{Black holes and the second law}",
    doi = "10.1007/BF02757029",
    journal = "Lett. Nuovo Cim.",
    volume = "4",
    pages = "737--740",
    year = "1972"
}

@article{Bekenstein:1973ur,
    author = "Bekenstein, Jacob D.",
    title = "{Black holes and entropy}",
    doi = "10.1103/PhysRevD.7.2333",
    journal = "Phys. Rev. D",
    volume = "7",
    pages = "2333--2346",
    year = "1973"
}

@article{Iizuka:2008hg,
    author = "Iizuka, Norihiro and Polchinski, Joseph",
    title = "{A Matrix Model for Black Hole Thermalization}",
    eprint = "0801.3657",
    archivePrefix = "arXiv",
    primaryClass = "hep-th",
    doi = "10.1088/1126-6708/2008/10/028",
    journal = "JHEP",
    volume = "10",
    pages = "028",
    year = "2008"
}

@article{Gibbons:1976ue,
    author = "Gibbons, G. W. and Hawking, S. W.",
    title = "{Action Integrals and Partition Functions in Quantum Gravity}",
    reportNumber = "PRINT-76-0995 (CAMBRIDGE)",
    doi = "10.1103/PhysRevD.15.2752",
    journal = "Phys. Rev. D",
    volume = "15",
    pages = "2752--2756",
    year = "1977"
}

@article{Bardeen:1973gs,
    author = "Bardeen, James M. and Carter, B. and Hawking, S. W.",
    title = "{The Four laws of black hole mechanics}",
    doi = "10.1007/BF01645742",
    journal = "Commun. Math. Phys.",
    volume = "31",
    pages = "161--170",
    year = "1973"
}

@article{Hawking:1982dh,
    author = "Hawking, S. W. and Page, Don N.",
    title = "{Thermodynamics of Black Holes in anti-De Sitter Space}",
    reportNumber = "PRINT-83-0019 (CAMBRIDGE)",
    doi = "10.1007/BF01208266",
    journal = "Commun. Math. Phys.",
    volume = "87",
    pages = "577",
    year = "1983"
}

@article{Witten:1998zw,
    author = "Witten, Edward",
    editor = "Bergstrom, L. and Lindstrom, U.",
    title = "{Anti-de Sitter space, thermal phase transition, and confinement in gauge theories}",
    eprint = "hep-th/9803131",
    archivePrefix = "arXiv",
    reportNumber = "IASSNS-HEP-98-21",
    doi = "10.4310/ATMP.1998.v2.n3.a3",
    journal = "Adv. Theor. Math. Phys.",
    volume = "2",
    pages = "505--532",
    year = "1998"
}

@article{Kraus:2016nwo,
    author = "Kraus, Per and Maloney, Alexander",
    title = "{A cardy formula for three-point coefficients or how the black hole got its spots}",
    eprint = "1608.03284",
    archivePrefix = "arXiv",
    primaryClass = "hep-th",
    doi = "10.1007/JHEP05(2017)160",
    journal = "JHEP",
    volume = "05",
    pages = "160",
    year = "2017"
}

@article{PhysRevLett.120.200604,
  title = {Numerical Large Deviation Analysis of the Eigenstate Thermalization Hypothesis},
  author = {Yoshizawa, Toru and Iyoda, Eiki and Sagawa, Takahiro},
  journal = {Phys. Rev. Lett.},
  volume = {120},
  issue = {20},
  pages = {200604},
  numpages = {6},
  year = {2018},
  month = {May},
  publisher = {American Physical Society},
  doi = {10.1103/PhysRevLett.120.200604},
  url = {https://link.aps.org/doi/10.1103/PhysRevLett.120.200604}
}

@article{Kleban:2004rx,
    author = "Kleban, M. and Porrati, M. and Rabadan, R.",
    title = "{Poincare recurrences and topological diversity}",
    eprint = "hep-th/0407192",
    archivePrefix = "arXiv",
    doi = "10.1088/1126-6708/2004/10/030",
    journal = "JHEP",
    volume = "10",
    pages = "030",
    year = "2004"
}

@article{Barbon:2014rma,
    author = "Barbon, Jose L. F. and Rabinovici, Eliezer",
    title = "{Geometry And Quantum Noise}",
    eprint = "1404.7085",
    archivePrefix = "arXiv",
    primaryClass = "hep-th",
    reportNumber = "IFT-UAM-CSIC-14-034",
    doi = "10.1002/prop.201400044",
    journal = "Fortsch. Phys.",
    volume = "62",
    pages = "626--646",
    year = "2014"
}

@article{Barbon:2003aq,
    author = "Barbon, J. L. F. and Rabinovici, E.",
    title = "{Very long time scales and black hole thermal equilibrium}",
    eprint = "hep-th/0308063",
    archivePrefix = "arXiv",
    reportNumber = "CERN-TH-2003-167, RI-03-07-007",
    doi = "10.1088/1126-6708/2003/11/047",
    journal = "JHEP",
    volume = "11",
    pages = "047",
    year = "2003"
}

@article{Hayden:2007cs,
    author = "Hayden, Patrick and Preskill, John",
    title = "{Black holes as mirrors: Quantum information in random subsystems}",
    eprint = "0708.4025",
    archivePrefix = "arXiv",
    primaryClass = "hep-th",
    reportNumber = "CALT-68-2659",
    doi = "10.1088/1126-6708/2007/09/120",
    journal = "JHEP",
    volume = "09",
    pages = "120",
    year = "2007"
}

@article{Sekino:2008he,
    author = "Sekino, Yasuhiro and Susskind, Leonard",
    title = "{Fast Scramblers}",
    eprint = "0808.2096",
    archivePrefix = "arXiv",
    primaryClass = "hep-th",
    reportNumber = "SU-ITP-08-18, OIQP-08-08",
    doi = "10.1088/1126-6708/2008/10/065",
    journal = "JHEP",
    volume = "10",
    pages = "065",
    year = "2008"
}

@article{Shenker:2013pqa,
    author = "Shenker, Stephen H. and Stanford, Douglas",
    title = "{Black holes and the butterfly effect}",
    eprint = "1306.0622",
    archivePrefix = "arXiv",
    primaryClass = "hep-th",
    reportNumber = "SU-ITP-13-08",
    doi = "10.1007/JHEP03(2014)067",
    journal = "JHEP",
    volume = "03",
    pages = "067",
    year = "2014"
}

@article{Maldacena:2015waa,
    author = "Maldacena, Juan and Shenker, Stephen H. and Stanford, Douglas",
    title = "{A bound on chaos}",
    eprint = "1503.01409",
    archivePrefix = "arXiv",
    primaryClass = "hep-th",
    doi = "10.1007/JHEP08(2016)106",
    journal = "JHEP",
    volume = "08",
    pages = "106",
    year = "2016"
}

@article{Cotler:2016fpe,
    author = "Cotler, Jordan S. and Gur-Ari, Guy and Hanada, Masanori and Polchinski, Joseph and Saad, Phil and Shenker, Stephen H. and Stanford, Douglas and Streicher, Alexandre and Tezuka, Masaki",
    title = "{Black Holes and Random Matrices}",
    eprint = "1611.04650",
    archivePrefix = "arXiv",
    primaryClass = "hep-th",
    reportNumber = "SU-ITP-16-19, SU-ITP-16/19, YITP-16-124",
    doi = "10.1007/JHEP05(2017)118",
    journal = "JHEP",
    volume = "05",
    pages = "118",
    year = "2017",
    note = "[Erratum: JHEP 09, 002 (2018)]"
}

@article{Chesler:2008hg,
    author = "Chesler, Paul M. and Yaffe, Laurence G.",
    title = "{Horizon formation and far-from-equilibrium isotropization in supersymmetric Yang-Mills plasma}",
    eprint = "0812.2053",
    archivePrefix = "arXiv",
    primaryClass = "hep-th",
    doi = "10.1103/PhysRevLett.102.211601",
    journal = "Phys. Rev. Lett.",
    volume = "102",
    pages = "211601",
    year = "2009"
}

@article{Balasubramanian:2011ur,
    author = "Balasubramanian, V. and Bernamonti, A. and de Boer, J. and Copland, N. and Craps, B. and Keski-Vakkuri, E. and Muller, B. and Schafer, A. and Shigemori, M. and Staessens, W.",
    title = "{Holographic Thermalization}",
    eprint = "1103.2683",
    archivePrefix = "arXiv",
    primaryClass = "hep-th",
    reportNumber = "HIP-2011-07-TH, UUITP-06-11",
    doi = "10.1103/PhysRevD.84.026010",
    journal = "Phys. Rev. D",
    volume = "84",
    pages = "026010",
    year = "2011"
}

@article{Abajo-Arrastia:2010ajo,
    author = "Abajo-Arrastia, Javier and Aparicio, Joao and Lopez, Esperanza",
    title = "{Holographic Evolution of Entanglement Entropy}",
    eprint = "1006.4090",
    archivePrefix = "arXiv",
    primaryClass = "hep-th",
    doi = "10.1007/JHEP11(2010)149",
    journal = "JHEP",
    volume = "11",
    pages = "149",
    year = "2010"
}

@article{Aparicio:2011zy,
    author = "Aparicio, Joao and Lopez, Esperanza",
    title = "{Evolution of Two-Point Functions from Holography}",
    eprint = "1109.3571",
    archivePrefix = "arXiv",
    primaryClass = "hep-th",
    reportNumber = "IFT-UAM-CSIC-11-59",
    doi = "10.1007/JHEP12(2011)082",
    journal = "JHEP",
    volume = "12",
    pages = "082",
    year = "2011"
}

@article{Bhattacharyya:2009uu,
    author = "Bhattacharyya, Sayantani and Minwalla, Shiraz",
    title = "{Weak Field Black Hole Formation in Asymptotically AdS Spacetimes}",
    eprint = "0904.0464",
    archivePrefix = "arXiv",
    primaryClass = "hep-th",
    reportNumber = "TIFR-TH-09-10",
    doi = "10.1088/1126-6708/2009/09/034",
    journal = "JHEP",
    volume = "09",
    pages = "034",
    year = "2009"
}

@article{Hartman:2013qma,
    author = "Hartman, Thomas and Maldacena, Juan",
    title = "{Time Evolution of Entanglement Entropy from Black Hole Interiors}",
    eprint = "1303.1080",
    archivePrefix = "arXiv",
    primaryClass = "hep-th",
    doi = "10.1007/JHEP05(2013)014",
    journal = "JHEP",
    volume = "05",
    pages = "014",
    year = "2013"
}

@article{Liu:2013iza,
    author = "Liu, Hong and Suh, S. Josephine",
    title = "{Entanglement Tsunami: Universal Scaling in Holographic Thermalization}",
    eprint = "1305.7244",
    archivePrefix = "arXiv",
    primaryClass = "hep-th",
    reportNumber = "MIT-CTP-4475",
    doi = "10.1103/PhysRevLett.112.011601",
    journal = "Phys. Rev. Lett.",
    volume = "112",
    pages = "011601",
    year = "2014"
}

@article{Anous:2016kss,
    author = "Anous, Tarek and Hartman, Thomas and Rovai, Antonin and Sonner, Julian",
    title = "{Black Hole Collapse in the 1/c Expansion}",
    eprint = "1603.04856",
    archivePrefix = "arXiv",
    primaryClass = "hep-th",
    doi = "10.1007/JHEP07(2016)123",
    journal = "JHEP",
    volume = "07",
    pages = "123",
    year = "2016"
}

@book{Concentration_Inequalities:A_Nonasymptotic_Theory_of_Independence,
    author = {Boucheron, Stéphane and Lugosi, Gábor and Massart, Pascal},
    title = "{Concentration Inequalities: A Nonasymptotic Theory of Independence}",
    publisher = {Oxford University Press},
    year = {2013},
    month = {02},
    isbn = {9780199535255},
    doi = {10.1093/acprof:oso/9780199535255.001.0001},
    url = {https://doi.org/10.1093/acprof:oso/9780199535255.001.0001},
}

@book{Vershynin_2018, place={Cambridge}, series={Cambridge Series in Statistical and Probabilistic Mathematics}, title={High-Dimensional Probability: An Introduction with Applications in Data Science}, publisher={Cambridge University Press}, author={Vershynin, Roman}, year={2018}, collection={Cambridge Series in Statistical and Probabilistic Mathematics}}

@misc{Simmons-Duffin:2019TASI,
        author  = {David Simmons-Duffin},
        title   = {TASI Lectures on Conformal Field Theory
 in Lorentzian Signature},
        howpublished = "\url{https://www.desy.de/~bargheer/string-journal-club/presentations/2023-05-16_Sebastian-Harris_Simmons-Duffin:_TASI-Lorentzian-CFT.pdf/}",
        year    = "2019"
}

@misc{incomplete_gamma_bound2,
        author  = {},
        title   = { },
        howpublished = "\url{https://dlmf.nist.gov/8.11}",
        year    = {   }
}
\end{document}